\documentclass[fleqn,usenatbib]{mnras}

\usepackage{newtxtext,newtxmath}

\usepackage[T1]{fontenc}
\usepackage{ae,aecompl}
\usepackage[dvipsnames]{xcolor}

\usepackage{graphicx}	
\usepackage{amsmath}	

\usepackage{amssymb}	
\usepackage{ulem}
\usepackage{verbatim}   

\title[The low-end of the BH mass function]{The low-end of the black hole mass function at cosmic dawn}

\author[Trinca et al.]{Alessandro Trinca$^{1,2,3}$\thanks{E-mail: alessandro.trinca@inaf.it}, Raffaella Schneider$^{1,2,3,5}$, Rosa Valiante$^{2,3}$, Luca Graziani$^{1,3,4}$,
\newauthor
Luca Zappacosta$^2$, Francesco Shankar$^6$
\\
$^{1}$Dipartimento di Fisica, ``Sapienza'' Universit$\grave{a}$ di Roma, Piazzale Aldo Moro 2, 00185 Roma, Italy \\
$^{2}$INAF/Osservatorio Astronomico di Roma, Via di Frascati 33, 00040 Monte Porzio Catone, Italy \\
$^{3}$INFN, Sezione Roma1, Dipartimento di Fisica, ``Sapienza'' Universit$\grave{a}$ di Roma, Piazzale Aldo Moro 2, 00185, Roma, Italy \\
$^{4}$INAF/Osservatorio Astrofisico di Arcetri, Largo E. Fermi 5, 50125 Firenze, Italy \\
$^{5}$Sapienza School for Advanced Studies, Viale Regina Elena 291, 00161 Roma, Italy\\
$^{6}$Department of Physics and Astronomy, University of Southampton, Highfield, SO17 1BJ, UK\\
}

\date{Accepted 2022 January 5. Received 2022 January 5; in original form 2021 July 31}

\pubyear{2022}

\begin{document}
\label{firstpage}
\pagerange{\pageref{firstpage}--\pageref{lastpage}}
\maketitle

\begin{abstract}

Understanding the formation and growth of supermassive black holes (SMBHs) at high redshift represents a major challenge for theoretical models.
In this work we investigate the early evolution of the first SMBHs by constraining their distribution in mass and luminosity at $z > 4$. In particular, we focus on the poorly explored low-mass end of the nuclear black hole (BH) distribution down to $z \simeq 4$, and explore its connection with the nature of the first BH seeds and the processes governing their mass growth. To this aim, we have developed CAT (Cosmic Archaeology Tool), a new semi-analytic model that describes the formation of the first stars and black holes in a self-consistent way and follows the co-evolution of nuclear BHs and their host galaxies for a representative population at $z > 4$. 
We find that current observational constraints favour models where the growth of BH seeds is Eddington limited and occurs at the Bondi-Hoyle-Lyttleton rate or where super-Eddington accretion occurs via a slim disk during gas rich galaxy mergers. The main difference between these two model variants lies at the low-end of the predicted mass and luminosity functions at $4 \le z \le 6$, where a clear gap appears in the first model, reflecting the stunted growth of {\it light} BH seeds formed as remnants of the first stars. Detecting this signature will be extremely challenging even for the future generation of space observatories, such as \textit{JWST}, \textit{Athena} and \textit{Lynx}.
\end{abstract}



\section{Introduction}

In the last decade hyper-luminous quasars ($L \geq 10^{47} \, \rm{erg/s}$) have been observed up to $z > 7$, suggesting that Super Massive BHs (SMBHs) with masses as high as ten billion solar have already formed when the Universe was less than $1$ Gyr old. The most massive detected system, \textsc{SDSS J0100+2802}, is a quasar observed at $z \sim 6.3$ \citep{wu2015} with an estimated mass of M$_{\rm BH} =(1.2 \pm 0.19) \, \times \, 10^{10} \rm \, M_{\odot}$, while the most distant quasars are ULAS J1342+0928 ($\rm{z} = 7.54$, \citealt{banados2018}) powered by a central BH with $\rm{M_{BH}} \sim 8 \times 10^{8} \, M_{\odot}$,  {\it P$\bar{o}$niu$\bar{a}$'ena} (J1007+2115) at $z = 7.52$ (M$_{\rm BH} = 1.5 \times 10^9 M_\odot$, \citealt{yang2020}) and the recently reported J0313-1806 at $z = 7.64$ (M$_{\rm BH} = 1.6 \times 10^9 M_\odot$, \citealt{wang2021}).

Explaining how these extreme objects form and grow during the first billion years of cosmic history still represents a major challenge for theoretical models (for thorough reviews see \citealt{volonteri2010,valiante2017,inayoshi2020}). 
Therefore, it is crucial to understand what is the nature of SMBH first progenitors (generally referred to as BH "seeds") and how these efficiently grow through gas accretion and mergers. 
Many possible scenarios for the formation of BH seeds have been proposed in the literature so far.

\textit{Light seeds}, with a BH mass $\sim 100 \, \rm{M_{\odot}}$, are supposed to be the remnants of the first generation of metal-free (Population III/Pop III) stars and are expected to form with masses ranging between a few 10s to a few 100s M$_\odot$, inside dark matter (DM) \textit{minihalos} ($M_{\rm halo} \sim 10^5 - 10^6 \, \rm{M_{\odot}}$) at very high redshifts ($z\geq 20$, see \citealt{bromm2013} for a review, and \citealt{hirano2014, hirano2015, hosokawa2016, sugimura2020} for more recent works).

Collisions between stars inside dense star clusters are instead supposed to give birth to \textit{intermediate mass} or \textit{medium-weight} BH seeds, with a mass $\sim 1000 \, \rm{M_{\odot}}$ \citep{omukai2008, devecchi2009, katz2015, sakurai2017, reinoso2018, reinoso2019, sassano2021}, although even more massive BH seeds could originate from merger episodes between stellar mass BHs (aided by strong gas inflows) inside similarly dense environments \citep{lupi2014, boco2020}. 

Recent studies have also suggested that very massive BHs ($\sim 10^5 - 10^6 \rm{M_{\odot}}$), usually referred to as \textit{heavy seeds}, can form via the direct collapse of a supermassive star. The above  mechanism is supposed to take place inside dark matter halos with virial temperatures $T_{\rm vir} > 10^4 \, \rm K$, where gas cooling and fragmentation is suppressed by their metal-poor composition and H$_2$ photo-dissociation caused by Lyman-Werner (LW) radiation \citep{omukai2001, bromm2003, wise2008, regan2009, hosokawa2012, latif2013, inayoshi2014, latif2016b, chon2016, becerra2018} or by dynamical heating \citep{wise2019, lupi2021} associated with strong gas inflows \citep{lodato2006, lodato2007, mayer2010, mayer2015, mayer2019a, haemmerle2019}.

Although the formation rate of heavy BH seeds is yet to be established (see \citealt{valiante2017, inayoshi2020} and references therein) and their growth efficiency might depend on the formation site inside the host galaxy \citep[e.g.][]{chon2021}, they represent one of the most promising formation scenarios to explain the existence of SMBHs at $z \geq 6$ without invoking super-Eddington accretion rates. It has been shown, in fact, that starting from a seed of $\rm{M_{BH}} \sim 10^5 \, \rm{M_{\odot}}$, a BH can reach the observed mass of high-redshift SMBHs through classical Eddington-limited growth \citep[][]{valiante2017, sassano2021}. Many studies also pointed out that less massive seeds would require persistent accretion of gas at the Eddington rate during all their existence to reach a billion solar masses in less than $\sim 1 \, \rm{Gyr}$ \citep{volonteri2010, madau2014a, banados2018}. 
Alternatively, intermittent gas accretion at super- or hyper-Eddington rates may efficiently grow lower mass seeds \citep{pezzulli2016, inayoshi2016, pezzulli2017b, pacucci2017, takeo2018}, provided that suitable conditions are met in their circum-nuclear regions at high-$z$ (see the discussion in \citealt{mayer2018} and references therein).

Despite the significant progresses made in their theoretical description, discriminating the nature of BH seeds by electromagnetic observations may be very challenging, for a number of reasons: BH seeds keep memory of their birth conditions and genetic origin as long as they live in isolation, accreting gas from their host galaxy. For seed progenitors of $z \sim 6 - 7$ SMBHs, these conditions are only met at $z > 10$ for $< 100 - 150$ Myr since their formation \citep{valiante2018statistics}.
In addition, \textit{light} BH seeds may be too faint to be detectable, even with upcoming facilities, at least in Eddington-limited growth scenarios \citep{valiante2018observability}. Hence, the only chance to constrain their nature would be to detect the gravitational waves emitted during their binary coalescence through third generation ground-based detectors, such as the Einstein Telescope\footnote{http://www.et-gw.eu/} \citep{valiante2020}.
Detecting the electromagnetic emission from rapidly growing \textit{heavy} BH seeds appears more promising, and photometric
selection techniques have been envisaged to help identify the more promising candidates \citep{pacucci2015, pacucci2016, natarajan2017, valiante2018observability}.

An alternative way to constrain the early evolution of black holes is by shedding light onto the low-mass end of the black hole mass
function at high-$z$ \citep{volonteri2008, volonteri2009, ricarte2018b, piana2021}, which is supposed to be very sensitive to the nature of BH seeds and their growth mode.

Several optical/Near-Infrared surveys such as the Sloan Digital Sky Survey (SDSS\footnote{https://www.sdss.org/}), the UKIRT Infrared Deep Sky Survey (UKIDSS\footnote{https://www.nottingham.ac.uk/astronomy/UDS/}), the DECam Legacy Survey (DECaLS\footnote{https://www.legacysurvey.org/decamls/}) and the Panoramic Survey Telescope and Rapid Response System (Pan-STARRS\footnote{https://panstarrs.stsci.edu/}) gave a fundamental contribution to the detection of luminous quasars at redshift $z \geq 6$, characterizing the bright end of their luminosity function (LF, \citealt{willott2010a, banados2016}). At very high redshift, however, these objects are rare and  presumably trace extremely biased regions of the Universe. 

Large-scale cosmological simulations predict that a much larger population of fainter AGNs, powered by less massive BHs, is assembling and growing together with their host galaxies at these cosmic epochs \citep{dimatteo2012, vogelsberger2014, schaye2015, feng2016}. This population has eluded direct detection until recent surveys, in particular the Subaru High-z Exploration of Low-Luminosity Quasars (SHELLQS), have started to explore fainter magnitudes, unveiling a large sample of low-luminosity quasars. These findings have raised the total number of observed AGNs at $z > 6$ above $200$, constraining for the first time the faint-end of their LF \citep{matsuoka2018, matsuoka2019}.

In recent years, a number of semi-analytical models
have been applied to investigate observational signatures of BH seeds and of their early
co-evolution with their host galaxies \citep{volonteri2008, somerville2008, devecchi2012, salvaterra2012, bin2013, Bonoli2014, valiante2016, valiante2018observability, pezzulli2016, pezzulli2017a, ricarte2018a, ricarte2018b, piana2021, dayal2020}. In these  models the evolution of the baryonic components of dark matter halos is described through physically- and/or observationally-motivated prescriptions and the growth of dark matter halos can be either generated analytically or extracted from numerical simulations \citep{baugh2006,delucia2019}.

In this work we investigate the mass function (MF) of super-massive black holes, and the corresponding AGNs LF, by following their redshift evolution from the epoch of BH seeds formation down to $z = 4$. In particular, we focus on the large population of low-mass faint systems that will be targeted by upcoming facilities, such as the \textit{James Webb Space Telescope} (JWST), the \textit{Advanced Telescope for High Energy Astrophysics} (ATHENA) and the \textit{Lynx X-ray Observatory}. We also investigate whether initial BH seed masses and their growth rates leave an imprint on the low-mass (faint) end of the MF (LF) that may potentially be used to discriminate among different scenarios. To this aim, we developed \textsc{cat} (Cosmic Archaeology Tool),  a semi-analytical model that describes structure formation in the first billion years of cosmic evolution following the hierarchical growth of dark matter halos, their stellar and gas content, and their nuclear black holes. \textsc{cat} can account for a wide range of halo masses: from the $10^6 \rm \, M_\odot$ mini-halos hosting the collapse of the first stars at $z = 20 - 30$, up to the largest galaxies with $M_h \sim 10^{12} - 10^{14} \, \rm M_\odot$ where the most powerful quasars at $z = 4 - 7$ are supposed to reside. \textsc{cat} can explore a statistics which is still prohibitively expensive for cosmological hydro-dynamical simulations \citep{springel2005, dimatteo2005, dimatteo2017, sijacki2007, sijacki2015, degraf2012, degraf2015, hirschmann2014, mcalpine2017, weinberger2017, habouzit2019}, and can be achieved only through zoom-in techniques on a small number of systems pre-selected in very large lower-resolution simulations \citep{regan2019, lupi2019,zhu2020}. 

\textsc{cat} builds on our semi-analytical model \textsc{gameteQSOdust} (\textsc{gqd}), that was successfully applied to study the co-evolution of SMBHs and their host galaxies at $z \geq 6$ \citep{valiante2011, valiante2014, valiante2016}. In the present version, \textsc{cat} enables to follow the formation of both \textit{light} and \textit{heavy} BH seeds depending on the environmental properties and to explore their contribution to the BH MF (AGNs LF) across a wide range of redshifts, mass (luminosity) scales and physical parameters, quantifying the conditions for seed formation and their mass growth rate. This kind of investigation is certainly beyond the modelling capabilities of current large-scale cosmological hydrodynamic simulations, which usually adopt simplified prescriptions for BH seeding and assume a fixed BH seed mass that is planted in dark matter halos above a given threshold mass, independently of their internal properties \citep{dimatteo2012, vogelsberger2014, schaye2015, khandai2015, feng2016}.
More physically sound BH seeding prescriptions have been adopted in smaller-scale or zoom-in simulations \citep{bellovary2011, habouzit2017, tremmel2017, huang2020}, but at the price of being unable to simultaneously
explore the low- and high-mass (luminosity) ends of the BH MF (LF). 

To explore how different high-$z$ formation scenarios leave their imprints on the low-mass (luminosity) end of the BH MF (LF) at $z \geq 4$, we have run a large set of simulations on a grid of halo merger histories extending over several orders of magnitude in mass at $z = 4$ using the galaxy formation model \textsc{galform} \citep{cole2000, parkinson2008}. 
Our model predictions are then compared with the observed properties of galaxies and AGNs in view of future observations which could shed some light in discriminating among different BH seeding and growth scenarios.

The paper is organized as follows. In Section 2 we introduce and describe the model while in Section 3 we show how we calibrate the model to set the free parameters that govern star formation and black hole growth. The results are presented in Section 4, where we also compare our findings with the most recent observations and theoretical models. Finally, in Section 5 we discuss and summarize our main results.

\section{The Cosmic Archaeology Tool}
\label{sec:model}
In this section we illustrate the Cosmic Archaeology Tool (\textsc{cat}) adopted in the present work. 
First, we present the galaxy formation model \textsc{galform} and how its
dark matter halo merger tree algorithm  was properly 
adapted to generate the sample of dark matter halos. Second, we describe the main features of \textsc{gqd} that has been imported in \textsc{cat} in order to follow the evolution of halo baryonic components (gas, stars, and nuclear black holes), along with the major improvements we introduced in the present work.

In what follows we assume a Lambda cold dark matter ($\Lambda \rm{CDM}$) cosmological model with the following parameters: $\Omega_\Lambda = 0.685 $, $\Omega_{\rm m} = 0.315$, $h = 0.674$,  $\Omega_{\rm b} = 0.05$ \citep{planck2018} so that the age of the Universe at the final redshift $z = 4$ is $t_{\rm H} \sim 1.53 \, \rm{Gyr}$.

\subsection{Halo Merger Trees}
\textsc{galform} is a semi-analytic model of galaxy formation that reconstructs the hierarchical merger history of a given dark matter halo, also referred to as halo merger tree. We adopted the improved Monte Carlo algorithm developed by \citet{parkinson2008}, based on the Extended Press Schechter theory (EPS) and properly tuned to obtain an accurate agreement with N-body simulations, in particular with the results of the Millennium Simulation (MS, \citealt{springel2005}).
Here we briefly summarize the merger trees reconstruction algorithm. 

Starting from a target halo with a given mass at redshift $z_0$, \textsc{galform} follows its evolution backward in time reconstructing its progenitors. The key point of this process is the conditional mass function given by the EPS theory \citep{cole2000}:

\begin{eqnarray}
\lefteqn{\displaystyle{
f(M_1 \vert M_2)\, d\ln M_1 = 
\sqrt{\frac{2}{\pi}} \,
\frac{\sigma_1^2
  (\delta_1-\delta_2)}{[\sigma_1^2-\sigma_2^2]^{3/2}} } \, \times} &&
\nonumber \\
&&\displaystyle{ \exp\left[ - \frac{1}{2} \frac{(\delta_1-\delta_2)^2}{(\sigma_1^2-\sigma_2^2)}\right]
\left\vert \frac{d\ln\sigma}{d\ln M_1} \right\vert \,
d\ln M_1} ,
\label{eq:cmff}
\end{eqnarray}
where $f(M_1 \vert M_2)$ represents the fraction of mass of halos of mass $M_2$ at redshift $z_2$ that is contained in progenitor halos of mass $M_1$ at an earlier redshift $z_1$. The values $\delta_1$ and~$\delta_2$ are instead the linear density thresholds for collapse at these two redshifts and $\sigma(M)$ represents the rms linear density fluctuation in spheres containing a mass $M$, extrapolated to $z=0$, with $\sigma_{1/2} \equiv \sigma(M_{1 /2})$. 
Starting from Eq. \eqref{eq:cmff} it is possible to obtain the mean number of halos of mass $M_1$ into which a halo of mass $M_2$ splits after a step up in redshift $dz_1$, that is
\begin{equation}
\frac{dN}{d M_1} =  \frac{1}{M_1} \ \frac{df}{dz_1} \frac{M_2}{M_1} dz_1
\qquad (M_1 < M_2) .
\label{eq:dnp}
\end{equation}
Hence, the halo is decomposed into its progenitors and the process is repeated on each new halo at previous redshift steps up to a final value $z_{\rm max}$, building up a complete tree.
Although the above algorithm produces merger trees with statistical properties in good agreement with those obtained through detailed N-body simulations, it should be noted that the classical EPS theory systematically underestimates the mass of the most massive progenitor halos with increasing redshift \citep{cole2008}. For this reason a perturbing function $G(\sigma_1 / \sigma_2, \, \delta_2/\sigma_2)$  was introduced, leading to a modification of Eq. \eqref{eq:dnp} as follows:
\begin{equation}
\frac{dN}{d M_1} \rightarrow \frac{dN}{d M_1}G(\sigma_1 / \sigma_2, \, \delta_2/\sigma_2), 
\label{eq:dnp_n}
\end{equation}
where 
\begin{equation}
G(\sigma_1 / \sigma_2, \, \delta_2/\sigma_2) = G_o \, \biggl( \frac{\sigma_1}{\sigma_2} \biggl)^{\gamma_1} \, \biggl( \frac{\delta_2}{\sigma_2} \biggl)^{\gamma_2} .
\end{equation}
$G_0$, $\gamma_1$ and $\gamma_2$ are free parameter calibrated to reproduce the MS conditional mass function.

We used the \textsc{galform} algorithm to simulate the formation histories of DM halos with masses in the range $[10^9 - 10^{14}] \, \rm{M_{\odot}}$ at $z_{\rm min} = 4$. 
This allows us to explore both the low-mass end of halo population and halos as large as 
$M_{\rm halo} \sim 10^{13} \, \rm{M_{\odot}}$ at $z \geq 6$, that are assumed to host the first SMBHs that power the observed highly luminous quasars \citep{fan2003,valiante2011,valiante2016}. 
We divide this mass interval into $11$ logarithmically spaced bins with size $0.5$. For each bin, we consider a final halo of mass equal to the central bin value and we use it as a starting point for the \textsc{galform} code to simulate $10$ independent halo merger trees.

Once the total merger tree sample has been generated, the resulting redshift dependent mass distributions of each mass bin are weighted according to the number density of DM halos at redshift $z = 4$, as given by the Sheth and Tormen mass function \citep[][]{sheth2001}. In this way, the normalized sample is representative of the halo population at redshifts $z \geq 4$.
It is important to stress that the limited number of merger tree realizations simulated for each final halo mass might, in principle, limit the allowed variability in the evolution of the corresponding galaxy properties. This is especially true for the lowest final halo masses. By doubling the number of merger tree simulations we find that the redshift evolution of mean galaxy properties does not change,  showing that the current sampling scheme provides enough statistical variance even for the lowest halo mass bins.

\subsubsection{The Dark Matter Mass Resolution}
In the $\Lambda$CDM cosmological model, where larger structures form hierarchically through successive mergers of smaller ones, the first stars are expected to form inside the so-called minihalos, i.e. small dark matter halos with $M_{\rm halo} \sim 10^5 - 10^6 \, \rm{M_{\odot}}$, at redshift $z \sim 20 - 30$. Here we classify as minihalos systems with virial temperature in the range $1200 \, {\rm K} \leq T_{\rm vir} \leq 10^4 \, {\rm K}$,  where $T_{\rm vir}$ can be expressed as a function of redshift $z$ as \citep{bromm2013}:
\begin{equation}
T_{\rm vir} \simeq 2 \times 10^3 \, \rm{K} \, \biggl( \frac{M_{\rm halo}}{10^6 \, \rm{M_{\odot}}} \biggl)^{2/3} \, \biggl( \frac{1+z}{20} \biggl)   .
\label{eq:Tvir}
\end{equation}
\\
\noindent In these objects, where the gas temperature is below the threshold of $10^4 \, \rm{K}$ for efficient cooling due to atomic hydrogen, the pristine gas can still cool down and fragment through the roto-vibrational emission of molecular hydrogen ($\rm{H_2}$), giving birth to the first generation of Pop III stars. Due to their birth conditions, Pop III stars are expected to be massive and a large fraction of these
will terminate their life as a stellar-mass BH, providing the first \textit{light} BH seeds \citep{valiante2016}. Hence, any model
attempting to describe the build-up of SMBHs from growing BH seeds must be able to describe the formation of the first stars and BHs in high-$z$ minihalos. For this reason, while running \textsc{galform}, we
set a minimum DM halo mass of $M_{\rm res} = M_{\rm halo} \, ( T_{\rm vir} = 1200 \, \rm{K})$. We assumed $1200 \, \rm{K}$ as the minimum virial temperature for the onset of efficient $\rm H_{2}$ cooling following the work of \citet[][]{haiman1996}.\footnote{Note, however, that in our model not all mini-halos are equally efficient at forming stars as the fraction of available cold gas depends on the strength of the illuminating UV background, the redshift and the metallicity (see eq. \ref{eq:SFR}).}

\begin{figure}
\includegraphics[width=8.50cm]{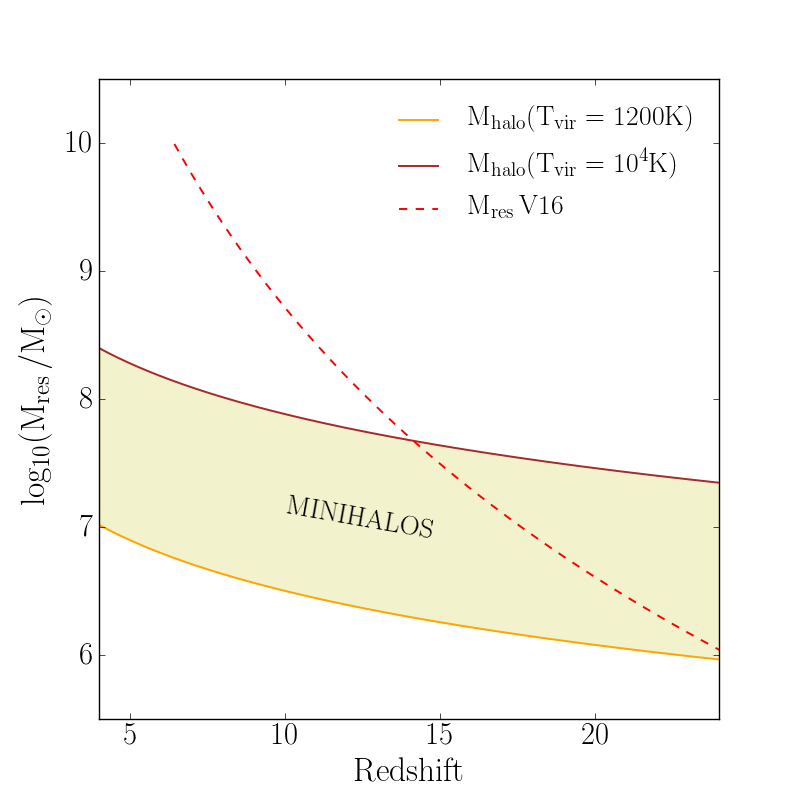}
\caption{The mass resolution adopted when generating \textsc{GALFORM} merger trees (orange solid line) is compared to the merger tree mass resolution of \citet[][red dashed line]{valiante2016}, where a final halo mass of $10^{13} \, \rm M_\odot$ at $z = 6.4$ is assumed. For comparison, we also show the redshift dependent minimum mass of atomic cooling halos (brown solid line) so that the yellow shaded region illustrates the masses of dark matter \textit{minihalos} with ${\rm  1200 \, K} \leq T_{\rm vir} \leq {\rm 10^4 \, K}$.}
\label{fig:Mres}
\end{figure}

In Fig. \ref{fig:Mres} the mass resolution as a function of redshift adopted in \textsc{cat} (solid orange line) is compared to the one adopted in \textsc{GQD} (red dashed line) for a final halo mass of $10^{13} \, \rm M_\odot$ at $z = 6.4$.
The solid brown line shows the redshift dependent minimum mass of \textit{atomic-cooling halos}, $M_{\rm halo} \, ( T_{\rm vir} = 10^4 \, \rm{K})$, so that the yellow shaded region represents the range of masses of the minihalos population. It is clear that the increased resolution of \textsc{cat} largely improves the statistics of minihalos with respect to \textsc{gqd} and allows us to follow their evolution all the way down to $z = 4$. We obtain maximal improvements especially in the description of the most massive halos since in \textsc{cat} the mass resolution at a given redshift is fixed, while in \textsc{GQD} it depends on the final halo mass.

\subsubsection{Multiple Mergers}
\label{sec:MultMerg}
In its original setup, \textsc{GALFORM} merger trees are generated according to an 
adaptive number of steps that ensures their binarity \citep{cole2000}. However, to
follow the baryonic evolution, we re-grid all the merger trees, according to \citet{valiante2016}, on $N_{\rm t} = 800$ time-steps logarithmically spaced in expansion factor between  $z = 24$ and $z = 4$, so that the time interval between two simulation snapshots is $\sim 0.5 \, \rm Myr$ at $z \sim 20$,
and $\sim 4 \, \rm Myr$  at $z \sim 4$. 
A consequence of this process is that the binary structure of the merger trees is no longer ensured since a given dark matter halo at a redshift step $dz_i$ can be formed as a result of the merger of more than two halos that were in place at a higher redshift $dz_{i+1}$, enabling the occurrence of what we call \textit{multiple mergers}.

It is common to define as \textit{major merger} between two DM halos a merger event where the mass ratio $\mu$ between the least and most massive halo is higher than a chosen threshold value. Because of the
loss of binarity, here we classify as \textit{major mergers} interactions where the mass ratio between the first and the second most massive halos among the merging ones is $\mu > 1/10$. This choice appears to be
conservative, possibly leading to an underestimation of the number of major mergers, due to the lack of information about the order of mergers in multiple interactions occurring in a single timestep.

\subsection{Baryonic evolution}
 In \textsc{cat}, the evolution of the baryonic component is governed by the same physical prescriptions adopted in \textsc{gqd}, that we briefly summarize below. 
We will focus on BH evolution and on the new features introduced in this work, referring the reader to \citet{valiante2014, valiante2016} for a more detailed description.

\textsc{gqd} was developed to investigate the formation history of the most extreme quasars observed at $z \geq 6$ and their host galaxies. To this aim, it follows the evolution of gas, stars, metals, and dust in each progenitor galaxy along a merger tree, tracking the process of star formation and the enrichment of the interstellar medium (ISM) due to Asymptotic Giant Branch (AGB) stars and Supernovae (SNe) by means of mass- and metallicity-dependent yields. The model follows a two-phase ISM  environment where dust grains can both be destroyed by SN shocks expanding in the diffuse hot medium and can grow in mass by accreting gas-phase metals in warm dense gas \citep[see][for details]{valiante2014, debennassuti2014}\footnote{The same physical prescriptions for metal and dust enrichment have
been adopted in semi-numerical models \citep{mancini2015, mancini2016} as well as in cosmological hydrodynamical simulations \citep{graziani2020} and provide good agreement with the observed dusty galaxy population at $z \geq 4$.}
Mechanical feedback due to SN explosions and energy deposition associated with BH growth is also considered. The energy released by these processes couples with the gas, eventually removing a significant fraction of the galactic reservoir though energy-driven galactic scale winds \citep{valiante2012}. 

So far, \textsc{gqd} has been applied to study the formation of single $z>6$ quasars, hosting $>10^9 \, M_\odot$ SMBHs, in association with the evolution of their host galaxies \citep{valiante2011, valiante2014}, BH seeds origin/properties \citep[][]{valiante2016,valiante2018statistics, valiante2018observability, sassano2021}, earliest binary BHs formation \citep{valiante2020} and different gas accretion regimes \citep[][]{pezzulli2016, pezzulli2017a, pezzulli2017b}. In the present work we are instead interested in studying a population of galaxies and their nuclear BHs across a broad range of masses and populating a less biased region of the Universe down to $z = 4$.

\subsubsection{Star formation and feedback}
In the same spirit of \textsc{gqd}, \textsc{cat} follows the evolution of the gas mass inside each galaxy during the whole DM halo assembly. Following halo virialization, the gas is accreted onto the newly collapsed halo and cools down to efficiently trigger star formation. The resulting fraction of gas mass is set by the balance between gas cooling and dynamical times. Inside each galaxy the star formation rate (SFR) is computed as:
\begin{equation}
{\rm SFR} = f_{\rm cool} \, M_{\rm gas} \, \epsilon_{\rm SF} / \tau_{\rm dyn},
\label{eq:SFR}
\end{equation}
where $M_{\rm gas}$ is the available gas mass reservoir, $\epsilon_{\rm SF}$ is the star formation efficiency per unit of time and $\tau_{\rm dyn} = [R_{\rm vir}^3 / (G \, M_{\rm halo}) ]^{1/2}$ is the dynamical time of the system. The SF efficiency $\epsilon_{\rm SF}$ represents a free parameter of the model and is calibrated as discussed in Section \ref{sec:Model Cal}. $f_{\rm cool}$ quantifies the reduced cooling efficiency in minihalos, where it depends on the halo virial temperature, redshift, gas metallicity and intensity of the illuminating LW radiation, as previously implemented by \citet{valiante2016, debennassuti2017}. Conversely, in atomic cooling halos we set $f_{\rm cool} = 1$. 
As described in \citet[][]{valiante2016}, we compute at each redshift $z_{\rm obs}$ the global LW cumulative background flux at the observed frequency $\nu_{\rm obs}$ as
\begin{equation}
J(\nu_{\rm obs},z_{\rm obs}) = \frac{(1+z_{\rm obs})^3}{4\pi} \int_{z_{\rm obs}}^{z_{\rm max}} dz \, c \, \Big| \frac{dt}{dz}\Big| \, \epsilon(\nu_{z},z) \,  e^{-\tau_{\rm H_2}(\nu_{\rm obs},z_{\rm obs},z)} 
\label{eq:Jlw}
\end{equation} 
where $\epsilon(\nu_{z},z)$ is the comoving emissivity in the LW band at redshift $z$, which is obtained summing over all the emitting sources, both stars and accreting BHs. In Eq. \ref{eq:Jlw}, $\tau_{\rm H_2}$ is the $\rm H_2$ optical depth in the LW band \citep[for a detailed description of its calculation see][]{valiante2016}, while $z_{\rm max}$ is the maximum redshift from which a LW photon emitted by a source at $z > z_{\rm obs}$ can reach the observer before being redshifted outside the LW band.

Finally, we account for the effects of photo-heating feedback by suppressing star formation in haloes with virial temperatures below the temperature of the intergalactic medium (IGM), $T_{\rm vir} < T_{\rm IGM}$. We consider $T_{\rm IGM} = Q_{\rm HII} \, T_{\rm reio} + (1 - Q_{\rm HII}) \, T_{\rm HI}$, where $T_{\rm reio} = 2 \times 10^4 \rm \, K$, $T_{\rm HI} = 0.017 (1+z)^2$ and the filling factor of HII regions, $Q_{\rm HII}$, is computed as in \citep{valiante2016}. 

The abundance of gas inside each galaxy is also affected by mechanical feedback associated with SN explosions and BH accretion, whose released energy drives massive outflows of gas out of the galaxy. The total gas ejection rate, $\dot{M}_{\rm ej}$ is described as:

\begin{equation}
\dot{M}_{\rm{ej}} = \dot{M}_{\rm{ej, SN}} + \dot{M}_{\rm{ej, AGN}} 
\label{eq:fbk}
\end{equation}
where $\dot{M}_{\rm ej,SN}$ and $\dot{M}_{\rm ej, AGN}$ are the SN- and AGN-driven outflow rates, respectively, defined as:
\begin{equation}
\dot{M}_{\rm{ej, SN}} = \frac{2 E_{\rm SN} \epsilon_{\rm w,SN} R_{\rm SN} (t)}{v_{\rm e}^2} ,
\label{eq:SNfbk}
\end{equation}
and
\begin{equation}
\dot{M}_{\rm{ej, AGN}} = 2 \, \epsilon_{\rm{w, AGN}} \, \epsilon_r \, \dot{M}_{\rm accr} \, \biggl( \frac{c}{v_{\rm e}} \biggl)^2.
\label{eq:AGNfbk}
\end{equation}
\noindent
In Eq.~\ref{eq:SNfbk} $R_{\rm SN} (t)$ is the SN explosion rate, which depends on the SF history and on the nature of the stellar populations hosted by each galaxy. For Pop III stars, we stochastically sample the Initial Mass Function (IMF) in each SF episode and $R_{\rm SN} (t)$ depends on the number of stars formed at each time in each galaxy. For Pop II stars, instead, we assume a fully sampled IMF and $R_{\rm SN} (t) = 1.25 x 10^{-2} \, M_{\odot}^{-1}$ (see Section \ref{sec:LightBHseeds} for more details). In Eq.~\ref{eq:SNfbk} $E_{\rm SN}$ represents the explosion energy per SN, assumed to be $2.7\times 10^{52}$\, erg for Pop III stars and $1.2\times 10^{51}$ erg for Pop II stars.

The terms $\, \dot{M}_{\rm accr}$ and $\epsilon_r$, in Eq.~\ref{eq:AGNfbk}, are instead the gas accretion rate and the AGN radiative efficiency, described in Section \ref{sec:BHaccretion}.
In both expressions, $v_{\rm e} = (2 G M/R_{\rm vir} )^{1/2}$ is the escape velocity of the galaxy while $\epsilon_{\rm w,SN}$ and $\epsilon_{\rm w,AGN}$ are free parameters representing the SN- and AGN-driven wind efficiencies, respectively. The adopted values are discussed in Section~\ref{sec:Model Cal}. 

The ejection of metal-enriched gas and dust due to SN explosions and AGN activity enriches the IGM, increasing its metallicity and dust content. This leads to a corresponding increase in the initial gas metallicity and dust-to-gas mass ratio of DM halos collapsing at later times, affecting their star formation history, as well as the formation of nuclear black holes, as will be outlined in the next section.

\subsection{Black Hole formation and evolution}

Supermassive black holes are supposed to grow via both gas accretion and mergers starting from their \textit{seeds}, less massive progenitors whose nature has a crucial role in understanding the origin of SMBHs. 

\subsubsection{Light BH Seeds}
\label{sec:LightBHseeds}
In \textsc{cat} the mass distribution of \textit{light BH seeds} depends on the Pop III stellar initial mass function (IMF), that is still highly uncertain  \citep{bromm2013}. According to the most recent numerical simulations of metal-free star forming regions hosted by minihalos at $z \sim 20 - 30$, the Pop III mass distribution ranges from a few 10s to a few 100s $\rm M_\odot$ \citep{greif2011, hirano2014, stacy2016, hosokawa2016, sugimura2020}. Following \citet{valiante2016}, we assume that Pop III stars form according to a Larson IMF:
\begin{equation}
\Phi (m_*) \propto m_*^{\alpha -1} \, e^{-m_*/m_{\rm ch}}
\label{eq:IMF}
\end{equation}
where $\alpha = -1.35$, $m_{\rm ch} = 20\, M_\odot$ and the possible range of stellar mass is $10 \, M_\odot \leq m_* \leq 300\, M_\odot$. This choice is motivated by stellar archaeology studies 
and appears to best match the observed Galactic halo metallicity distribution function and the properties of C-enhanced and C-normal stars at [Fe/H] $< -3$ \citep{debennassuti2014, debennassuti2017}.

In our model, we stochastically sample the Pop III IMF untill we saturate the total stellar mass formed in each star formation episode. To consistently compute the BH remnants mass distribution, we assume that Pop III stars with masses in the range $[40 - 140] \, \rm{M_\odot}$ and $[260 - 300] \, M_\odot$ collapse directly to BHs of comparable mass \citep{heger2002}. Since these \textit{light BH seeds} are expected to wander through the host galaxy, it is very unlikely that they will undergo mergers \citep{volonteri2010}, unless they form in binary systems \citep{sugimura2020}. Moreover, dynamical effects like three-body scattering or gravitational recoil following BH mergers could lead to the ejection of the merging objects from the host galaxy \citep{campanelli2007}, especially inside smaller dark matter halos \citep{dunn2020}. For this reason, here we assume that only the most massive BH settles in the center of the galaxy and is considered as its \textit{light BH seed}.

Pop III star formation can be sustained until the gas metallicity of the star forming region  remains below a critical value $\rm{Z \leq Z_{cr}}$, where we assume $\rm{Z_{cr}} = 10^{-3.8} \, \rm{Z_\odot}$ \citep{valiante2016}. Above this threshold value, metal-fine structure lines and dust cooling increase the cooling efficiency \citep{omukai2001, schneider2002, omukai2005, schneider2006, schneider2012b}, leading to a transition in the characteristic stellar masses. We therefore assume that above $\rm{Z_{cr}}$, Pop II stars form in the mass range $0.1 \, M_\odot \leq m_* \leq 100\, M_\odot$ according to a Larson IMF with $m_{\rm ch} = 0.35\, M_\odot$ \citep{debennassuti2014, debennassuti2017}.

\subsubsection{Heavy BH Seeds}
The second viable scenario for BH seed formation implemented in  \textsc{cat} is the so called Direct Collapse (DC) mechanism. Inside atomic-cooling halos (where $T_{\rm vir} \geq 10^4 \, \rm{K}$), where metal and dust cooling is still inefficient ($\rm{Z} \leq \rm{Z_{cr}}$), if the abundance of molecular hydrogen is suppressed by LW photons ($11.2 - 13.6 \, \rm{eV}$) inducing H$_2$ photo-dissociation, the gas collapses almost iso-thermally with no fragmentation, leading to the formation of a single super massive star that becomes unstable, due to nuclear exhaustion or GR instabilities \citep{hosokawa2012, inayoshi2014}, and forms a \textit{heavy BH seed}, with mass in the range $[10^4 - 10^6] \, \rm{M_\odot}$ \citep{latif2013, ferrara2014, becerra2015, latif2016a, becerra2018}.

The importance of \textit{heavy BH seeds} for the formation of high redshift SMBHs strongly depends on their birth rate that is still subject to large uncertainties \citep{inayoshi2020}. 
If one neglects the effects of dynamical heating associated with structure formation \citep{wise2019} or with major mergers \citep{mayer2010}, the abundance of \textit{heavy BH seeds} in the family tree of SMBHs depends on the adopted value of $\rm{Z_{cr}}$ and on the critical value of the LW flux ($\rm{J_{cr}}$) above which H$_2$ remains photo-dissociated. The latter condition is usually expressed as $\rm{J_{LW}} \geq \rm{J_{cr}}$, where $\rm{J_{LW}}$ is the cumulative flux into the LW energy band in units of $10^{-21} \, \rm{ erg \, s^{-1} \, cm^{-2} \, Hz^{-1} \,sr^{-1}}$. The value of $\rm{J_{cr}}$ is still very uncertain and depends on (i) the total spectral energy distribution (SED) of the radiation background created by the various sources \citep{agarwal2015}, (ii) the efficiency of H$_2$ self-shielding, and (iii) the increase of the free electron fraction due to the presence of intense ionizing radiation
which increases the H$_2$ formation rate \citep{inayoshi2015suppression}. As a result, values of $\rm{J_{cr}}$ ranging in a wide interval between $\sim 30$ and $\sim 10^4$ have been proposed in the literature (see \citealt{woods2019} and \citealt{inayoshi2020} and references therein). In addition, it has been recently suggested that the strong gas accretion rates may favour super-massive star formation event at higher metallicities than usually assumed \citep{chon2020}, through the so-called super-competitive accretion scenario. This has been shown to increase the number of \textit{heavy BH seeds} by a factor ranging from 2 \citep[][]{sassano2021} to 4 \citep{regan2020}.

Following \citet{valiante2016}, here we adopt as threshold values $\rm{Z_{cr}} =10^{-3.8} \, \rm{Z_\odot}$ and $\rm{J_{cr}} = 300$ to identify regions in atomic-cooling halos where 
\textit{heavy BH seeds} can form. If the conditions $\rm Z < Z_{cr}$ and $\rm J_{LW} \geq J_{cr}$ 
are satisfied, we set in the center of the galaxy a \textit{heavy BH seed} with a mass of $10^5 \, \rm{M_\odot}$.
 
\subsubsection{Black Hole Mergers}
\label{sec:BH_mergers}
Once formed, BH seeds are expected to grow via both gas accretion and coalescences with other BHs, eventually forming the SMBHs that power high redshift AGNs \citep[see][for complete reviews]{volonteri2010, Johnson16, Inayoshi2019}.

Following \citet{valiante2011} we assume that two BHs coalesce only during major halo-halo mergers, i.e. if the mass ratio of their interacting host DM halos is $\mu > 1/10$ \citep[as defined in Section \ref{sec:MultMerg};][]{tanaka2009}. In our model, both the host galaxies and their nuclear BHs merge within the characteristic time interval of the simulation ($\Delta t \sim 0.5 - 4 \, \rm Myrs$) and the merger product settles in the nuclear region of the final galaxy. Conversely, in minor mergers ($\mu < 1/10$), only the most massive BH is assumed to migrate in the center of the newly formed galaxy.
The least massive one is instead considered as a satellite, wandering in the outskirts of the main galaxy \citep[e.g.][]{callegari2009, tamfal2018}, and its subsequent evolution is no longer followed within the model.

Although oversimplified, our assumption is based on the common expectation that Keplerian BH binaries form promptly in interacting galaxies of "similar" mass and shrink to sub-pc separations (comparable to the primary BH influence radius) on relatively short timescales \citep[within about a million years in the most optimistic cases; e.g.][]{mayer2007, tanaka2009}. We will return to this point in Section \ref{sec:discussion}.

\subsubsection{Black Hole Accretion}
\label{sec:BHaccretion}
Nuclear BHs are assumed to grow by accreting gas from the surrounding medium. The growth is regulated by the processes of star formation and mechanical feedback, which both lead to a depletion of gas inside the host galaxy. Following the original \textsc{gqd} model, in our {\it reference} model we assume that nuclear BHs accrete gas according to the Bondi-Hoyle-Lyttleton (BHL) accretion rate \citep{hoyle1941, bondi1952}, given by:
\begin{equation}
\dot{M}_{\rm BHL} =  \alpha \frac{4 \pi G^2  M_{\rm BH}^2 \rho_{\rm gas}(r_A)}{c_s^3} \,\, .
\label{eq:BHL}
\end{equation}
In the above equation $c_s$ is the sound speed, which is estimated assuming a gas temperature $T_{\rm gas} = T_{\rm vir}$, and $\rho_{\rm gas}(r_A)$ is the gas density evaluated at the Bondi radius, i.e. the radius of gravitational influence of the black hole, $r_A = 2G M_{\rm BH}/c_s^2$. Following \citet{valiante2011}, the gas density distribution is approximated as a singular isothermal sphere with a flat core:
\begin{equation}
\rho_{\rm gas}(r) = \frac{\rho_{\rm norm}}{1 + ( r/r_{\rm core})^2}
\label{eq:rho_prof}
\end{equation}
where $r_{\rm core} = 0.012 \, R_{\rm vir}$ and $\rho_{\rm norm}$ represents a normalization constant that ensures that, at each time step, the gas is distributed within the halo virial radius.

The parameter $\alpha$ in Eq. \ref{eq:BHL} does not appear in the original BHL formula and it is usually introduced in numerical simulation as a correction factor to take into account the enhanced gas density in the inner regions around the central BH. In fact, due to the lack of resolution of the simulations, the actual BHL accretion rate tends to be strongly underestimated \citep{dimatteo2012, schaye2015}. As will be discussed in Section \ref{sec:Model Cal}, the $\alpha$ parameter is a free parameter of the model\footnote{Note that since the Bondi radius is always much smaller than the core radius, the parameter $\alpha$ quantifies the unknown gas density enhancement around the black hole.}. 

In our {\it reference} model, we assume that the BH accretion rate, $\dot{M}_{\rm BH}$, can not exceed the Eddington limit,
so that:
\begin{equation}
\dot{M}_{\rm BH} = (1-\epsilon_r) \, \dot{M}_{\rm accr} = {\rm min}(\dot{M}_{\rm BHL}, \dot{M}_{\rm Edd}), 
\end{equation}
\noindent
where:
\begin{equation}
\dot{M}_{\rm Edd} =  \frac{L_{\rm Edd}}{\epsilon_{\rm r} \, c^2},
\label{eq:Edd_rate}
\end{equation}
$\epsilon_{\rm r}$ is the radiative efficiency, i.e. the efficiency at which the accreting gas is converted into radiated luminosity,
\begin{equation}
L_{\rm Edd} =  \frac{4 \pi c G M_{\rm BH} \, m_p}{\sigma_{\rm T}}
\end{equation}
is the Eddington luminosity, $c$ is the speed of light, and $\sigma_{\rm T}$ is the Thomson cross-section. Here we assume that $\epsilon_{\rm r} = 0.1$ \citep{shakura1973}, unless otherwise specified.
The bolometric luminosity of the accreting BH can be expressed as:
\begin{equation}
L_{\rm bol} = \epsilon_{\rm r}\, \dot{M}_{\rm accr} \, c^2.
\label{eq:erad}
\end{equation}
\noindent

\begin{table}
	\centering
	\caption{Adopted set of parameters characterizing our \textsc{cat} {\it reference} model and the {\it super-Edd} and {\it merger-driven} model variants (see Section \ref{sec:Model Cal}): the star formation efficiency $\epsilon_{\rm SF}$, the SN and AGN wind efficiencies $\epsilon_{\rm w, SN}$, $\epsilon_{\rm w, AGN}$ and the BH accretion parameter $\alpha$.}
	\label{tab:cal}
	\begin{tabular}{llccccr}
		\hline
		  Model & $\rm{\epsilon_{\rm SF}}$ & $\rm{\epsilon_{\rm w, SN}}$ & $\rm{\epsilon_{\rm w, AGN}}$ & $\rm{\alpha}$ & $\epsilon_{\rm r}$\\
		\hline
		 {\it reference} & 0.05 & 1.6 $\times$ $10^{-3}$ & 2.5 $\times$ $10^{-3}$ & 90 & 0.1\\
		 {\it super-Edd}  & 0.05 & 1.6 $\times$ $10^{-3}$ & 2.5 $\times$ $10^{-3}$ & 40 & Eq.\eqref{eq:slim_radeff} & \\
		 {\it merger-driven} & 0.05 & 1.6 $\times$ $10^{-3}$ & 2.5 $\times$ $10^{-3}$ & 1 & Eq.\eqref{eq:slim_radeff} & \\
		\hline
	\end{tabular}
\end{table}

\subsection{Model variants}
\label{sec:modelvariants}
As will be discussed in detail in Sections \ref{sec:results} and \ref{sec:discussion}, our {\it reference} model, presented in the previous Section, shows the best agreement with different independent theoretical and observational constraints in reproducing the BH mass and luminosity distribution at $z \gtrsim 4$. However, we also decided to explore two additional model variants, in order to understand how the evolution of the AGN population depends on the assumed accretion paradigm. In fact, these alternative models do not change the BH seeding prescriptions but consider different scenarios for their mass growth, as described below.

\subsubsection{Exceeding the Eddington limit}
\label{sec:Eddington limit}
In the \textsc{cat} {\it reference} model BHs are assumed to experience spherical accretion according to the Bondi-Hoyle-Lyttleton rate, with a maximum allowed limit at the Eddington rate (see \ref{sec:BHaccretion}). The BH mass dependence of both rates may represent a strong limitation to early BH growth, particularly in the case of \textit{light} BH seeds. 

Provided that the gas reservoir can be efficiently replenished, through large-scale accretion or galaxy mergers, BH seeds at high redshift may quickly reach accretion rates with Eddington ratios $\eta_{\rm Edd} \equiv \dot{M}_{\rm BHL}/\dot{M}_{\rm Edd} \rightarrow 1$. In order to check whether the restriction $\eta_{\rm Edd} \leq 1$ provides a limitation to \textit{light} BH seeds mass growth, we have explored an alternative model, that we dubbed {\it super-Edd}, where 
BHs are allowed to grow at super-Eddington rates (see \citealt{mayer2019b} and \citealt{inayoshi2020} for a
thorough presentation of the main super-critical accretion models applied to BH seeds growth). Here we adopt the optically-thick, geometrically slim disk solution developed by \citet{abramowicz1988}, where part of the generated heat remains trapped within the accreting flow and is advected into the BH, leading to a low radiative efficiency.
In this model variant, BHs accrete gas according to the BHL formula (see Eq. \ref{eq:BHL}), so that:
\begin{equation}
\dot{M}_{\rm BH} = \dot{M}_{\rm BHL} , 
\end{equation}
\noindent
but their bolometric luminosity is computed according to the fitting formula for the radiative efficiency proposed by \citet{madau2014a} which, in turn, is based on the numerical solution obtained by \citet{sadowski2009}:

\begin{equation}
\frac{L_{\rm bol}} {L_{\rm Edd}} = \ A(a) \left[ \frac{0.985}{\dot{M}_{\rm Edd}/\dot{M}_{\rm accr} + B(a)} + \frac{0.015}{\dot{M}_{\rm Edd}/\dot{M}_{\rm accr} + C(a)} \right]      .
\label{eq:SL}
\end{equation}
\noindent 
where $A(a)$, $B(a)$, $C(a)$ are three functions of the black hole spin parameter $a = a_{\rm BH}$

\begin{eqnarray*}
A(a) & = & (0.9663 - 0.9292 a)^{-0.5639} ,\\
B(a) & = & (4.627 - 4.445 a)^{-0.5524} ,\\
C(a) & = & (827.3 - 718.1 a)^{-0.7060}.
\end{eqnarray*} 
\noindent
Note that in the above expression, the Eddington rate is defined as $\dot{M}_{\rm Edd} \equiv 16 \, L_{\rm Edd}/c^2$, i.e. it is a factor $1.6$ larger than the definition given by Eq. \eqref{eq:Edd_rate}.
Hence, in this model variant, the radiative efficiency is computed as:
\begin{equation}
\epsilon_{r} = \frac{L_{\rm bol}}{\dot{M}_{\rm accr} c^2}.
\label{eq:slim_radeff}
\end{equation}
\noindent
To allow a better comparison with the {\it reference} model, the spin parameter has been assumed to be $a_{\rm BH} = 0.572$ for all the BHs. This ensures that in the limit $\dot{M}_{\rm accr} << \dot{M}_{\rm Edd}$ the radiative efficiency $\epsilon_r \rightarrow 0.1$, the same value adopted in the {\it reference} model.

With the above assumptions, large accretion rates up to $\dot{M}_{\rm accr} \sim 100 \, \dot{M}_{\rm Edd}$ lead to a radiated luminosity that remains only slightly super-Eddington, $L_{\rm bol} \leq 5 L_{\rm Edd}$. Therefore, this enables super-Eddington accretion while - at the same time - the effects of AGN feedback is still limited, reducing the energy injected in the surrounding medium, and favouring BH growth.

\subsubsection{Merger-driven black hole accretion}
\label{sec:mergerdriven}
As shown by \citet{pezzulli2016} and  \citet{pezzulli2017b}, super-Eddington accretion through a radiatively inefficient slim disk can be triggered by gas-rich galaxy mergers at high redshift. 
By implementing these prescriptions in \textsc{gqd}, it was found that - in highly biased regions, such as those that will later host bright quasars - episodic intense accretion is capable to grow $\sim 100 \, M_\odot$ \textit{light} BH seeds to masses $\geq 10^4 \, M_\odot$, comparable to the mass of \textit{heavy} BHs, by $z \sim 20$.
Since these biased regions are growing their mass at a faster pace, it is interesting to investigate the impact of this scenario on the general galaxy and AGN populations at $z \geq 4$.

Here we adopt a simplified description, that we refer to as {\it merger-driven} model, where we assume that, following a major merger, the gas inside the newborn galaxy is able to quickly loose angular momentum and the nuclear BH undergoes a period of 
enhanced accretion, with a rate given by: 
\begin{equation}
\dot{M}_{\rm BH} = \frac{\epsilon_{\rm BH} \, M_{\rm gas}}{\tau_{\rm accr}} 
\end{equation}
\noindent
where $M_{\rm gas}$ is the gas mass inside the newly formed galaxy, $\epsilon_{\rm BH}$ is the BH accretion efficiency and $\tau_{\rm accr}$ is the accretion timescale. We assumed an efficiency $\epsilon_{\rm BH} = 1/3 \, \, \epsilon_{\rm SF}$ and a constant value of $\tau_{\rm accr} = 10 \, \rm Myr$, that is comparable to the maximum dynamical timescale of the bulges found by \citet{pezzulli2016} in their simulations. To be consistent with this merger-driven gas inflow scenario, the SFR in the newly formed galaxy is also enhanced by:
\begin{equation}
{\rm SFR} = \frac{\epsilon_{\rm SF} \, M_{\rm gas}}{\tau_{\rm dyn, SF}} 
\end{equation}
\noindent
where we assume a fixed star formation timescale $\tau_{\rm dyn, SF} = \tau_{\rm accr}$. The enhancement in BH accretion and SF is assumed to terminate
when either the ratio between the gas and BH masses, $M_{\rm gas}/ M_{\rm BH}$, becomes lower than 10 or after a time interval $\Delta \rm t_{\,burst} = \tau_{\rm dyn}/100$. At that point, both star formation and BH accretion turn back to the quiescent mode, where the SFR is described by
Eq. \eqref{eq:SFR} and the BH accretion rate is described according to the BHL formula expressed by Eq. \eqref{eq:BHL}, but assuming $\alpha = 1$ (see Table \ref{tab:cal} for the set of assumptions and parameters characterizing this {\it merger-driven} growth model).

\begin{figure*}
\vspace{\baselineskip}
\includegraphics[width=8.50cm]{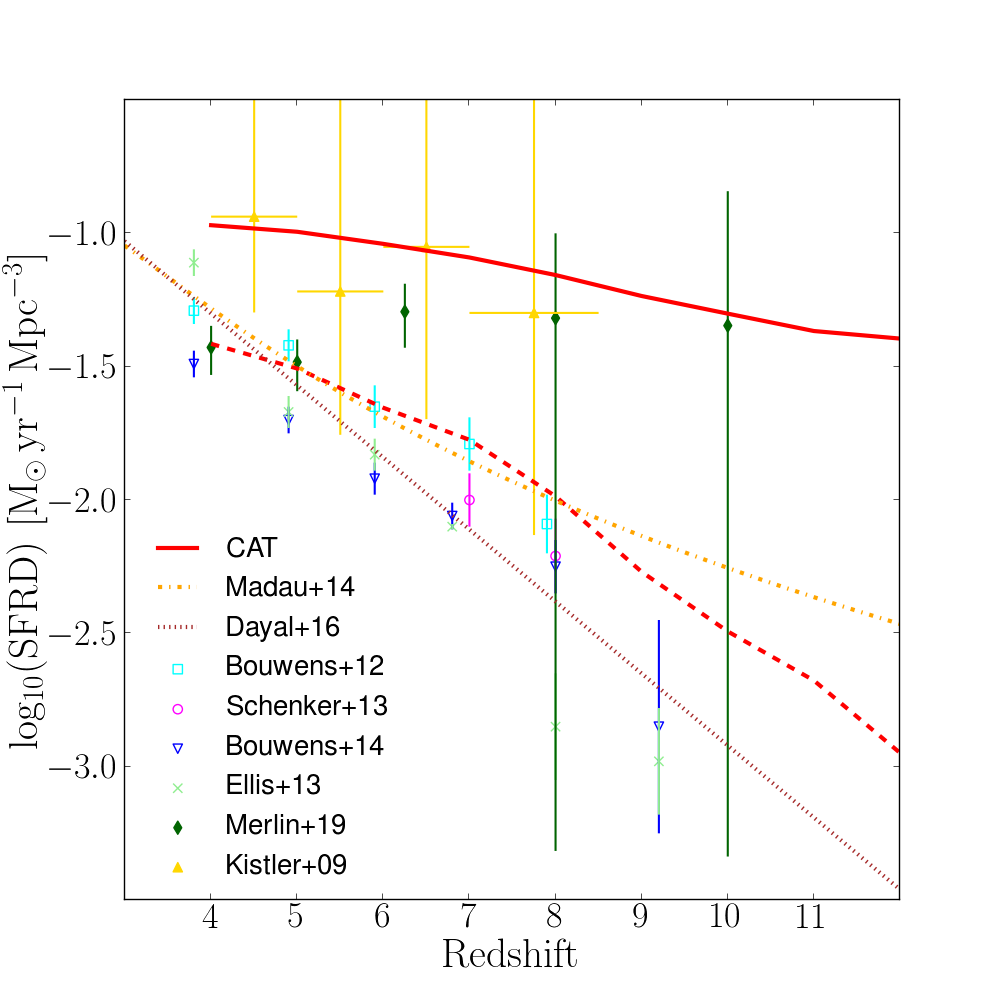}
\includegraphics[width=8.50cm]{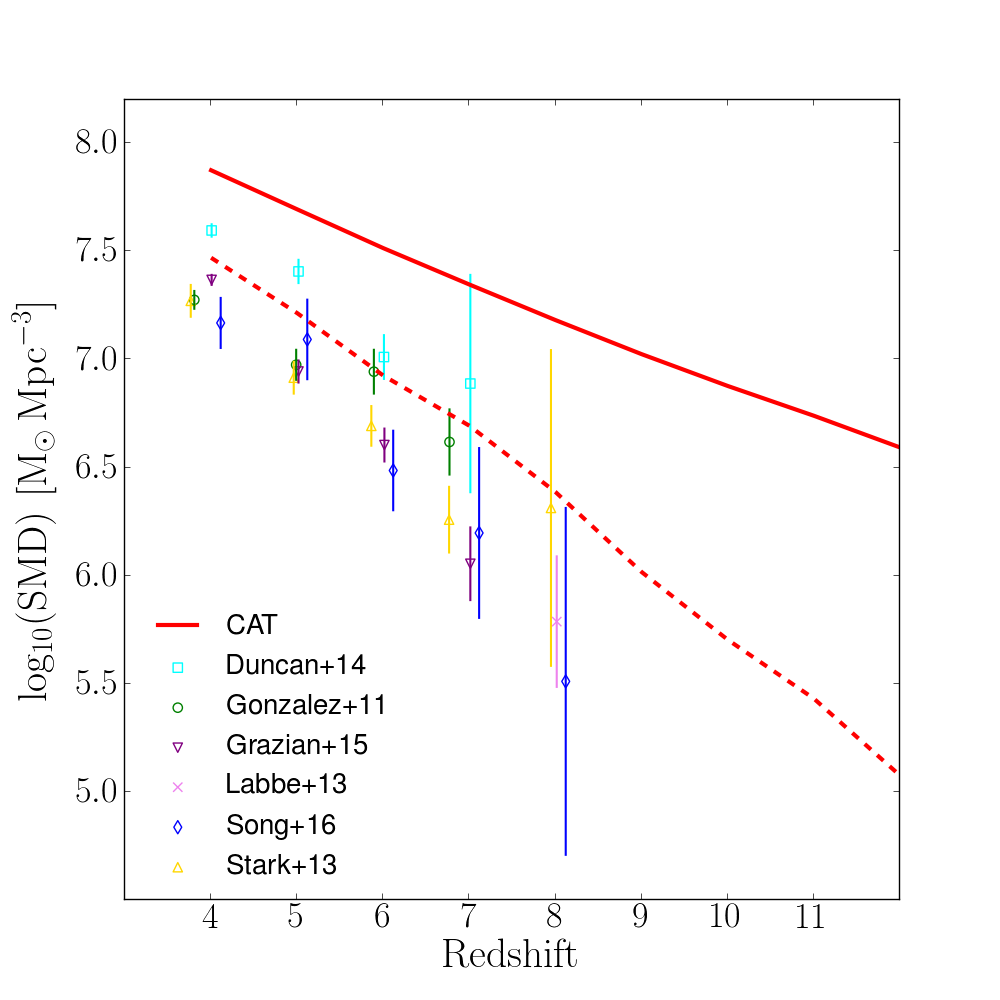}
\caption{Redshift evolution of the comoving SFRD (left panel) and stellar mass density (right panel) predicted by \textsc{cat}, when the entire galaxy population is considered (red solid lines) and when only galaxies with $M_{\rm UV} < - 17.7$ are accounted for (red dashed lines). In each panel, the model predictions are compared with different observational results, as indicated in the legenda:  \citet{kistler2009}, \citet{gonzalez2011},  \citet{bouwens2012}, \citet{labbe2013},  \citet{stark2013}, \citet{schenker2013},  \citet{ellis2013}, \citet{duncan2014}, \citet{bouwens2014}, \citet{oesch2014}, \citet{grazian2015},  \citet{song2016}, \citet{merlin2019}. In the right panel, the yellow dashed-dotted line shows the empirical SFRD by \citet{madau2014b} while the brown dotted line shows the model prediction by \citet{dayal2016}.}
\label{fig:SFRD}
\end{figure*}

\section{Model Calibration}
\label{sec:Model Cal}
As discussed in Section \ref{sec:model} our model presents four parameters which are tuned to regulate the evolution of the baryonic component: the star formation efficiency ($\epsilon_{\rm SF}$), the SN and AGN wind efficiencies ($\epsilon_{\rm w, SN};  \epsilon_{\rm w, AGN}$) and the BH accretion parameter ($\alpha$). In \citet{valiante2011,valiante2016} these free parameters were tuned to reproduce the SMBH mass and the properties of the host galaxy of SDSSJ1148+5251 at $z=6.4$ \citep{fan2002}, a well known source, assumed to represent a prototypical example of luminous  quasars at $z > 6$. In this work we target the observed properties of high redshift quasars, as well as a realistic cosmic star formation history down to $z \geq 4$. 
Here we discuss the main observables used to calibrate the reference model, checking the consistency between predictions and observations. Although not reported, the same calibration has been performed for the model variants described in Section~\ref{sec:modelvariants}, see Table \ref{tab:cal} for a summary of the adopted parameters.

\subsection{Cosmic star formation and stellar mass density}
We first tuned the model predictions on the observed total star formation rate density (SFRD), i.e. the total stellar mass formed per unit time and comoving volume. Measurements of the cosmic SFRD are mainly inferred from galaxy UV luminosity, currently limited to bright sources (UV magnitudes $M_{\rm UV} \leq - 17.7$), which are also affected by dust extinction. 
To compare the model predictions with observations, we therefore consider the contribution from sources with UV luminosity above the observed threshold ($M_{\rm UV} \leq - 17.7$)\footnote{The same correction is adopted to compare the cosmic stellar mass density (SMD) predicted by the model with observations.}.
To apply this luminosity cut, we convert the SFR of each galaxy to an intrinsic UV luminosity following the relation \citep{madau2014b}:
\begin{equation}
L_{\rm  UV} = \biggl( \frac{\rm  SFR}{\rm M_{\odot} \, yr^{-1}} \biggl) \,\, 7.14 \times 10^{27} \, \rm{erg \, s^{-1} \, Hz^{-1} \,}
\end{equation}
\noindent
Fig. \ref{fig:SFRD} shows that the {\it reference} model (dashed red line) is in good agreement with the observed star formation rate density and stellar mass density evolution, from $z = 4$ out to $z = 8$. Note that \textsc{cat} predicts a large number of currently undetected  sources ($M_{\rm UV} > - 17.7$), which dominate the star formation rate density and stellar mass density at $z > 4$ (solid red lines).
The impact of dust extinction on the galaxy luminosity might increase even further the contribution coming from undetected sources, particularly at high redshift, where faint galaxies account for the majority of the total SFRD.

\subsection{Mass and luminosity of quasars at $z>5$}
\begin{figure}
\vspace{\baselineskip}
\includegraphics[width=9.00cm]{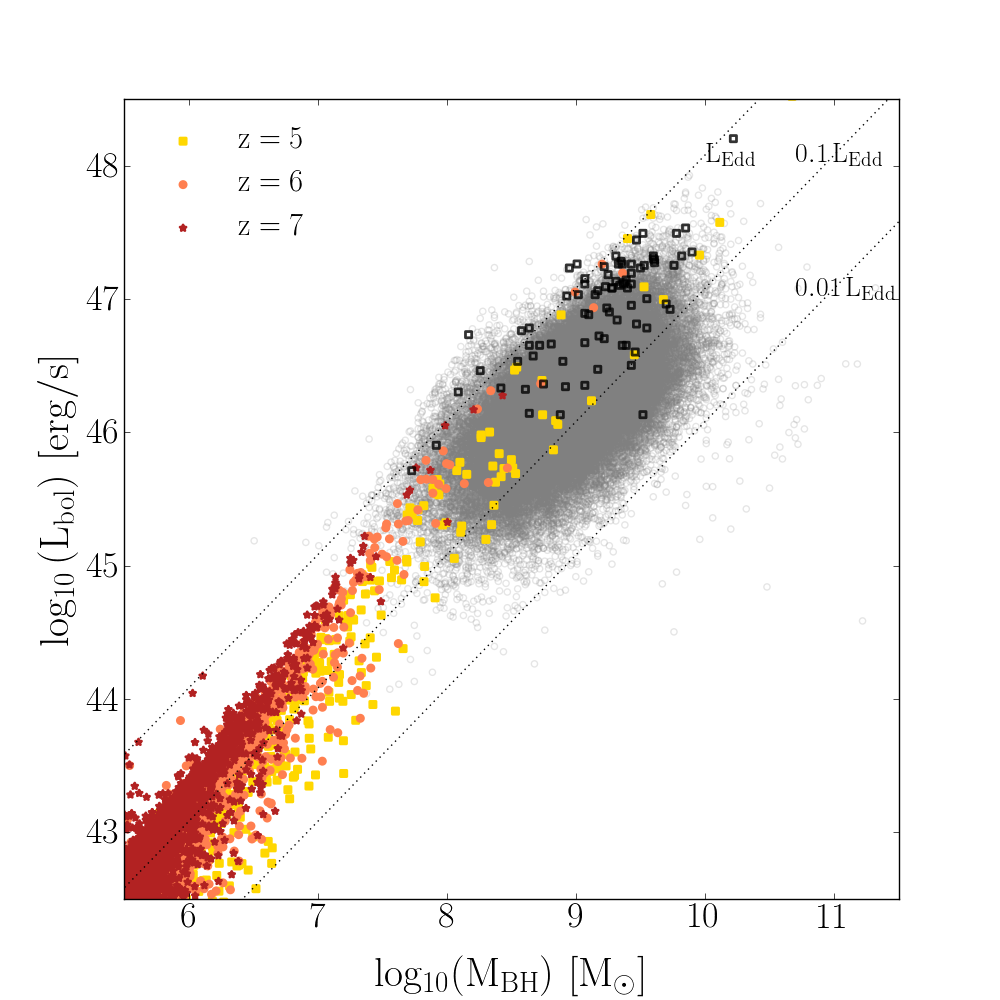}
\centering
\caption{Bolometric luminosity as a function of the black hole mass for high redshift AGNs. Colored data points represent the quasar sample of the \textsc{cat} reference model at $z=5$ (yellow), 6 (orange), and 7 (red). Black empty squares represent a collection of observed quasars at redshift $5.8 < z < 7.5$, while grey data show a large sample of quasars between $0.6 < z< 2$ drawn from the SDSS-DR7 quasar catalog by \citet[][]{shen2011} and for which MgII-based BH masses have been derived. Dotted lines marks the position of BHs with $L_{\rm bol}=L_{\rm Edd}$, $0.1 L_{\rm Edd}$ and $0.01 L_{\rm Edd}$.
}

\label{fig:Lbol}  
\end{figure}

In order to check whether our model, once calibrated, properly reproduces the properties of the observed population of high redshift quasars, here we compare the mass and bolometric luminosity of the most massive systems predicted by \textsc{cat} at $z > 5$ with the values inferred from quasar observations at similar redshifts. The results of the \textit{reference} model at $z = 5, 6$ and 7 are shown, respectively, as yellow, orange, and dark red points in Fig. \ref{fig:Lbol}. The empty black data points show a sample including all $z>5.8$ quasars for which BH masses have been derived via MgII line single epoch virial estimator\footnote{We adopted the virial BH mass relation involving the MgII emission line full width half maximum and continuum luminosity at 3000\AA, using the calibration from \citet[][]{shen2012}. The compilation of sources considered here has been assembled collecting data from the following works: \citet[][]{willott2010a,derosa2011,wu2015,mazzucchelli2017,reed2017,shao2017,chehade2018,eilers2018,eilers2020,banados2018,wang2018,wang2020,sheng2019,reed2019,onoue2019,pons2019,andika2020,yang2020}.}. The diagonal dotted lines indicate the Eddington luminosity as a function of the black hole mass.

We find that our simulated sample is composed mainly by low-mass objects, $\lesssim 10^{7.5} \, \rm M_\odot$, shining close to $0.1 \, \rm L_{Edd}$ regardless of their redshift. More massive BHs show instead higher Eddington ratios, especially at $z=6$ and 7, where a large amount of gas is available for accretion.
The comparison with observational data at $5.8 < z < 7.5$ (empty squares) shows that, despite our limited statistics due to the low number density of such objects, the most massive BHs predicted by \textsc{cat} at similar redshifts populate the observed range of quasar masses and luminosities.

In conclusion, Figs.\, \ref{fig:SFRD} and \ref{fig:Lbol} show that the parameter values assumed in our \textit{reference} model (see Table \ref{tab:cal}) guarantee a good agreement with the observed quasar population at $z > 5$ and lead, at same time, to a galaxy population characterized by an evolution of the SFR density and stellar mass density in agreement with the available data. 

Finally, a similar analysis has been carried out also for the two model variants, and the corresponding model parameters are reported in Table \ref{tab:cal}.

\begin{figure*}
\vspace{\baselineskip}
\includegraphics[width=17.00cm]{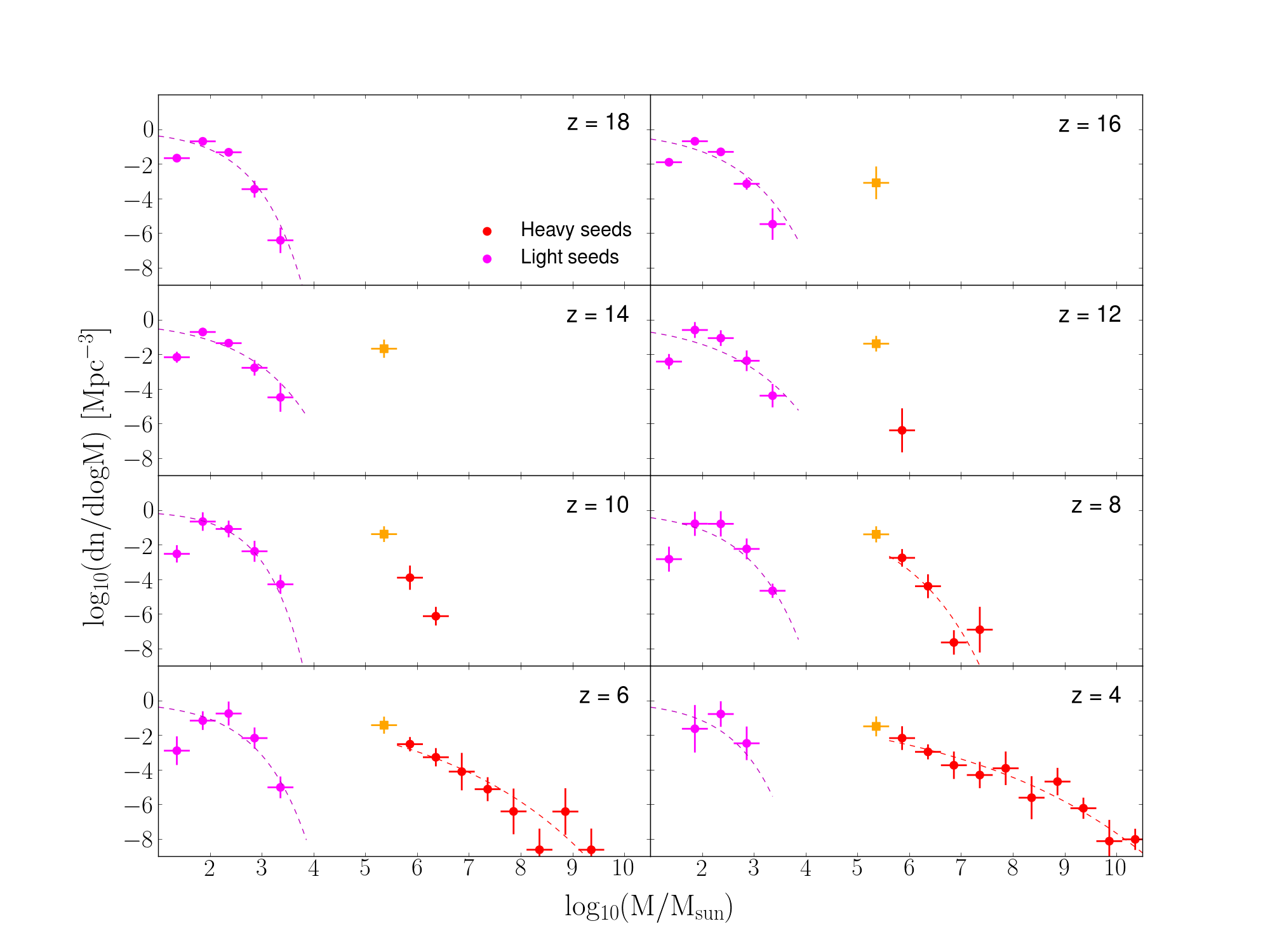}
\centering
\caption{The BH MF predicted by the {\it reference} model at different redshifts, ranging from $z = 18$ down to $z=4$. In each panel, we show the separate contributions of BHs descendants of \textit{light} and \textit{heavy seeds} with magenta and red points, respectively. The $1 \sigma$ Poisson error bars are shown in each data point. Where possible, the best fits of the two distributions are also shown with dashed lines. The orange data point represents the mass bin populated by newly formed and/or not grown heavy black holes seeds and it is not considered to produce best fit curves since its occupation is closely related to the adopted seed birth mass.}
\label{fig:SMBH_MFevo}  
\end{figure*}

\section{Results}
\label{sec:results}

The main motivation of our study is to investigate if, when and how 
the nature of BH seeds and their mass growth affect the population of BHs at $z > 4$. 
In this Section, we first analyze the redshift evolution of the BH mass function 
predicted by the \textit{reference} model and discuss how it changes in the two model variants.
Second, we investigate whether the differences across models are revealed in the predicted luminosity functions, exploring at the same time their accordance with observational constraints obtained by current and future facilities.

\subsection{Black Hole Mass Function}
\label{sec:BHMF}

In Fig. \ref{fig:SMBH_MFevo} we show the BH mass function (MF) predicted by the {\it reference} model at different redshifts, ranging from $z = 18$ down to $z = 4$. 
Mass bins populated by \textit{light} and \textit{heavy} (i.e. at least with one DCBH progenitor) \textit{seed} descendants are shown in magenta and red points, respectively. Here, as in the rest of the paper, binned data points are shown with the corresponding $1 \sigma$ Poisson error bars. 
The mass bin which includes newly formed and/or not grown heavy BH seeds has been highlighted in orange and it is not considered when fitting the distributions.
In fact, the high number density of $10^5 \, \rm{M_\odot}$ BHs reflects the adopted seeding prescription, while a more realistic heavy seed birth mass function \citep[e.g.][]{ferrara2014} and/or the inclusion of the so-called medium-weight seed formation channel \citep[][]{sassano2021} would probably result in a lower number density, spread across the intermediate mass range $(10^3-10^5 \, \rm M_\odot)$.

It is immediately evident that the distribution is characterized by two well defined regions: BH descendants of \textit{light} BH \textit{seeds}, are confined below $10^5 \, \rm{M_\odot}$, while BHs grown from \textit{heavy} BH \textit{seeds} dominate the high-mass end of the MF, i.e. above $10^5 \, \rm{M_\odot}$. These two populations appear well separated across all the redshifts explored by our model. 
Early on ($z \geq 18$) only \textit{light} BH seeds are formed, some of which have already grown up to $\sim 10^{3.5} \, \rm{M_\odot}$. 
At $14 \leq z \leq 17$ the first \textit{heavy} BH seeds start to form, but they do not significantly grow above their formation mass scale ($10^5 \, \rm{M_\odot}$).
At these same redshifts, the distribution of \textit{light} BH seed descendants has already reached its maximum extension in mass, well below $10^4 \, \rm{M_\odot}$.
At $z < 14$ \textit{heavy} BH seeds descendants continue to grow, and since their formation becomes progressively rarer, their mass distribution shifts to larger masses, covering the mass range $10^6 M_\odot - 10^{10} M_\odot$ at $z \lesssim 6$. Hence, the MFs of \textit{light} and \textit{heavy} BH descendants remain completely segregated in mass as a consequence of the 
inefficient growth of  \textit{light} BH seeds. Some of these may be involved in major mergers contributing to the growth of the \textit{heavy} seed descendants, others may instead participate in minor mergers, eventually becoming satellites (or wandering) BHs, which, in turn, explains why the highest mass bin of the distribution shifts to lower values with increasing time (decreasing redshift).

The gap in the BH number density just below the minimum adopted mass of \textit{heavy} BH seeds indicates thus that in the \textsc{cat} \textit{reference} model \textit{light} BH seeds fail to grow efficiently all the way down to $z \sim 4$. This could be a consequence of the host environmental properties of this class of seeds, namely the abundance of gas available to fuel BH growth, or the adopted mass accretion model. For this reason, we explored the redshift evolution of the BH MFs predicted by two model variants: the \textit{super-Edd} model and the  the \textit{merger-driven} model (see Section \ref{sec:modelvariants} and Table \ref{tab:cal} for more details).

\begin{figure*}
\vspace{\baselineskip}
\includegraphics[width=17.00cm]{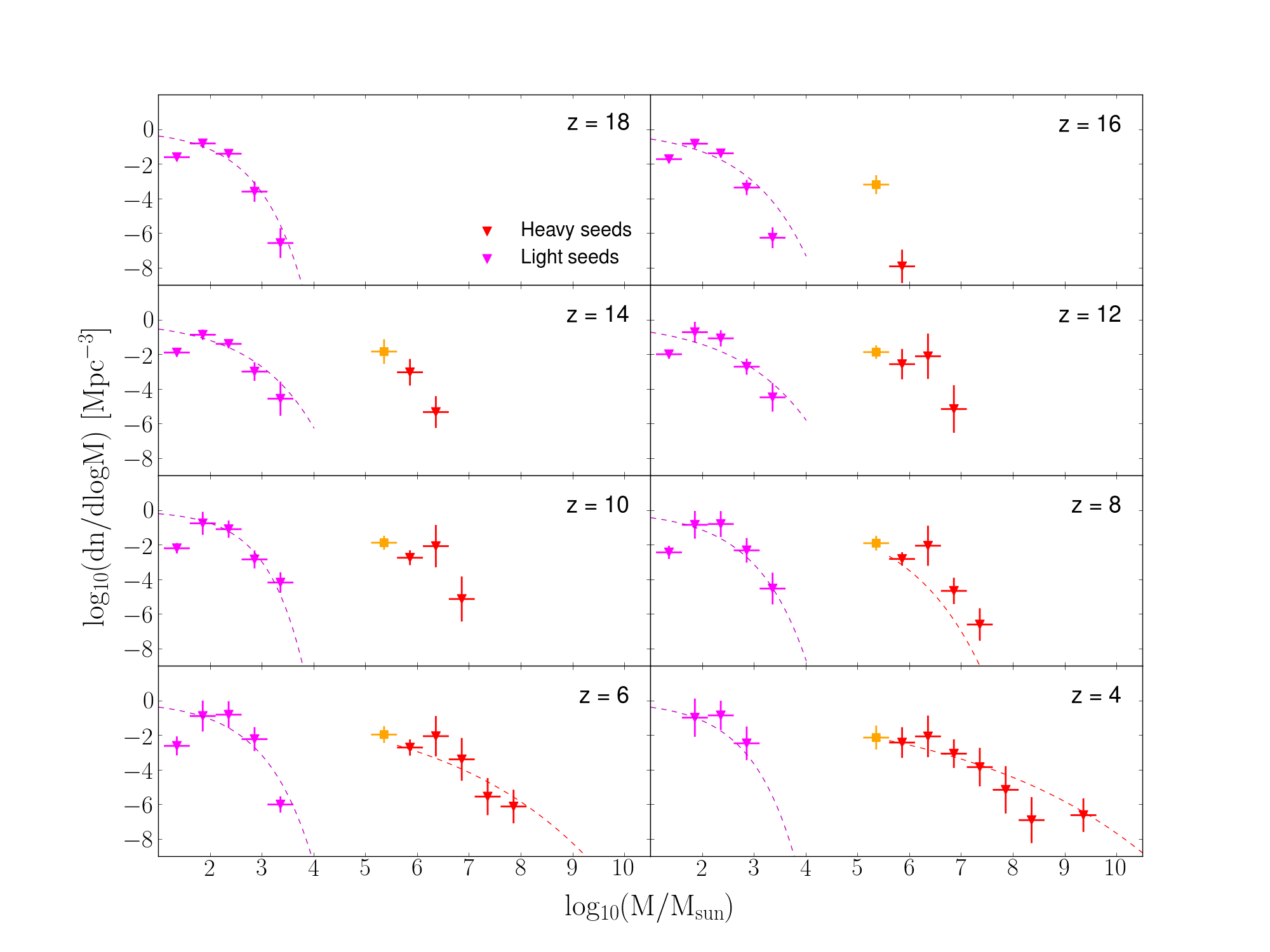}
\centering
\caption{Same as Fig. \ref{fig:SMBH_MFevo} but for the \textit{super-Edd} model (see Section \ref{sec:Eddington limit}). As a reference, we show with magenta and red dashed lines the best fits of the \textit{light} and \textit{heavy} BH seed descendant distributions obtained for the \textit{reference} model.}
\label{fig:SMBH_MFSE}
\end{figure*}

Fig. \ref{fig:SMBH_MFSE} illustrates the results for  the \textsc{cat} \textit{super-Edd} model. To aid the comparison with the \textit{reference} model, the magenta and red dashed lines indicate the best fit to the MFs of \textit{light} and \textit{heavy} BH seeds predicted by the \textit{reference} model (as in Fig.\ref{fig:SMBH_MFevo}). 
We find a significantly faster growth of \textit{heavy} BH descendants, especially in the first phase after their formation, at redshifts $ 16 \lesssim z \lesssim 10$. At lower redshift this accelerated growth seems to slow down quickly and, despite the early build up of the global mass function, the maximum BH masses reached are $\sim 1 \, \rm dex$ below the results of the reference model. The rapid consumption of gas and the associated feedback in the early BH growth phase might be the reason for this behaviour. In fact, super Eddington accretion onto massive seeds leads to a quick depletion of gas from the host galaxy, which affects the subsequent BH evolution at later times.
Hence, this \textit{super-Edd} model well reproduces the global star formation history but struggles in reproducing the formation of the billion solar mass BHs observed to power quasars at $z > 6$.

\begin{figure*}
\vspace{\baselineskip}
\includegraphics[width=17.00cm]{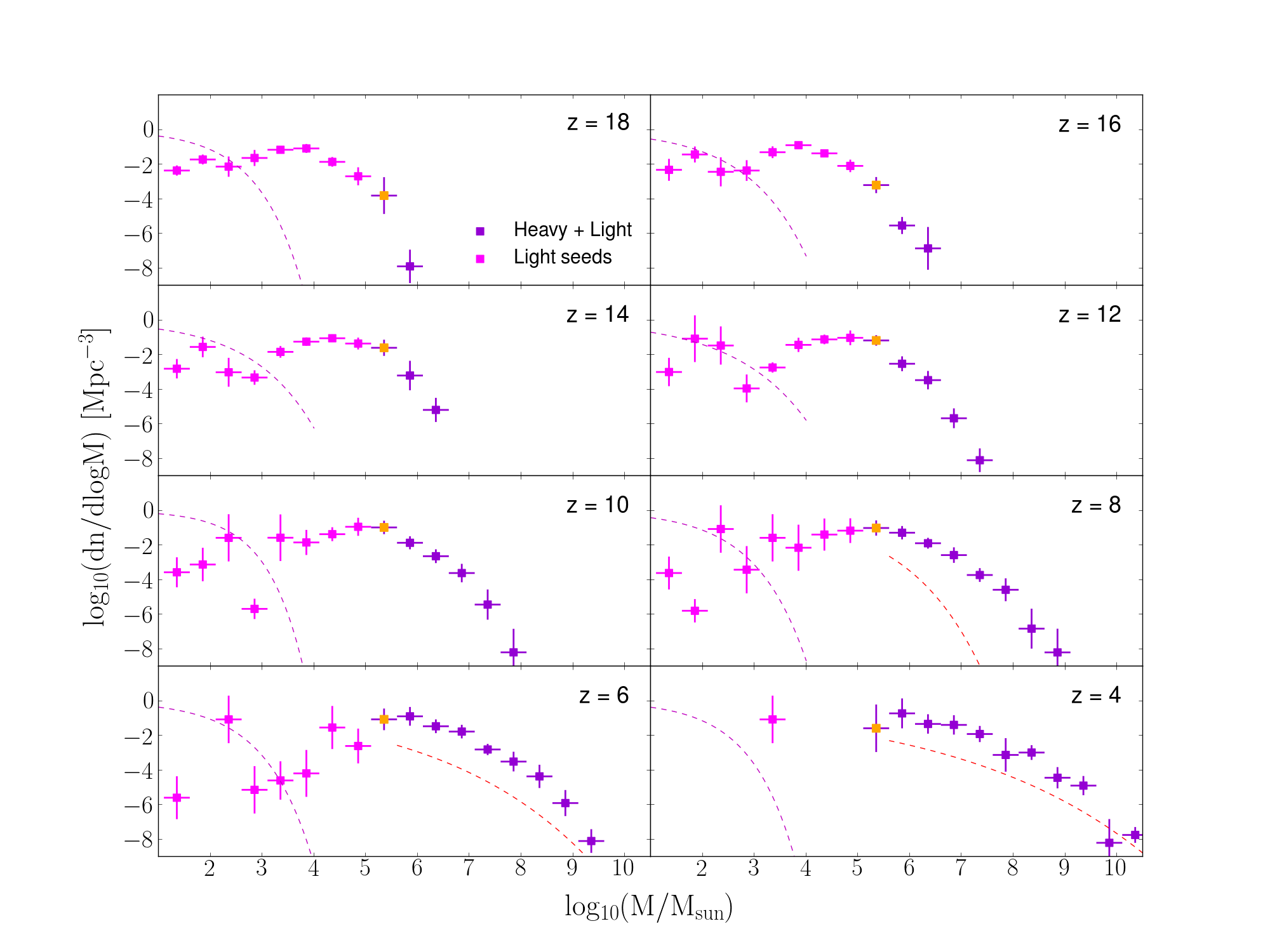}
\centering
\caption{Same as Fig.\ref{fig:SMBH_MFSE} but for the \textit{merger-driven} accretion model (see Section \ref{sec:mergerdriven}). Here we show with violet data points the mass bins above $\rm 10^5 \, M_\odot$ since, unlike the previous models, they are now populated by both \textit{light} and \textit{heavy} BH seed descendants. Distributions (data points with error bars) are compared with best fits obtained for the \textit{reference} model (dashed lines).}
\label{fig:SMBH_MF_brs}  
\end{figure*}

However, the \textit{super-Edd} results show the same characteristic feature of the \textit{reference} model observed in Fig. \ref{fig:SMBH_MFevo}, i.e. a persisting gap between the \textit{light} and \textit{heavy} BH seeds distribution.
This implies that the reason for the stunted growth of \textit{light} BH seeds descendants does not reside in imposing an Eddington limit to the BH accretion. Instead, it could be due to a limited content of gas to fuel BH growth inside their hosts, or it could be a consequence of the adopted BHL accretion rate. 
In order to discern between these two possibilities, we investigate the predictions of the \textit{merger-driven} model, shown in Fig. \ref{fig:SMBH_MF_brs}.
In this figure we indicate again with magenta data points BH mass bins contributed only by \textit{light} BH seeds descendants. However, for BH masses $\geq 10^5 M_\odot$, the mass bins are now populated also by BHs grown from \textit{light} BH seeds, in addition to those coming from \textit{heavy} BH seeds progenitors and we show this mixed population in violet. In the previous \textit{reference} and \textit{super-Edd} models, instead, the high mass end of the BH mass function is entirely populated by BHs with at least one \textit{heavy} BH seed progenitor.
As a comparison, we also show with magenta and red dashed lines the best fit to the MFs of \textit{light} and \textit{heavy} BH seeds in the the \textit{reference} model, respectively.

The \textit{merger-driven} model shows a very rapid and early growth of \textit{light} seeds, which reach BH masses as high as $10^{6} \rm M_\odot$ already at $z \sim 18$. The resulting distribution evolves shifting towards higher mass values, with minor changes in the global shape. This leads to a BH MF at $z \lesssim 6$ that continuously ranges between $\sim 10^4$ and $10^{10} \, \rm M_\odot$. 
At the low-mass end, the original gap has now disappeared since \textit{light} BH seeds and their descendants can efficiently grow, competing with \textit{heavy} seeds in SMBHs growth.
The high-mass end of the distribution, instead, reaches values consistent with the reference model, predicting however a slightly higher number density of objects between $\sim 10^6$-$10^8 \, \rm M_\odot$.

It is important to note that the peak in the BH number density is shifted towards higher masses at later times (lower redshifts). This reflects the progressive depletion of low-mass objects as BH seeds are able to efficiently increase their masses through accretion and mergers, while newly formed seeds become rarer.
The decreased number density of low-mass objects is clearly noticeable at $z \lesssim 6$.

\begin{figure*}
\vspace{\baselineskip}
\includegraphics[width=17.00cm]{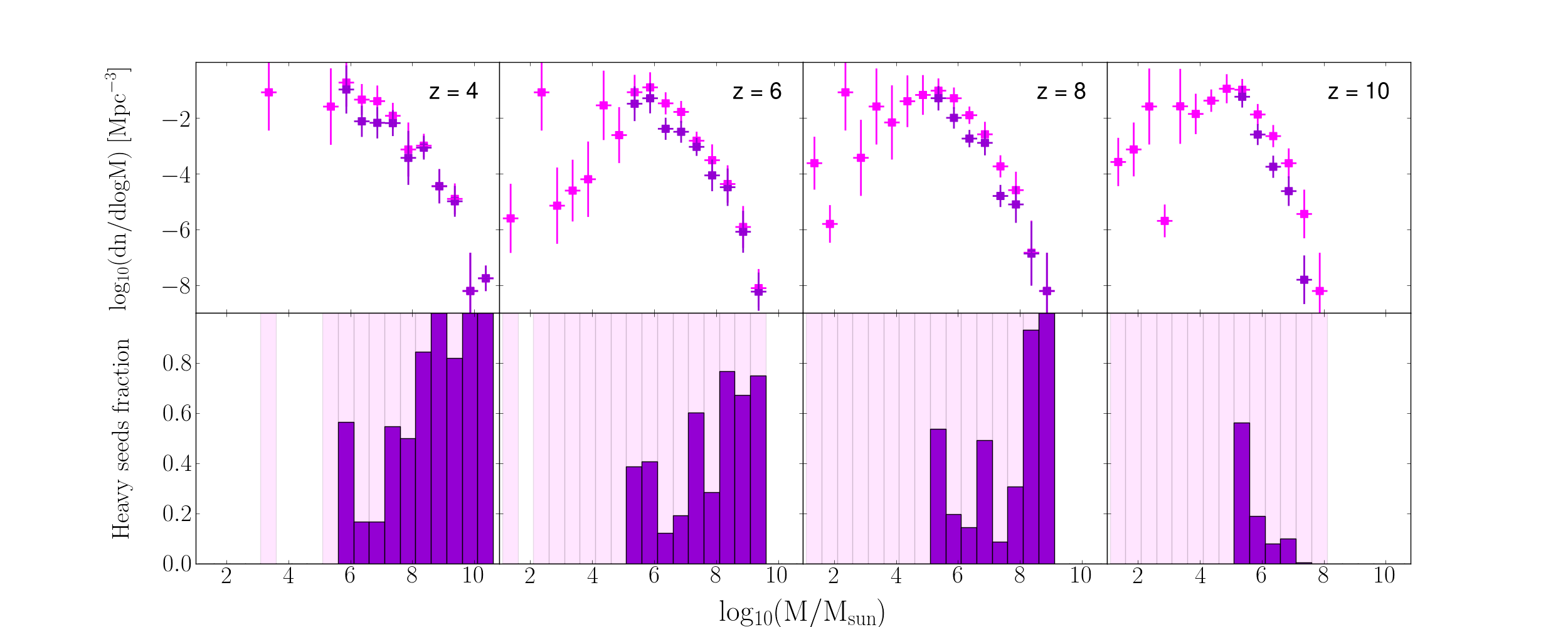}
\centering
\caption{The Figure shows, for the \textit{merger-driven} model, the relative contribution to the final BH mass function of BHs descending from at least one \textit{heavy} or only \textit{light} BH seeds. Top row: the total BH mass function (magenta points, same as in Fig. \ref{fig:SMBH_MF_brs}) and the contribution of only \textit{heavy}-BH seeds descendants (violet) at $z = 4, 6, 8,$ and $10$ (from left to right). Bottom row: the corresponding percentage of \textit{heavy} BH seed descendants in different mass bins. BHs descending only from \textit{light}-BH seeds (represented by lighter magenta regions of the histograms) clearly provide, in this model, a significant contribution to the global MF, even at the high-mass end.}
\label{fig:HeavySeeds}
\end{figure*}

In order to emphasize the contribution to the mass function of BHs grown only from \textit{light} BH seeds in the \textit{merger-driven} model, in Fig. \ref{fig:HeavySeeds} we show the mass function of \textit{heavy} seed descendants at $z = 4,6,8$ and $10$, along with their percentage of occupation of each BH mass bin. It is clear that, in this model variant, BHs descending only from \textit{light} seeds represent a large fraction of the BH population. In particular, besides largely dominating the entire mass function at $z \gtrsim 10$, BHs grown from  \textit{light} seeds represent $\sim 50 \%$ of the SMBH population ($M_{\rm BH} > 10^5 \, M_{\odot}$) down to redshift $z \sim 6$.

Hence, these results lead us to the conclusion that the main limiting factor to the growth of a BH seed is the mode of gas accretion rather than the gas mass reservoir around the black hole. The successful growth of \textit{light} seeds predicted by the \textit{merger-driven} scenario suggests that the key element affecting the efficiency of BH growth is the way in which the gas reaches the central regions, and therefore the physical mechanisms driving BH mass accretion.

However, we found that the large-scale mechanism of merger-driven infall of gas toward the central regions around the nuclear BH is not sufficient, alone, to ensure an efficient growth of the smaller seeds. A black hole accretion model that allows super-Eddington growth, as the \textit{slim disk} model assumed in the \textit{merger-driven} variant, is also needed to fill up the gap in the MF. 
In fact, assuming an Eddington limited growth in the merger-driven scenario, $\dot{M}_{\rm BH} \leq \dot{M}_{\rm Edd}$, we open-up again a clear gap in the mass distribution below $10^5 \, \rm M_\odot$, similarly to that shown in Figure \ref{fig:SMBH_MFSE} and \ref{fig:SMBH_MF_brs}. In addition, such a model predicts an overall reduction of the BH growth due to the low BH duty cycle assumed in the merger-driven scenario.

This suggests that the low-mass end of the BH mass function
at $z \leq 6$ may provide important indications on the mass scale and growth rate of BH seeds at high redshift, while, at the high mass end, SMBHs with masses $\geq 10^6 M_\odot$ are relatively insensitive
to the nature of their BH progenitors.
Still, the depopulation of the low-mass region (below $\sim 10^5 \, \rm M_\odot$) predicted by the \textit{merger-driven} model at $z\lesssim 6$ might resembles the clear gap in the MF predicted by a Bondi-like accretion mechanism, worsening the chances to clearly identify the distinctive features of different evolutionary scenarios.

The question is then whether current or future surveys may be able to discriminate among the various models, as it will be discussed in the next section. 

\begin{figure*}
\vspace{\baselineskip}
\includegraphics[width=18.5cm]{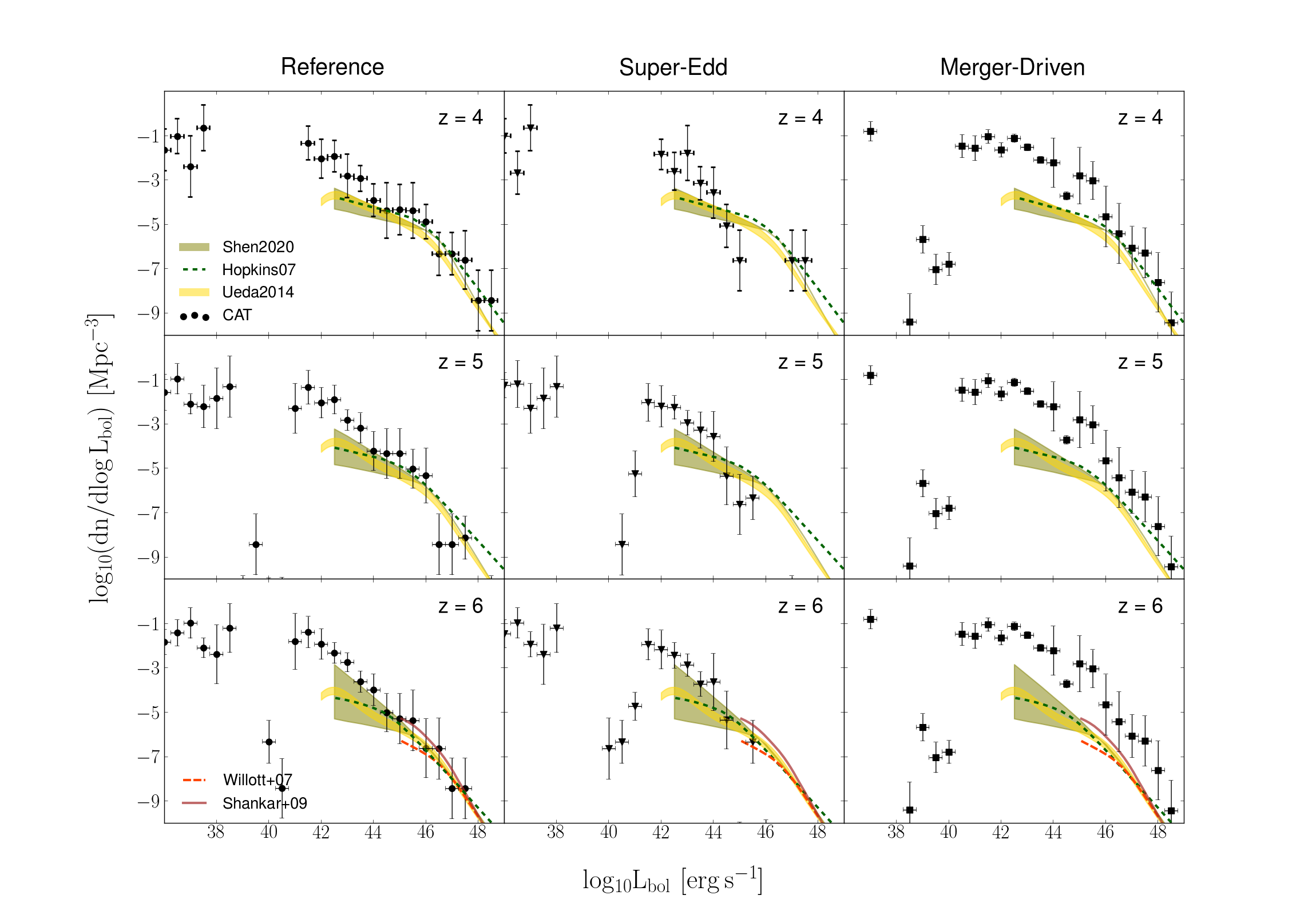}
\caption{Bolometric luminosity function of AGNs at $z = 4, 5$ and $6$ (from top to bottom). Black data points show the luminosity distribution of \textsc{cat} AGN population. Each column shows the results obtained in one of the three models examined: the \textit{reference model} (left panels), the \textit{super-Edd} model (central panels) and the \textit{merger-driven} model (right panels). Our results are compared with the predictions of \citet[][green dashed lines]{hopkins2007}, \citet[][red dashed line]{willott2010b}, \citet[][brown solid line]{shankar2009} and \citet[][green shaded area]{shen2020}. We also show as a comparison the constraints on the bolometric LF derived from the X-ray data by \citet{ueda2014} (yellow shaded area, see text).}
\label{fig:BolLF}
\end{figure*}

\subsection{The AGN Luminosity Functions}
\label{sec:AGN_LF}

\textsc{cat} model allows us to estimate the accretion rate of each nuclear black hole during its evolution through cosmic time, as determined by the environmental conditions present in its host galaxy. Hence, we are able to infer the luminosity of all active BHs (see Sections \ref{sec:BHaccretion}, \ref{sec:Eddington limit}, and \ref{sec:mergerdriven}) and to reconstruct their luminosity distribution, i.e. the AGN luminosity function (LF). Here we compare the bolometric LF predicted by \textsc{cat} with several independent estimates based on multi-band observational data. These comparisons represent a fundamental benchmark for our BH evolution model, in order to determine which of the considered accretion scenarios better reproduce the observed trends. At the same time, they also allow us to check whether current or future surveys may be able to detect signatures of BH seeds populations and of their growth mode.
A comparison between \textsc{cat} predictions and the findings of the most recent AGN UV and X-ray surveys is discussed in Section \ref{sec:LFs} (see Figures \ref{fig:observedUVLF}, \ref{fig:XLFs}.)

\subsubsection{Bolometric Luminosity Function}

In Figure \ref{fig:BolLF} we show the bolometric luminosity function predicted by all models in Table 1 at $z = 4, 5$ and 6. In the \textit{reference} and \textit{super-Edd} models a clear gap in the luminosity function is visible below a threshold luminosity of $\sim 10^{41} \, \rm erg\, s^{-1}$ at all redshifts. 
This reflects the behaviour of the BH mass functions shown in Figs. \ref{fig:SMBH_MFevo} and \ref{fig:SMBH_MFSE}, and it is a consequence of the failed growth of \textit{light} black hole seeds that characterize these two models, as previously discussed. While \textit{heavy} seeds descendants continue to grow, populating the bright-end of the LF, the empty region in the luminosity distribution widens with decreasing redshift.
In fact, as already described in Section \ref{sec:BHMF}, \textit{light} seeds descendants are progressively more involved in galaxy and black hole merger events, eventually becoming satellites BHs or merging with \textit{heavy} descendants. 

In the \textit{merger-driven} model instead the efficient growth of light BH seeds leads to a continuous luminosity distribution at all redshifts. However, at $z \sim 6$ the mass distribution shown in Fig. \ref{fig:SMBH_MF_brs} results in a LF which peaks around $\sim 10^{40 - 42} \, \rm erg \, s^{-1}$ and then quickly drops down at fainter luminosities. This behaviour, as mentioned before, reflects the progressively lower number of lighter seeds, due to both the involvement in galaxy mergers or their efficient growth toward higher masses.

Note that this third model is characterized by brief periods of enhanced accretion onto BHs. If we infer the LF at a single given time, thus, we might under-sample rare objects with high luminosity that are active only for a short interval of time. Therefore, following \citet{griffin2019}, when computing the LF for the \textit{merger-driven} model we average the LF over a time window $\Delta t_w$ of $\sim 50 \,\rm Myr$ around each redshift of interest. To reconstruct the global luminosity distribution, each active object is then assigned a weight 
\begin{equation}
w = t_{\rm Q}/\Delta t_{\rm w} \,\, ,
\label{eq:timeAvLF}
\end{equation}
where $t_{\rm Q}$ is the time spent in the enhanced accretion mode during the considered time window.

The model predictions are compared to the results obtained by \citet{willott2010b}, based on SDSS and CFHQS data, and with the LF evolutionary models proposed by \citet{shankar2009} and \citet{hopkins2007}. 
We also show the LF constraints presented by \citet{shen2020}, which are based on a large compilation of AGNs observations in different energy bands, from the rest-frame IR to the X-rays. 
Finally, we compare our findings with the results of the X-ray studies by \citet[][]{ueda2014}, applying the bolometric correction proposed by \citet[][]{duras2020}.
The shaded  region shows the range of uncertainties associated to obscuration effects, with the  upper  and lower bounds illustrating the LF obtained  with and without obscuration correction, respectively. \footnote{Here the obscuration correction has been applied assuming a fraction of Compton-thick AGNs with column density $24<\rm{N_H}<26$ four times higher that the one considered in \citet[][]{ueda2014}. This is in agreement with the recent work of \citet[][]{ananna2019}, which found a much higher number density of Compton-thick objects with respect to the original work of \citet[][]{ueda2014}.}

The LFs predicted by the \textit{reference} model are remarkably consistent with the observed bright-end distributions ($ \gtrsim 10^{42} \rm \, erg \, s^{-1}$) at all the considered redshift, $z = 4 - 6$. The number of sources is instead slightly over-predicted at fainter luminosities, $\sim 10^{40-42} \rm erg \, s^{-1}$, where, however, the observations should be largely affected by obscuration effects. 

The distribution of low-luminosity sources predicted by the \textit{super-Edd} model (middle panels of Figure \ref{fig:BolLF}) is similar to that described in the \textit{reference} case. Conversely, the number density of brighter sources, above $\sim 10^{44} \, \rm erg \, s^{-1}$, dramatically drops, especially at $z=5$ and $6$, as a consequence of the lack of massive/luminous AGN predicted in the model.\\
This result suggests that the accelerated growth of BH seeds at early times, driven by the super-Eddington accretion (see Fig. \ref{fig:SMBH_MFSE}), strongly affects the BH evolution at later times. In particular, the enhanced accretion rate and the associated feedback significantly deplete the gas reservoir available around the BH. On the other hand, new infalling material from the external medium is mainly consumed by star formation that is quite efficient in massive halos hosting growing heavy seeds. Hence, the growth in mass of the initial seed remains shortly stuck and it will undergo major accretion episodes only as a consequence of gas-rich mergers, failing to reproduce both masses and luminosities of the extreme quasars observed at $z \sim 6$.

In the \textit{merger-driven} model (right panels of Figure \ref{fig:BolLF}) the bright-end of the luminosity function ($L_{bol} \gtrsim 10^{46} \rm erg \, s^{-1}$) is consistent with the observational data at $4 \leq z \leq 5$. At $z=6$ the LF is instead over-estimated as a consequence of the early mass growth of BHs driven by halo mergers.
At the faint-end of the distribution, the \textit{merger-driven} model over-predicts the number of objects at all redshift within the $10^{41} \rm erg \, s^{-1}\lesssim L_{bol} \lesssim 10^{45} \rm erg \, s^{-1}$ luminosity range. 
Compared to the {\it reference} and {\it super-Edd} models, this is due to the population of efficiently grown \textit{light} seed descendants that were unable to significantly grow in mass in the classic Bondi-like accretion scenario. 
At even fainter luminosities, below $\sim 10^{40} \rm erg \, s^{-1}$, the LF drops down similarly to what we observe in the \textit{reference} model, but as a result of the exact opposite process. 
In fact, while in the BHL accretion scenario the lack of sources below this threshold is due to the low mass of \textit{light} seed descendants, here most of them grows efficiently, depopulating this luminosity region at low redshift. 

It is worth considering that if the bursts of accretion driving BH growth take place on very short timescales, $\lesssim 2 \rm \, Myrs$, the $t_{Q}$ parameter in Eq. \ref{eq:timeAvLF} will be over-estimated, explaining, at least partially, the higher number density of sources predicted by the merger-driven model compared to observations, as will be highlighted in detail in Section \ref{sec:LFs}.

Note that, for the \textit{merger-driven} scenario, we considered a simplified version of the original model proposed by \citet[][]{pezzulli2016} mainly to investigate the impact of the accretion model on the overall shape of the BH luminosity and mass distributions. 

A more sophisticated description of BH growth triggered by gas-rich galaxy mergers will be developed in future works to further explore its comparison to observations.

It is important to emphasize that the agreement of the \textit{reference} model with observations of the bright-end of the LF is an important result, as it implies that \textsc{cat} is capable to reproduce reliable mass and luminosity distributions of the population of massive ($\gtrsim 10^{7}M_\odot$) nuclear BHs down to $z \sim 4$, despite the BH seeding and growth model was originally aimed at reproducing the evolution of the most extreme SMBHs ($\sim 10^9 \rm \, M_\odot$) at $z \geq 6$.
The above result suggests that the mechanisms which drive the formation and evolution of these class of massive BHs across cosmic times remain the same, regardless of the mass of the final object and of its host galaxy.

However, as already observed, if we extrapolate the empirical LF at luminosities $\lesssim 10^{42} \rm erg \, s^{-1}$, the \textit{reference} model seems to predict an excess of faint AGNs compared to observations. This may be a hint that the accretion mechanism assumed in \textsc{cat} leads to an overgrowth of less massive BHs.
In order to properly reproduce the faint end of the BH LF, in fact, similar semi-analytic models artificially shut off BH growth below a given halo mass \citep[e.g.][]{piana2021}, assuming that this is caused by the impact of SN feedback \citep[][]{habouzit2017, angles-alcazar2017}. Hence, an improved treatment of the accretion process of less massive nuclear BHs may be required to provide a better agreement with the empirical relations at low luminosities, as will be discussed in Section \ref{sec:discussion}. 

Nevertheless, we should note that the empirical LF predicted at $z \gtrsim 4$ are constrained mainly by luminous sources with $\rm L_{bol} \gtrsim 10^{43} \, erg \, s^{-1}$. The distributions at lower luminosities are thus highly uncertain, especially since the population of fainter AGNs might be heavily obscured, as suggested by several studies \citep[][]{matsuoka2018, giallongo2019}.
The improved agreement at faint luminosities observed in the UV and X-ray LFs between \textsc{cat} predictions and several observations (see Section \ref{sec:LFs}) supports the idea that these tensions might be due, at least partially, to observational limits.

Overall, this comparison suggests that the \textit{reference} model provides the closest match to current observations, which however are able to probe only the bright-end of the predicted LF, dominated by {\it heavy} BH seeds descendants. Yet, at the very faint-end, below the gap, the LF is due to the more numerous population of inefficiently growing \textit{light} BH seeds which remains completely invisible, as they lie at luminosities that are 4 orders of magnitudes lower than the faintest luminosity probed by current data. 
We also find that a significantly different evolutionary scenario, such as the \textit{merger-driven} model, might lead to comparable results for the bright-end of the AGN luminosity distribution. 
The main differences show up instead in the low luminosity region, making therefore very challenging to discern between different accretion models through observational campaigns.

A more detailed comparison of the model predictions with current and forthcoming observational facilities, in particular with the most promising deep sky surveys, is carried out in Section \ref{sec:LFs}.

The improved sensitivity of the forthcoming generation of surveys might nevertheless push down the observational limit toward the threshold of $\rm 10^{41} \, erg \, s^{-1}$, which is less than two orders of magnitude below the actual empirical constraints. In this range of luminosity the number density of accreting BHs is maximally sensitive to the assumed model for black hole growth, as shown in Fig. \ref{fig:BolLF}, despite host contamination might become significant for such faint sources. 
Still, even at these luminosities, different scenarios for black hole fueling and accretion might result in similar luminosity distributions, characterized by a decreasing number of sources below $\sim 10^{40} \rm erg \, s^{-1}$. 

\subsubsection{BH UV and X-ray Luminosity Function}
\label{sec:LFs}
In order to compare more extensively \textsc{cat} predictions with current and future observational facilities, and in particular with the most promising deep sky surveys, we derived, for the three model variants, the BH luminosity distribution predicted in the UV and X-ray energy bands.
The bolometric luminosity of each active BH can be converted into a B-band and X-ray luminosity adopting specific bolometric corrections, such as in \citet[][]{duras2020}:

\begin{equation}
\frac{L_{\rm bol}}{L_{\rm B}} = 5.13 \quad ; \qquad \frac{L_{\rm bol}}{L_{\rm X}} = a \biggl[ 1 + \biggl( \frac{log(L_{\rm bol}/L_{\odot})}{b} \biggl)^{c} \biggl]
\label{eq:bolcorr}
\end{equation}
where $L_{\rm B}$ and $L_{\rm X}$ are, respectively, the luminosity in the B ($4400$ \AA) and $(2\!-\!\!10\,)$ keV band, while $a = 10.96$, $b = 11.93$ and $c = 17.79$ are parameters calibrated on a population of both type 1 and type 2 quasars.
Then, the B-band luminosity has been converted into a UV luminosity at $1450$ \AA $\,$ assuming that $L_{\nu} \propto \nu^{-0.44}$ \citep{dayal2019}.

\begin{figure*}
\vspace{\baselineskip}
\includegraphics[width=18.5cm]{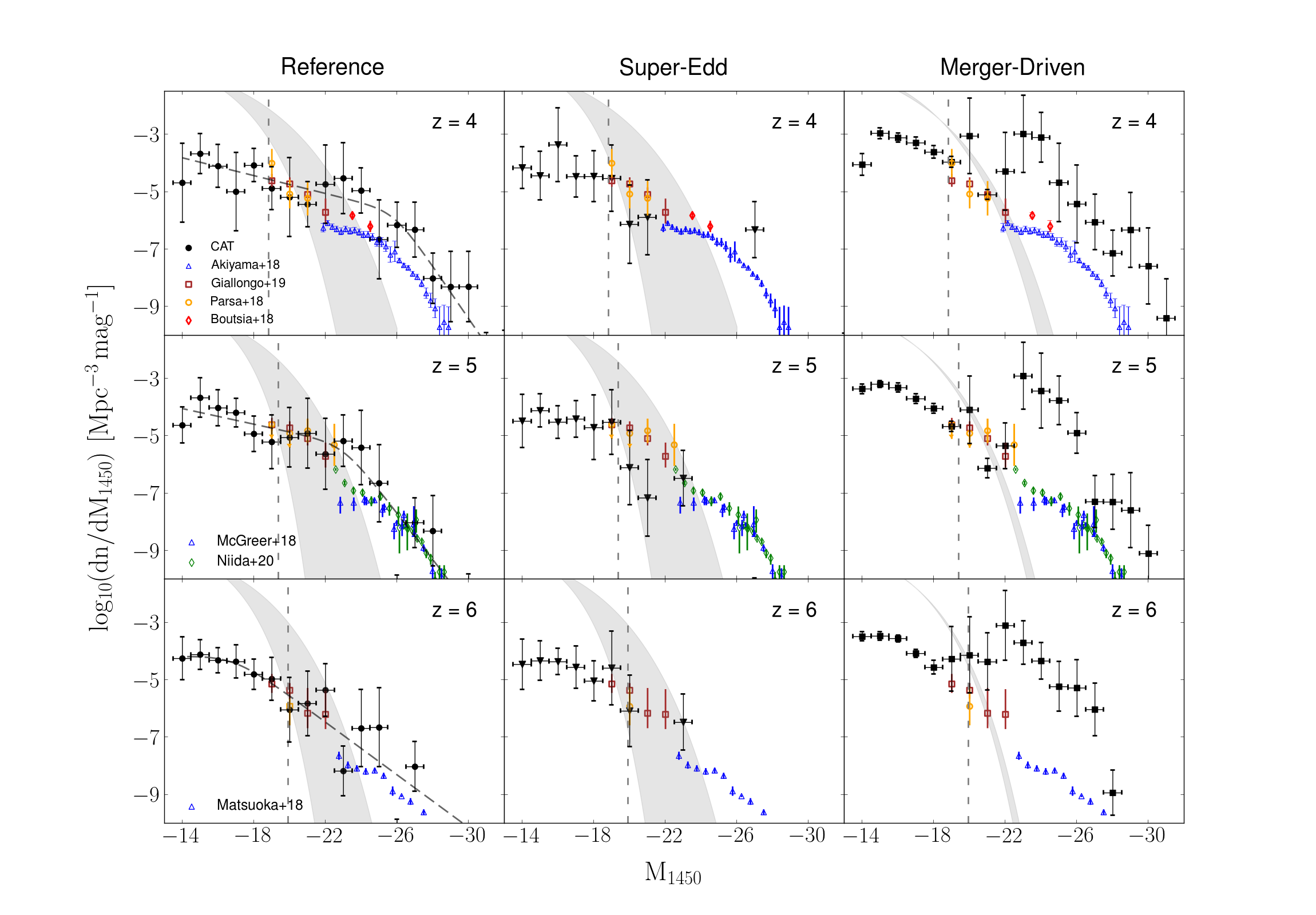}
\caption{The AGN UV LFs at $z = 4, 5,$ and $6$ (from top to bottom). Each column shows the results obtained in one of the three \textsc{cat} models: the \textit{reference} model (left panels), the \textit{super-Edd} model (central panels) and the \textit{merger-driven} model (right panels). Black data points represent the predicted AGN LFs, properly corrected for obscuration as described in the text. The black dashed line shows the best fit for the \textit{Reference} model. The grey shaded area encloses the region between the intrinsic and the dust corrected galaxy UV luminosity function predicted by each model at the corresponding redshift.
Coloured data represent the observational constraints on he AGN LF obtained by \citet[][orange circles]{parsa2018}, \citet[][red diamonds]{boutsia2018}, \citet[][brown squares]{giallongo2019}, 
by the SHELLQs survey \citep{matsuoka2018}, the CFHT Legacy Survey \citep{mcgreer2018}, and the Hyper Suprime-Cam Wide Survey \citep{akiyama2018, niida2020} (blue triangles and green diamonds). The vertical black dashed lines show the JWST luminosity limit at each redshift predicted by \citet[][]{griffin2020}.}
\label{fig:observedUVLF}
\end{figure*}

However, for a proper comparison with observational data we must account for obscuration effects. In fact, the intrinsic LFs predicted by \textsc{cat} have to be corrected for the fraction of obscured AGNs in each energy band.
Following \citet[][]{merloni2014}, we assume the fraction of obscured AGNs in the UV band to be a decreasing function of the intrinsic X-ray luminosity of the source.
Hence, we compute the observable fraction as: 
\begin{equation}
f_{\rm obs} = 1 - 0.56 - \frac{1}{\pi} \rm{arctan} \, \biggl( \frac{43.89 - log \,L_X} {0.46} \biggl)
\label{eq:UVObs}
\end{equation}
\noindent
We show the resulting LFs, that we call obscured AGN UV LFs, in Fig. \ref{fig:observedUVLF}. Here, the predictions obtained for the three different models at redshift $z = 4, 5$ and $6$ are represented by the black data points with error bars. The black long-dashed lines represent the best-fit to the distributions at each redshift, which are well represented by broken power-laws that flatten below a characteristic luminosity. For comparison, in each panel we also show the corresponding galaxy UV LFs predicted by \textsc{cat} model at the same redshift. In order to take into account dust extinction, we corrected the galaxy LFs as
\begin{equation}
L_{\rm UV, obs} = L_{\rm UV} \, \rm{exp}[-\Sigma_{\rm gas} \, {\cal D} \, k_{\rm UV}]
\label{eq:DustObs}
\end{equation}
where $\Sigma_{\rm gas}$ is the gas surface density, ${\cal D}$ the dust-to-gas mass ratio and $k_{\rm UV}$ is the extinction coefficient per unit mass in the energy band of interest. The value of $k_{\rm UV}$ has been inferred considering the extinction curve of the Small Magellanic Cloud (SMC) \citep[][]{weingartner2001}. 
We assumed here a simple screen model where the optical depth is computed considering the contribution of all the gas and dust mass inside the galaxy. This will possibly result in an overestimation of the impact of dust obscuration, if compared to more sophisticated two-phase dust extinction models \citep[see e.g.][]{mancini2016}. Therefore, in Fig. \ref{fig:observedUVLF}, we show as a grey shaded area the region enclosed between the best fit of the intrinsic and dust-corrected galaxy UV LFs.
In the \textit{reference} and \textit{super-Edd} models, we observe that dust extinction might heavily affect the galaxy UV luminosity. In the merger-driven scenario, instead, the lower abundance of gas due to the more competitive mechanisms of star formation and BH accretion translates into a smaller difference between the intrinsic and dust-corrected luminosity functions.

However, for all the models and at all redshifts, the UV LF of AGNs dominates at the bright-end, i.e. for magnitudes $M_{\rm 1450} \le -22$. At fainter luminosities, up to $M_{\rm 1450} \sim -20$, the AGN LF still dominates only if the UV emission from the host galaxy is heavily reduced by dust extinction. Finally, at even fainter magnitudes the UV LF is dominated by the stellar emission from the host galaxies. 

In the same figure, \textsc{cat} model predictions are compared to the observations by the SHELLQs survey \citep{matsuoka2018}, the CFHT Legacy Survey \citep{mcgreer2018}, and the Hyper Suprime-Cam Wide Survey \citep{akiyama2018, niida2020} (blue triangles and green diamonds). 
The figure also shows the results of the analysis by \citet[][]{parsa2018} and  \citet{giallongo2019}, who estimated the AGN UV LF using X-ray data, resulting in a LF which should be less affected by obscuration. We also report the results of the COSMOS spectroscopic survey conducted by \citet{boutsia2018}. 

Among the three models, the \textit{reference} one provides the best agreement with observational constraints.
In particular, at fainter magnitudes the predicted LF well trace the observations of \citet[][]{giallongo2019}, \citet[][]{parsa2018} and \citet[][]{boutsia2018}.
In the bright end of the LF, instead, the distribution predicted by the \textit{reference} model shows a more pronounced scatter due to the lower statistics, especially at $z \sim 6$. However, if we fit the obtained data with a broken power-law, we find again a close agreement with the empirical data, despite a slight over-prediction in the number density of bright sources at $z=4$.

The agreement with the data is worse for the \textit{super-Edd} model, which in fact, despite obtaining results similar to the \textit{reference} model at $M_{\rm UV} \gtrsim -19$, fails to reproduce the observed distributions at higher luminosities as pointed out also for the bolometric LF (Section \ref{sec:AGN_LF}).

Lastly, the \textit{merger-driven} model shows an overall distribution very similar to that predicted by the \textit{reference} model but shifted towards higher luminosities, which results in a higher number of sources at bright magnitudes $M_{\rm{1450}} \lesssim -23$ compared to the observed one.

Interestingly we observe that at lower luminosities in all the three model variants the predicted LFs appear to be in better agreement with the LFs obtained through X-ray selection techniques. This suggests, as already outlined by \citet[][]{matsuoka2018} and \citet[][]{giallongo2019}, that the apparent tension between different observational results might be due to an increasing incompleteness at fainter magnitudes or to an higher fraction of obscured AGNs toward lower luminosities.

It is important to note that in none of the models the UV LFs show the clear gap that was present in the bolometric LFs. This is because in the \textit{reference} and \textit{super-Edd} models,
the gap appears at UV magnitudes $M_{\rm{1450}}\ge -10$, that are several orders of magnitudes below the sensitivity limits of current observational facilities.  
Even with the sensitivity of JWST, whose luminosity limit is shown with vertical dashed lines in Fig. \ref{fig:observedUVLF} \citep[][]{griffin2020}, it will be impossible to observe such faint sources, unless with the help of gravitational lensing. 
In addition, at magnitudes $M_{\rm 1450} \gtrsim -22$ deblending techniques need to be applied to discriminate the AGN emission and the emission coming from star formation in the host galaxy.

Using Eq. \ref{eq:bolcorr}, we have also computed the LFs predicted by the three \textsc{cat} models in the $\rm [2 - 10] \, keV$ X-ray energy band.
The results are shown in Fig.\ref{fig:XLFs}. Black data points and error bars represent the binned intrinsic LFs at redshift $z = 4, 5$ and $6$ (from top to bottom) and different columns refer to the three \textsc{cat} model variants that we have considered (\textit{reference}, \textit{super-Edd}, and \textit{merger-driven} models from left to right). In each panel, the dashed lines represent the best-fit to the model predictions and the 
dotted lines show the X-ray LFs contributed by star formation in the host galaxies at the same redshift. The latter component has been estimated
using the empirical relation proposed by \citet{fornasini2018}:
\begin{equation}
L_{\rm XRBs} = 10^{29.98} (1+z)^{0.62} \; M_* + 10^{39.78} (1+z)^{0.2} \; \rm{SFR}^{0.84} 
\label{eq:XRBs}
\end{equation}
where $M_*$ and ${\rm SFR}$ are, respectively, the galaxy stellar mass in solar units and the star formation rate in solar masses per year. 
In Fig. \ref{fig:XLFs} \textsc{cat} predictions are compared with the X-ray LFs obtained by \citet{fiore2012}, \citet{ueda2014}, and \citet{miyaji2015}, which are some of the most complete studies of the AGN X-ray emission up to redshift $z \sim 4 - 6$. We also show the LF obtained by \citet{vito2018} through a wide sample of AGNs at redshift $3.6 > z > 6$. 

Since each of these works accounts differently for the fraction of absorbed and obscured AGNs, we decided to show with a shaded region, for the \textit{reference} and the \textit{merger-driven} models, how the best-fit distribution of the \textsc{cat} XLF changes considering absorption for the Compton-thin AGN population.
We assume here the fraction of un-absorbed quasars $\psi_{\rm X, unabs}$ as a function of the X-ray luminosity as proposed by \citet[][]{ueda2014}:
\begin{equation}
\psi_{\rm X, unabs} = 1 - \rm{min}\big[\psi_{\rm{max}} \; , \;\rm{max}[\psi^* - \beta \, (\rm{Log} \, L_{\rm X} - 43.75)\;,\; \psi_{\rm{min}}] \big] 
\label{eq:UedaObs}
\end{equation}
where $\psi_{\rm{max}}= 0.84$,  $\psi_{\rm{min}} = 0.2$,  $\psi^* \simeq 0.73$ and $L_{\rm X}$ is the $\rm [2-10] \, keV$ luminosity in $\rm erg \, s^{-1}$.
It important to note, however, that the distributions corrected for quasar absorption represent only an upper limit for the un-obscured AGN XLF, since we do not consider the contribution of the population of heavily-obscured Compton-thick AGNs, which might be relevant especially at lower luminosities.
In Fig. \ref{fig:XLFs}, the luminosity limits reported by \citet[][]{griffin2020} for the \textit{Athena} and \textit{Lynx} observatories are also shown, at each redshift, with dashed and dotted grey vertical lines, respectively.

In the X-ray band, the \textit{reference} model provides again a very close agreement with the empirical data. The predicted LF well matches the observations in the entire luminosity range explored, $\sim 10^{42}-10^{46} \, \rm erg \, s^{-1}$, and predicts similar values for the break magnitude of the double power-law distribution at all redshifts.

The X-ray LF produced by the \textit{super-Edd} model has a shape similar to the \textit{reference} one, but systematically under-predicts the number of bright sources above $10^{43} \rm erg \, s^{-1}$.

Finally, the \textit{merger-driven} model predicts a peculiar luminosity distribution. Below $\sim 10^{44} \rm erg \; s^{-1}$, the X-ray LF has a shape similar to the \textit{reference} model but with a slightly larger amplitude. At higher luminosities the distribution is dominated instead by the large number of systems undergoing the bursty post-merger accretion phase, leading to a large scatter and to an over-estimation of the bright-end of the X-ray LF. However, we have to be careful in comparing with empirical data the luminosity of systems during such brief phases of merger-driven accretion. In fact, if these bursts of accretion take place on timescales much shorter than the typical time-step of our model, the time interval $t_Q$ (presented in Eq. \ref{eq:timeAvLF}) during which the BH undergoes an enhanced accretion will be over-estimated, leading to an higher number of sources in the bright-end of the AGN LF.
In addition, these rapidly accreting systems are supposed to be highly affected by obscuration, which would furtherly shorten the duration of the observable burst of luminosity.

A gap in the X-ray LFs appears for both the \textit{reference} and the \textit{Super-Edd} models, just below a luminosity of $\sim 10^{41} \rm erg \; s^{-1}$, as a consequence of the inefficient accretion of \textit{light} black hole seeds.
Starting from the same luminosity, we observe also in the \textit{merger-driven} variant a declining number of sources toward the faint-end of the distribution. In contrast with the previous case, this is due to the very efficient growth of light seed descendants coupled with the increasingly less probable seed formation at lower redshift, as we already observed in Fig. \ref{fig:SMBH_MFevo}.
Nevertheless, the \textit{merger-driven} model shows a significantly smoother decline with respect to the marked gap observed in the \textit{reference} scenario. These different predictions might potentially be used to discern between different accretion mechanisms with future surveys exploring such faint luminosities, as will be discussed in the next section.

\subsubsection{Predictions for future surveys}
\begin{figure*}
\vspace{\baselineskip}
\includegraphics[width=18.5cm]{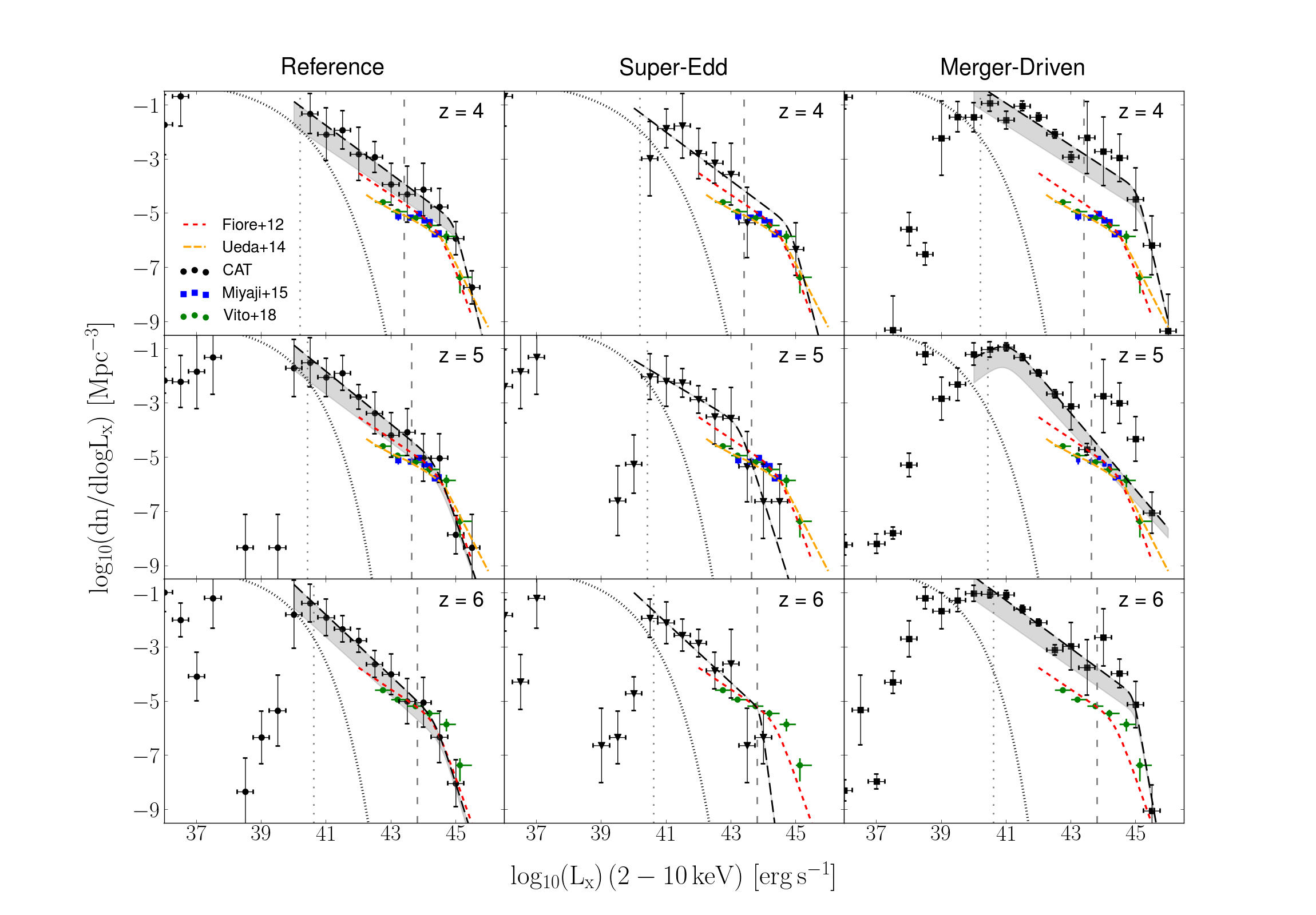}
\caption{The AGN LFs in the $\rm [2-10] \, keV$ X-ray energy band at $z = 4, 5,$ and $6$ (from top to bottom). Each column shows the results obtained in one of the three models examined: the \textit{reference model} (left panels), the \textit{super-Edd} model (central panels) and the \textit{merger-driven} model (right panels). In the left and right panels, the grey shaded regions show how the best-fit distributions change assuming the fraction of absorbed AGNs proposed by \citet[][]{ueda2014}. In each panel, the black dotted line represents the X-ray LF associated with star formation in the host galaxies at the same redshift, that we have estimated using the relation proposed by \citet[][see text]{fornasini2018}. Coloured data represent the observational results by \citet[][red dashed line, $z=4.5, \;6$]{fiore2012}, \citet[][orange dashed line, $4<z<5$]{ueda2014}, \citet[][blue squares, $3<z<5.8$]{miyaji2015} and \citet[][green circles, $3.6<z<6$]{vito2018}. The dashed and dotted grey vertical lines represent the luminosity limits estimated by \citet{griffin2020} for, respectively, \textit{Athena} and \textit{Lynx}.}
\label{fig:XLFs}
\end{figure*}

Distinctive features in the AGN X-ray LFs, characterizing the underlying model of accretion, might represent a key element to discern between different possible scenarios for early BH growth. According to \textsc{CAT} model predictions, such differences should affect primarily the evolution of lower mass BHs, requiring observational constraints at very faint luminosities, $L_{\rm X} \lesssim 10^{41} \, \rm erg \, s^{-1}$.
It is interesting to note that, as shown in Fig. \ref{fig:XLFs}, sources with an X-ray luminosity of $\sim 10^{41} \rm erg \; s^{-1}$ might still be observable with the next generation X-ray observatory \textit{Lynx}, which will explore the faint-end of the LFs, constraining the evolution of \textit{light} black hole seeds and their dominant accretion mode. \textit{Athena} observations will be restricted instead to brighter X-ray sources, constraining the distribution above $10^{43} \rm erg \; s^{-1}$ \citep[][]{griffin2020}.

Still, it will be very challenging to investigate the AGN LFs sufficiently in detail to discern between different BH evolutionary scenarios.
In fact, from Fig. \ref{fig:XLFs} we observe that at luminosities $\leq 10^{41} \rm erg \; s^{-1}$ the X-ray emission produced by stellar binaries formed in the host galaxies becomes comparable to the one contributed by AGNs. Therefore, the investigation of the X-ray luminosity distribution at such faint luminosities has to be carried out with methods that carefully take into account potential contamination from star formation in the host galaxies, in order to reliably discern between different BH growth modes.
Moreover, the presence of a clear gap in the observed AGN number density below a given luminosity, as predicted in our \textit{reference} model, might be covered, in future observations, by large scatters in the theoretical scaling relations or observational parameters assumed to estimate the empirical LFs.

\begin{figure*}
\vspace{\baselineskip}
\includegraphics[width=17.00cm]{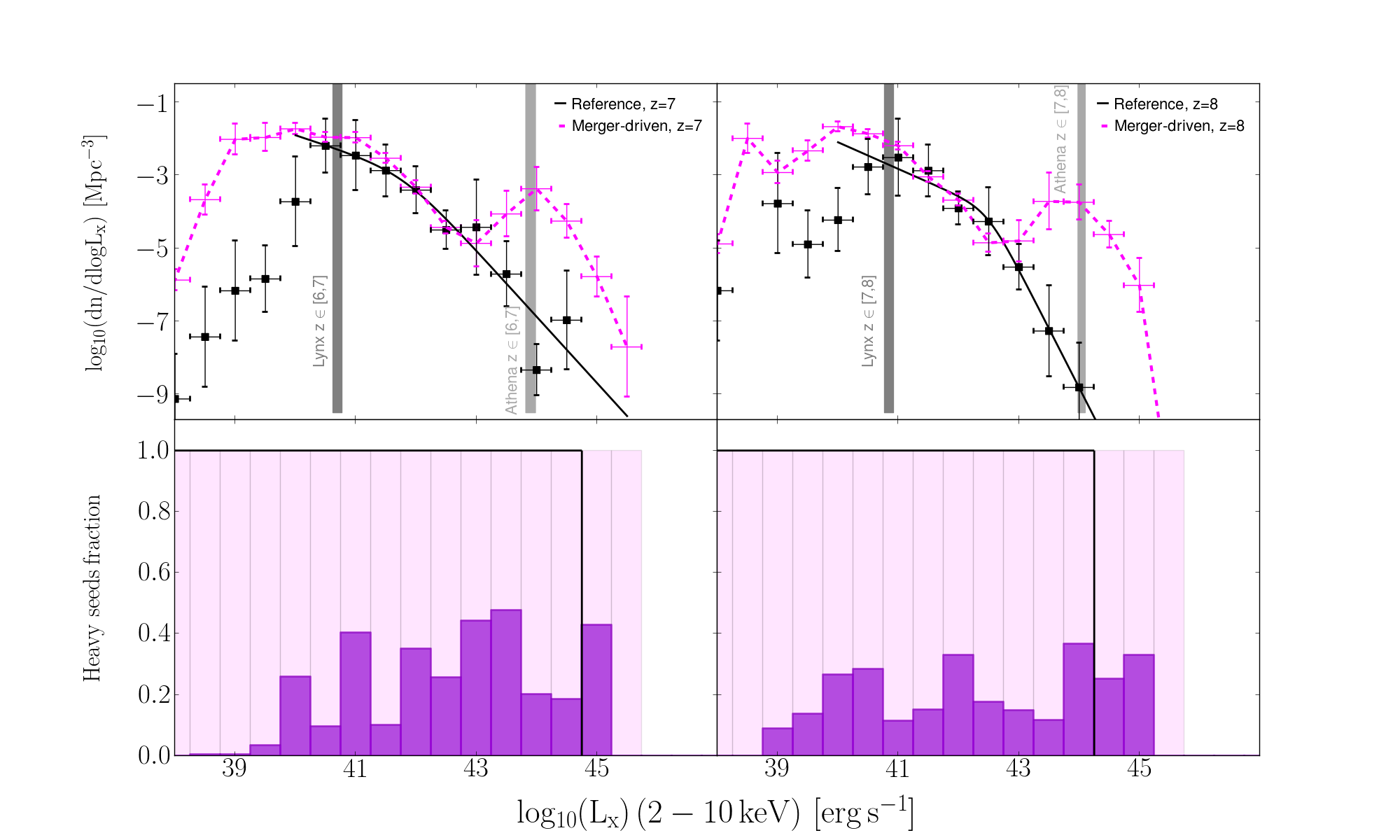}
\centering
\caption{Top panels: AGN X-ray LFs predicted by CAT at $z \sim 7,8$. The black data points are the results of our \textit{reference} model, with the black solid curves representing the best fit of the bright-end distribution at each redshift. As a comparison, we also show the results of the \textit{merger-driven} model with dashed magenta lines and error bars. The sensitivity limits proposed by \citep[][]{griffin2020} for a mission like Lynx and for Athena in the redshift ranges $\rm z \in [6,7], \, [7,8]$ are shown, respectively, as dark grey and light grey vertical thick lines. Bottom panels: percentage of heavy BH seed descendants in each luminosity bin, for the \textit{reference} (black lines) and \textit{merger-driven} (violet histograms) models. In the \textit{reference} model, all the AGNs with $\rm{log_{10}}(L_{\rm{X}}) > 38$ have at least one heavy seed progenitor.}
\label{fig:FutSurv}
\end{figure*}

In order to investigate the observational capability of the \textit{Athena} and \textit{Lynx} missions at even higher redshifts, in Fig. \ref{fig:FutSurv} we compare their forecast sensitivity in the redshift range $\rm z \in [6,7]$ and $[7,8]$ with the AGN X LFs predicted by the \textsc{CAT} \textit{reference} and \textit{merger-driven} models at $z \sim 7$ and $8$, accounting for obscuration.
Similarly to the LFs shown at $z \leq 6$, a major difference between the two models is noticeable at the faint-end of the LFs, where the \textit{reference} model shows a steeper decrease in the number density of fainter objects, due to the inefficient growth of light seeds descendants. This feature, however, appears just below the forecast sensitivity limit of \textit{Lynx} and might be challenging to identify even with such deep observations. Despite that, our predictions suggest that a mission with sensitivity comparable to \textit{Lynx} would have the potential to unveil a large population of AGNs, even at $z \ge 6- 8$. This would open up the possibility to explore BHs with mass $\sim 10^5 - 10^6 \, \rm M_{\odot}$, which dominate the AGN population at $z \leq 15$, regardless of the assumed BH accretion model.

At these high redshifts, the forecast sensitivity of \textit{Athena} would enable to explore only the bright-end of the distribution, at $L_{\rm X} > 10^{44} \, \rm erg \, s^{-1}$. While this hampers the possibility of detecting the dominant BH population at $z \sim 7-8$, 
an interesting feature appears $L_{\rm X} \gtrsim 10^{43} \, \rm erg \, s^{-1}$, where the \textit{merger-driven} model predicts a much larger number density of sources compared to the \textit{reference} one. This is a consequence of the growing rate of galaxy mergers with increasing redshift, which causes a larger fraction of AGNs to experience enhanced, super-Eddington accretion, increasing their luminosity. Such a distinctive feature could be potentially observable by \textit{Athena}. Therefore, the observation of a slower decline of the AGN X-ray LF at higher redshifts might be a hint that early BH evolution is strongly driven by short period of enhanced accretion occurring during galaxy mergers. An important caveat here is that the predicted luminosity (and observability) of such rapidly accreting BHs might be affected by the short timescale of the process and by additional gas obscuration in the nuclear regions, as discussed in Section \ref{sec:LFs}.

In the bottom panel of Fig. \ref{fig:FutSurv} we also show the percentage of BHs with at least one heavy seed progenitor in each luminosity bin. We observe that, in contrast with the results of the \textit{reference} model, where all the AGNs are predicted to descend from at least one heavy BH seed, in the \textit{merger-driven} scenario, more than $50\%$ of AGNs descend from light BH seeds, even at the brightest X-ray luminosities. When BH growth can exceed the Eddington limit and it is triggered by galaxy mergers, the fraction of AGNs descending from heavy BH seeds is sub-dominant and decreases with redshift.

\subsubsection{Black hole - galaxy scaling relations}
In addition to the BH mass and luminosity distribution, we explored \textsc{cat} predictions for the BH-galaxy scaling relations. In particular, it is interesting to understand if and how these relations are affected by the different BH growth scenario assumed in this work. 

\begin{figure*}
\vspace{\baselineskip}
\includegraphics[width=12.00cm]{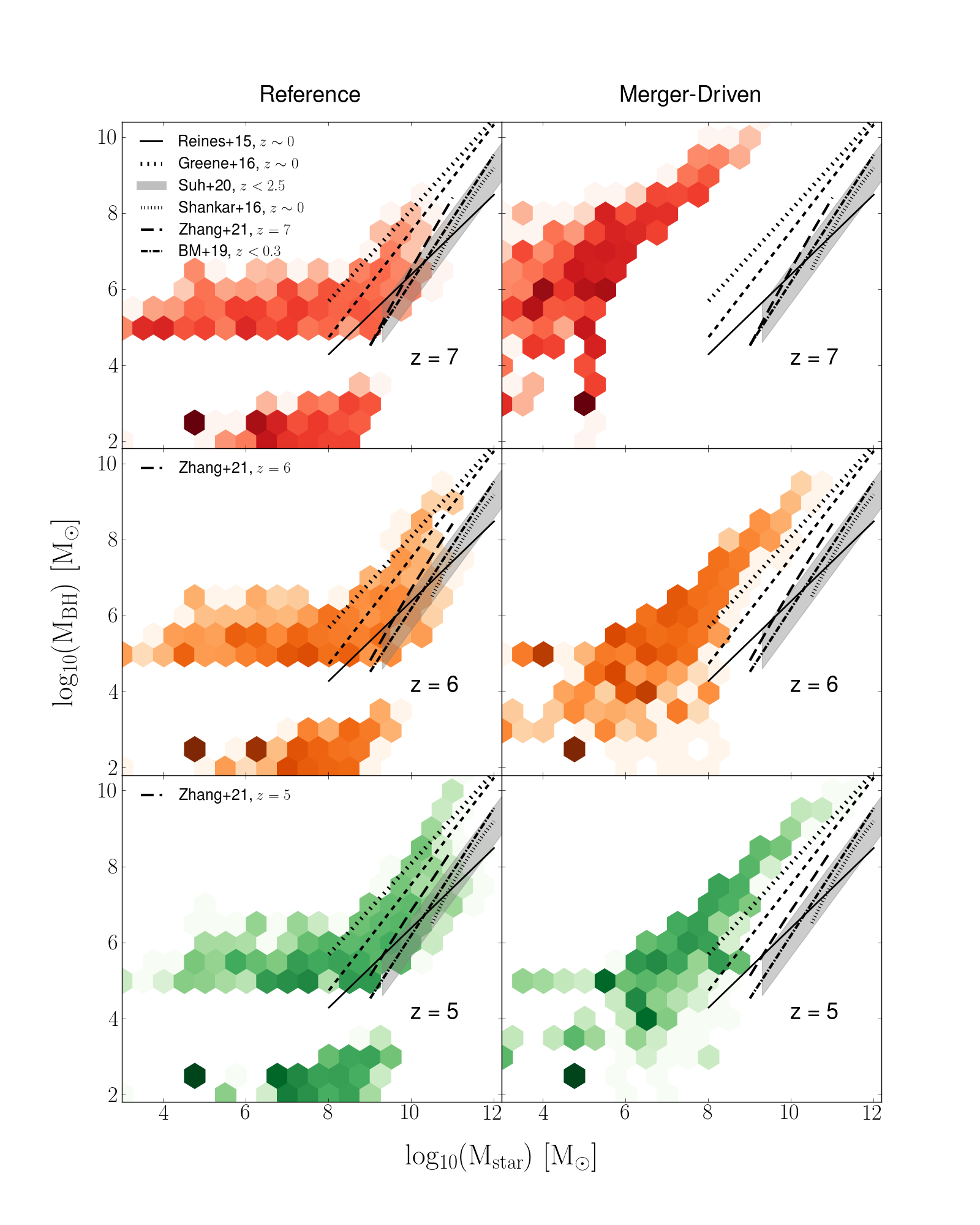}
\centering
\caption{Black hole mass as a function of the galaxy stellar mass for the \textsc{cat} galaxy population at different redshifts, $z = 5,6,7$. Left and right panels represent, respectively, the results obtained in the \textit{reference} and \textit{merger-driven} models. \textsc{cat} results are compared with the predictions of the empirical models proposed by \citet[][solid and dashed lines for, respectively, AGNs and elliptical galaxies]{reines2015}, \citet[][dotted line]{shankar2016}, \citet[][loosely dotted line]{greene2016}, \citet[][dash-dotted line]{baron2019}, \citet[][grey shaded area]{suh2020} and \citet[][long dashed line]{zhang2021}.}
\label{fig:Mbh_Mstar}
\end{figure*}

In Figure \ref{fig:Mbh_Mstar} we show the $M_* - M_{\rm BH}$ relation for our sample of galaxies at $z = 5, \,6$ and $7$. Left and right columns represent, respectively, the results obtained in the \textit{reference} and \textit{merger-driven} models.
We compare \textsc{cat} predictions with several empirical relations based on AGN and galaxy observations in the local Universe, investigating both the unobscured \citep[][]{reines2015, shankar2016, greene2016, suh2020} and obscured \citep[][]{baron2019} AGN population. We also assume as a reference the predictions obtained from the empirical model recently presented by \citet[][]{zhang2021} at the redshifts of interest.

The \textsc{cat} \textit{reference} model shows again a clear gap around $M_{\rm BH} \sim 10^4 \, \rm M_\odot$, which splits the two populations of galaxies hosting a \textit{light} or a \textit{heavy} BH seed descendant. Note that, in less massive galaxies, below $M_* \lesssim 10^{9} \, \rm M_\odot$, the stellar mass is largely independent of the nature of the nuclear BH seed.
The population of massive halos with $M_* \gtrsim 10^{9} \, \rm M_\odot$ hosts instead more massive BHs, showing a correlation between the two quantities. Remarkably, the galaxy population obtained by the \textsc{cat} \textit{reference} model closely reproduces the relations proposed by the empirically-constrained models assumed as comparison.

The \textit{merger-driven} model shows instead a continuous relation between the mass of the galaxy and that of its nuclear BH with an increasing scatter for lower mass galaxies. It is interesting to note that, also in this model, the galaxy population lies on a slope which is very similar to the ones predicted by numerical models. The very early growth of BHs, which characterize this model variant, is clearly noticeable at $z = 7$, where massive BHs populate less massive galaxies compared to the \textit{reference} model. This is probably a natural consequence of the more competitive BH accretion model assumed, which impacts on the efficiency of star formation especially at early times. However, at lower redshift galaxies quickly increase their stellar mass, leading to a final distribution at $z \sim 5$ which presents an offset of $\lesssim 1$ dex compared to the prediction of empirical models.

The above results show that even very different paradigms for BH growth, as the ones considered in the \textit{reference} and \textit{merger-driven} models, lead to similar properties for the galaxy and BH populations at $z \sim 4 - 5$. 
Deep sky observations at higher redshift will be hence crucial to better understand the nature of the first BH seeds and their co-evolution with the host galaxy.  

\section{Discussion}
\label{sec:discussion}

In this section, we first compare the results of \textsc{cat} with independent numerical and semi-analytical studies. We then discuss the main caveats of the model and how we plan to address these in the future. 

\subsection{Comparison with previous studies}
\label{sec:MF_comparison}
The formation and evolution mechanisms of supermassive black holes at high redshift have been the focus of several studies in the last few years. Despite very challenging, observational constraints on the BH mass function at redshift $z \gtrsim 4$ have been proposed by different works, often relying on the correlations between the black hole mass and the properties of the SMBH host \citep{shankar2009,shankar2010, merloni2008, willott2010b}. However, the scaling relations between the BH mass and the host galaxy are mainly determined in the local universe, while their evolution in redshift is still largely uncertain. Hence, in order to reconstruct the BH mass function at higher redshift, these works usually rely on the AGN luminosity distribution as a tracer of the accretion history of SMBHs, assuming the local BH distribution as a boundary condition. Unfortunately, this requires some assumptions on the efficiency of the BH accretion process, such as duty cycle, radiative efficiency, obscured AGN fraction, etc., leading to discrepancies between different results \citep[see][for a detailed discussion]{kelly2012}. This point has to be carefully taken into account in the comparison between the observational constraints and the intrinsic mass function obtained by theoretical models as \textsc{cat}. 
Important efforts on the SMBHs evolution have been carried out also through large-scale cosmological simulations. This approach starts usually with a given cosmological framework and follows the baryonic evolution by zooming in particularly dense regions where the massive nuclear black holes are supposed to form. This class of simulations has shown to be able to characterize properly the evolution of a wide range of massive black holes. However, it still hardly succeeds in reproducing the most extreme sources that we observe at high redshift, which reach masses above $10^9 \,\rm M_\odot$ already at $z \geq 7$ \citep{mortlock2011, yang2020, wang2021}. This is probably due to the large volume needed to resolve the rare overdense regions where the assembly of the most massive SMBHs can take place \citet[see e.g.][]{tenneti2018, tenneti2019}.
The limited volume of this class of simulations leads indeed to a significant underestimation of the number of rare and bright AGNs at very high redshifts, which consequentially affects the predicted mass and luminosity functions \citep[][]{amarantidis2019}.

\begin{figure*}
\vspace{\baselineskip}
\includegraphics[width=14.00cm]{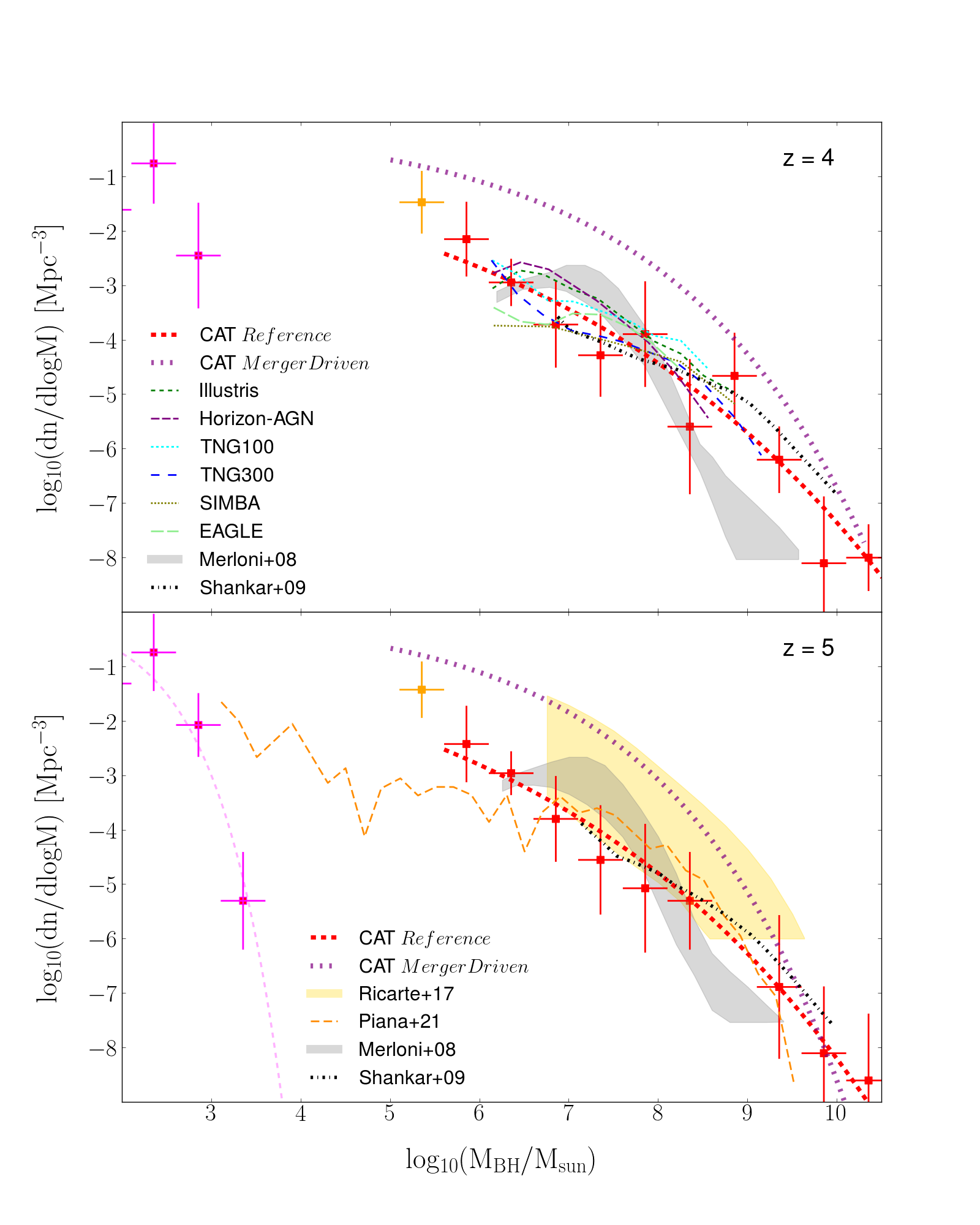}
\centering
\caption{Comparison between the mass function predicted by \textsc{cat} and the results of numerical simulations (upper panel, $z = 4$) and semi-analytical models (lower panel, $z = 5$).
As in Fig.\ref{fig:SMBH_MFevo}, the data points show the binned distribution obtained by \textsc{cat} \textit{reference} model for \textit{light} (magenta) and \textit{heavy} (red, orange) BH descendants, with the red dashed line representing the best-fit of the distribution of \textit{heavy} seed descendants. The violet dotted lines show the best-fit of the BH MF obtained by the \textit{merger-driven} model. In the upper panel, as a comparison, we present with colored dashed lines the results of large-scale cosmological simulations, namely: Illustris (dark green), IllustrisTNG100 (cyan), IllustrisTNG300 (blue), SIMBA (olive) and EAGLE (light green).  In the lower panel we show the mass function obtained by the semi-analytical models by \citet[][yellow shaded region]{ricarte2018a} and \citet[][orange dashed line]{piana2021}.
The grey shaded area and the dashed-dotted black lines represent instead the observational constraints proposed by \citet[][]{merloni2008} and \citet{shankar2009} respectively.}
\label{fig:BHMF_comp}  
\end{figure*}

In Fig. \ref{fig:BHMF_comp} we compare the binned mass function at redshift $z = 4$ obtained by \textsc{cat} \textit{reference} model with the results of some of the most recent and important large-scale cosmological simulations, namely the Illustris \citep{sijacki2015}, IllustrisTNG \citep{weinberger2017}, Horizon-AGN \citep{volonteri2016}, SIMBA \citep{dave2019, Thomas2019} and EAGLE \citep{mcalpine2017, mcalpine2018}. For completeness, we also show the observational constraints obtained from \cite{merloni2008} and \cite{shankar2009}.
We can see that the MF predicted by numerical simulations covers a smaller range of BH masses with respect to our semi-analytical approach.
While at the high-mass end of the distribution this is due to the limited simulation volume, below $\sim 10^6 \, \rm M_\odot$ the nuclear BH population can not be properly modelled due to the resolution limits of the simulations, which dictate the seeding prescription. In fact, since large-scale simulations are not able to well resolve lower mass galaxies, they are forced to seed more massive galaxies with $M_{*} \gtrsim 10^9 \, \rm M_\odot$
with nuclear BHs with mass $\sim 10^{5-6} \, \rm M_\odot$ \citep[][]{habouzit2020}.

The figure shows that numerical simulations provide consistent results at the high mass end of the distribution, above $10^8 \, \rm M_\odot$, while they differ mostly at lower masses, probably as a result of the different sub-grid physics and seeding prescriptions adopted. The BH mass function obtained in our \textit{reference} model is in broad agreement with the results of the numerical simulations, especially in the BH mass range $[ 10^6 - 10^8] M_\odot$, where it is consistent with the observationally constrained MF of \cite{merloni2008}. 
At higher mass values, where the contribution of the most massive and rarer systems becomes relevant, both \textsc{cat} and numerical simulations seem instead to slightly overestimate the number of SMBHs with respect to \cite{merloni2008}. Fitting our binned SMBH mass function with a power law, though, a better agreement is found with the observational constraints obtained by \cite{shankar2009} in the high-mass range, above $ \rm \sim 10^9 \, M_\odot$.

In Fig.\ref{fig:BHMF_comp}, for comparison, we also show the best fit of the BH mass function obtained by \textsc{cat} \textit{merger-driven} model. As anticipated in the previous sections, this model variant predicts a significantly larger number density of BHs with masses $\lesssim 10^9 M_\odot$ compared to the \textit{reference} model and to the results of numerical simulations and observational studies.
This is probably a consequence of the importance that \textit{light} seeds acquire in this alternative scenario, where they are able to efficiently grow across cosmic time and contribute in the building up of the entire BH mass function. 
It has to be pointed out, however, that the mass function predicted by our \textit{merger-driven} scenario at $z \sim 4$ seems in tension with the estimate local BH mass density inferred from the $M_{\rm BH}-M_{*}$ relation \citep[see e.g. the recent work of][]{shankar2020}. This suggests that a more refined modeling of the black hole accretion process, as the one originally proposed by \citet[][]{pezzulli2016}, which assumes a distinguished treatment for the galaxy bulge and disk environments, would be required in this scenario. That, in fact, might ensure a better consistency with the AGN mass and luminosity distributions, maintaining at the same time a good accordance with the global constraints.

In the bottom panel of Fig.\ref{fig:BHMF_comp}, we also compare our results at $z = 5$ with similar semi-analytical studies.
We show in particular the results of different SMBH evolutionary models studied by \citet[][]{ricarte2018a}, considering both burst and steady mode accretion. Despite their focus is on massive BH seeds, mainly DCBHs or extremely fast-growing Pop III remnants, their results in the mass range $[10^7 - 10^{10}] M_\odot$ are in very good agreement with \textsc{cat} predictions.
Interestingly, their results fall just between the best-fit of our \textit{reference} and \textit{merger-driven} models, as we would expect for a mixed accretion scenario as the one considered there.

\textsc{cat} predictions are also compared with the results of the \textit{Delphi} semi-analytic model proposed by \citet[][]{piana2021}. The two models implement similar approaches in investigating the BH mass distribution, following the evolution of both stellar and direct collapse BH progenitors in a cosmological framework. However, \textsc{cat} and \textit{Delphi} present considerably different prescriptions for BH seed formation and growth. In fact, in the latter, BH seeding is less sensitive to the galaxy environmental conditions, since \textit{all} dark matter halos are initially seeded with a stellar ($150 \, \rm M_\odot$) or a DCBH seed ($10^{3-4} \rm \, M_\odot$), depending only on the initial incident LW radiation. The newly formed BHs are then assumed to accrete a given fraction of the available gas, independently of their mass, without exceeding the Eddington limit.
Despite these differences, we find consistent results for the BH MF at $z \sim 5$ in the whole range of masses above $\sim 10^{6.5} \, M_\odot$. 
Interestingly, they do not find any gap in the mass function at lower BH masses. This is probably due to the different BH accretion mode adopted in \textit{Delphi}, which does not depend on the actual BH mass but only on the available gas mass in the host halo, at odds with our \textit{reference} and \textit{super-Edd} models, where BHs are assumed to grow at the Bondi-Hoyle rate. 
In addition, in \textit{Delphi} the initial mass assumed for \textit{heavy} (direct collapse) BHs is different from what we adopt in \textsc{CAT}: while we consider an initial mass of $10^5 \, \rm M_\odot$, in the center of the supposed mass range for this class of seeds \citep{latif2016b, becerra2018}, \citet[][]{piana2021} rely on a more conservative value of $10^{3-4} \, \rm M_\odot$. This difference translates into a continuous mass function ranging between $\sim 10^{3}-10^{10} \, \rm M_\odot$ and possibly accounts for the observed flattening in the mass function with respect to \textsc{cat} predictions at $M_{\rm BH} \lesssim 10^{6.5} \, \rm M_\odot$.
This comparison suggests that the initial mass of \textit{heavy} BH seeds has a strong influence on the shape of the AGN mass and luminosity distributions. In particular, a smaller birth mass of heavy seeds might reduce the predicted gap in the MF and, therefore, have an impact on the observability of the distinctive features characterizing different accretion models. However, it is important to note that predicting the initial mass function of \textit{heavy} BH seeds is very challenging as this likely depends on the adopted conditions for their formation \citep{ferrara2014, bhowmick2021}.

\subsection{Main caveats of the model}

Despite the successes of \textsc{cat} model predictions in reproducing a wide range of observational constraints (see Section \ref{sec:Model Cal})
and the good agreement with independent studies discussed above, we plan to overcome some of the limitations that we have already anticipated in the previous sections and that we discuss below. The first one is the adopted seeding prescription. Although our model allows us to describe the formation of \textit{light} and \textit{heavy} BH seeds depending on the properties of their birth environment, we plan to expand our model following the approach of \citet[][]{sassano2021} to include the formation of \textit{medium-weight} BH seeds by runaway stellar collisions in dense star clusters. This will allow us to track the formation and mass growth of three independent families of BH seeds in a full cosmological context and to predict their observational signatures. 

The paradigm assumed for the BH accretion in low mass galaxies represents an additional crucial point. As observed in Section \ref{sec:AGN_LF}, the LFs obtained in the \textsc{cat} \textit{reference} model predicts an excess of faint sources, below $\sim 10^{42} \rm erg \, s^{-1}$, if compared to observations.
Despite empirical data are not strongly constrained at such low luminosities, this might be a consequence of considering an accretion model which is too efficient for BHs growing in low mass galaxies. In similar semi-analytic models that are calibrated to reproduce the AGN LF at lower redshift, e.g. \citet[][]{piana2021}, BH accretion is in fact artificially inhibited below a typical halo mass of $\sim 10^{12} \, \rm M_\odot$ to mimic the effect of SN feedback suggested by numerical studies \citep[][]{habouzit2017,angles-alcazar2017}. As shown in Figure \ref{fig:BHMF_comp} this lead to a lower number density of BHs below $10^{7} \rm M_\odot$ compared to \textsc{cat} predictions.
In future works, an improved prescription for BH accretion, assuming in particular an $\alpha_{BH}$ parameter (Eq. \ref{eq:BHL}) which depends on the properties of the host galaxies, as proposed in \citet[][]{booth2009}, will be crucial to better understand the nature of these discrepancies. 

A further improvement of the model will be to include the effects of BH dynamics during galaxy mergers. As discussed in Section \ref{sec:BH_mergers},
in the current version of \textsc{cat}, two BHs are assumed to merge only during major mergers of their dark matter host halos, while in minor mergers the heaviest BH is assumed to migrate to the center of the newly formed galaxy while the lightest one remains a satellite and we do not follow further its evolution. A more physical description of BH dynamics 
requires to take into account the impact of processes dominating on different spatial scales/cosmic epochs \citep{begelman1980,armitage2002, armitage2005, colpi2014}:
dynamical friction against background gas and stars regulate the sinking timescale of BHs on kpc-to-pc scales, determining whether a bound system can form \citep[e.g.][and references therein]{capelo2015, tamburello2017,  pfister2017, pfister2019, tamfal2018, bortolas2020, barausse2020}.
On smaller scales, interactions with gaseous disks, stars and other BHs (triple/multiple interactions) instead control the duration of the binary BH hardening phase \citep[e.g.][and references therein]{bortolas2016, bortolas2018, biava2019, arcasedda2019, lima2020}. We plan to study these aspects in future works, building on the first implementation of BH dynamics (triple BH interactions) in \textsc{gqd} recently proposed by \citet{valiante2020}.

\section{Summary and conclusions}
\label{sec:conclusions}

In this work we have used the \textit{Cosmic Archaeology Tool} semi-analytical model to explore how different 
BH accretion and feedback prescriptions affect the formation and evolution of the first galaxies and their nuclear BHs from $z \sim 25$ to $z \sim 4$. In particular, our aim was to use \textsc{cat} has a laboratory to test whether the nature of the first BH seeds and their growth mode may leave observable imprints on the BH mass and luminosity functions. Note that, unlike the majority of numerical and semi-analytic models presented and used as a comparison in Section \ref{sec:MF_comparison}, \textsc{cat} follows the formation of both \textit{light} and \textit{heavy} BH seeds, with a seeding prescription that depends on the physical conditions at their formation sites. This enables us to investigate how these two seed populations contribute to the statistical properties of the BH population at $4 \le z \le 6$.

We explored both a classic BHL accretion scenario, assumed as our \textit{reference} model, and two model variants: the \textit{super-Edd} model, where we removed the Eddington limit for BH accretion, and the \textit{merger-driven} model, where - in addition - enhanced BH accretion episodes are triggered by galaxy mergers.

The most important results of this work are summarized below.
\begin{itemize}
    
    \item The \textit{reference} model shows the best agreement with observational data. The predicted BH luminosity function is in good agreement with several empirical constraints at $z = 4, 5, 6$, especially at higher luminosities. At the faint-end, close to the current observational limits, \textsc{cat} slightly over-predicts the number of sources, suggesting an overstated growth for less massive BHs. The predicted BH mass function is consistent with independent numerical models, as well as with different empirical constraints, although the agreement is limited to specific mass ranges as these empirical constraints do not always provide consistent results.\\
    
    \item The \textit{super-Edd} model seems instead to fail at reproducing the observations. The lack of an Eddington-limit leads to an accelerated early growth of BH seeds and thus to a quick depletion of gas inside their host galaxies. These conditions strongly affect the subsequent growth of nuclear BHs, which fail to reach both masses and luminosities of the most extreme quasars observed at $z \gtrsim 6$. \\
    
    \item The \textit{merger-driven} model predicts global trends for the mass and luminosity distributions very similar to the results of the \textit{reference} model, despite not entirely consistent with empirical constraints and observational data. This tension might be also due to an intrinsic limit, since we implemented a simplified version of the original model developed by \citet[][]{pezzulli2016}. In future works, refined prescriptions for BH accretion, as well as a more accurate treatment of the involved timescales, will enable us to derive tighter constraints.\\
    
    \item We find that the main difference between the \textit{reference} and \textit{merger-driven} accretion models lies in the evolution of the \textit{light} BH seed population. In the first model, their stunted growth leads to a clear gap in the resulting mass and luminosity functions, while in the second model they are able to grow in gas-rich galaxy mergers resulting into continuous BH mass and luminosity distributions where both \textit{light} and \textit{heavy} BH seeds can contribute to the same mass and luminosity bins.\\
    
    \item {The signature of the BH seeds growth mode is imprinted in the BH luminosity function at the very faint end, in a luminosity regime that will be extremely challenging to test observationally. In the X-ray, a mission with a sensitivity comparable to the \textit{Lynx} X-ray observatory might be able to probe the $z \sim 4 - 6$ luminosity function at $L_{\rm X} \leq 10^{41}$ erg/s, possibly unveiling precious hints on the BH accretion mechanism. Interestingly, we find that at $z \geq 6 - 8$ the forecast sensitivity of {\it Athena} could be enough to disentangle the signature of super-Eddington, \textit{merger-driven} BH growth at $L_{\rm X} > 10^{43}$ erg/s by detecting a larger number of AGNs or a milder evolution at the bright-end of the X-ray LF compared to the predictions of the \textit{reference} model.}\\
\end{itemize}

\section*{Acknowledgements}
We wish to thank the anonymous Referee for the constructive suggestions and insightful comments that improved the quality of the paper. We acknowledge support from the Amaldi Research Center funded by the MIUR program \lq \lq Dipartimento di Eccellenza \rq \rq (CUP:B81I18001170001), from the INFN TEONGRAV specific initiative, and the networking support by the COST Action CA16104. LZ acknowledges financial support under ASI/INAF contract 2017-14-H.0.

\section*{Data Availability}
The simulated data underlying this article will be shared on reasonable request to the corresponding author.


\bibliographystyle{mnras}
\bibliography{bibBHmassfunc} 

\begin{thebibliography}{}
\makeatletter
\relax
\def\mn@urlcharsother{\let\do\@makeother \do\$\do\&\do\#\do\^\do\_\do\%\do\~}
\def\mn@doi{\begingroup\mn@urlcharsother \@ifnextchar [ {\mn@doi@}
  {\mn@doi@[]}}
\def\mn@doi@[#1]#2{\def\@tempa{#1}\ifx\@tempa\@empty \href
  {http://dx.doi.org/#2} {doi:#2}\else \href {http://dx.doi.org/#2} {#1}\fi
  \endgroup}
\def\mn@eprint#1#2{\mn@eprint@#1:#2::\@nil}
\def\mn@eprint@arXiv#1{\href {http://arxiv.org/abs/#1} {{\tt arXiv:#1}}}
\def\mn@eprint@dblp#1{\href {http://dblp.uni-trier.de/rec/bibtex/#1.xml}
  {dblp:#1}}
\def\mn@eprint@#1:#2:#3:#4\@nil{\def\@tempa {#1}\def\@tempb {#2}\def\@tempc
  {#3}\ifx \@tempc \@empty \let \@tempc \@tempb \let \@tempb \@tempa \fi \ifx
  \@tempb \@empty \def\@tempb {arXiv}\fi \@ifundefined
  {mn@eprint@\@tempb}{\@tempb:\@tempc}{\expandafter \expandafter \csname
  mn@eprint@\@tempb\endcsname \expandafter{\@tempc}}}

\bibitem[\protect\citeauthoryear{{Abramowicz}, {Czerny}, {Lasota}  \&
  {Szuszkiewicz}}{{Abramowicz} et~al.}{1988}]{abramowicz1988}
{Abramowicz} M.~A.,  {Czerny} B.,  {Lasota} J.~P.,   {Szuszkiewicz} E.,  1988,
  \mn@doi [\apj] {10.1086/166683}, \href
  {https://ui.adsabs.harvard.edu/abs/1988ApJ...332..646A} {332, 646}

\bibitem[\protect\citeauthoryear{{Agarwal} \& {Khochfar}}{{Agarwal} \&
  {Khochfar}}{2015}]{agarwal2015}
{Agarwal} B.,  {Khochfar} S.,  2015, \mn@doi [\mnras] {10.1093/mnras/stu1973},
  \href {https://ui.adsabs.harvard.edu/abs/2015MNRAS.446..160A} {446, 160}

\bibitem[\protect\citeauthoryear{{Akiyama} et~al.,}{{Akiyama}
  et~al.}{2018}]{akiyama2018}
{Akiyama} M.,  et~al., 2018, \mn@doi [\pasj] {10.1093/pasj/psx091}, \href
  {https://ui.adsabs.harvard.edu/abs/2018PASJ...70S..34A} {70, S34}

\bibitem[\protect\citeauthoryear{{Amarantidis} et~al.,}{{Amarantidis}
  et~al.}{2019}]{amarantidis2019}
{Amarantidis} S.,  et~al., 2019, \mn@doi [\mnras] {10.1093/mnras/stz551}, \href
  {https://ui.adsabs.harvard.edu/abs/2019MNRAS.485.2694A} {485, 2694}

\bibitem[\protect\citeauthoryear{{Ananna} et~al.,}{{Ananna}
  et~al.}{2019}]{ananna2019}
{Ananna} T.~T.,  et~al., 2019, \mn@doi [\apj] {10.3847/1538-4357/aafb77}, \href
  {https://ui.adsabs.harvard.edu/abs/2019ApJ...871..240A} {871, 240}

\bibitem[\protect\citeauthoryear{{Andika} et~al.,}{{Andika}
  et~al.}{2020}]{andika2020}
{Andika} I.~T.,  et~al., 2020, \mn@doi [\apj] {10.3847/1538-4357/abb9a6}, \href
  {https://ui.adsabs.harvard.edu/abs/2020ApJ...903...34A} {903, 34}

\bibitem[\protect\citeauthoryear{{Angl{\'e}s-Alc{\'a}zar},
  {Faucher-Gigu{\`e}re}, {Quataert}, {Hopkins}, {Feldmann}, {Torrey}, {Wetzel}
  \& {Kere{\v{s}}}}{{Angl{\'e}s-Alc{\'a}zar} et~al.}{2017}]{angles-alcazar2017}
{Angl{\'e}s-Alc{\'a}zar} D.,  {Faucher-Gigu{\`e}re} C.-A.,  {Quataert} E.,
  {Hopkins} P.~F.,  {Feldmann} R.,  {Torrey} P.,  {Wetzel} A.,   {Kere{\v{s}}}
  D.,  2017, \mn@doi [\mnras] {10.1093/mnrasl/slx161}, \href
  {https://ui.adsabs.harvard.edu/abs/2017MNRAS.472L.109A} {472, L109}

\bibitem[\protect\citeauthoryear{{Arca Sedda}, {Berczik}, {Capuzzo-Dolcetta},
  {Fragione}, {Sobolenko}  \& {Spurzem}}{{Arca Sedda}
  et~al.}{2019}]{arcasedda2019}
{Arca Sedda} M.,  {Berczik} P.,  {Capuzzo-Dolcetta} R.,  {Fragione} G.,
  {Sobolenko} M.,   {Spurzem} R.,  2019, \mn@doi [\mnras]
  {10.1093/mnras/sty3458}, \href
  {https://ui.adsabs.harvard.edu/abs/2019MNRAS.484..520A} {484, 520}

\bibitem[\protect\citeauthoryear{{Armitage} \& {Natarajan}}{{Armitage} \&
  {Natarajan}}{2002}]{armitage2002}
{Armitage} P.~J.,  {Natarajan} P.,  2002, \mn@doi [\apjl] {10.1086/339770},
  \href {https://ui.adsabs.harvard.edu/abs/2002ApJ...567L...9A} {567, L9}

\bibitem[\protect\citeauthoryear{{Armitage} \& {Natarajan}}{{Armitage} \&
  {Natarajan}}{2005}]{armitage2005}
{Armitage} P.~J.,  {Natarajan} P.,  2005, \mn@doi [\apj] {10.1086/497108},
  \href {https://ui.adsabs.harvard.edu/abs/2005ApJ...634..921A} {634, 921}

\bibitem[\protect\citeauthoryear{{Ba{\~n}ados} et~al.,}{{Ba{\~n}ados}
  et~al.}{2016}]{banados2016}
{Ba{\~n}ados} E.,  et~al., 2016, \mn@doi [\apjs] {10.3847/0067-0049/227/1/11},
  \href {https://ui.adsabs.harvard.edu/abs/2016ApJS..227...11B} {227, 11}

\bibitem[\protect\citeauthoryear{Ba{\~n}ados et~al.,}{Ba{\~n}ados
  et~al.}{2018}]{banados2018}
Ba{\~n}ados E.,  et~al., 2018, Nature, 553, 473

\bibitem[\protect\citeauthoryear{{Barausse}, {Dvorkin}, {Tremmel}, {Volonteri}
  \& {Bonetti}}{{Barausse} et~al.}{2020}]{barausse2020}
{Barausse} E.,  {Dvorkin} I.,  {Tremmel} M.,  {Volonteri} M.,   {Bonetti} M.,
  2020, arXiv e-prints, \href
  {https://ui.adsabs.harvard.edu/abs/2020arXiv200603065B} {p. arXiv:2006.03065}

\bibitem[\protect\citeauthoryear{{Baron} \& {M{\'e}nard}}{{Baron} \&
  {M{\'e}nard}}{2019}]{baron2019}
{Baron} D.,  {M{\'e}nard} B.,  2019, \mn@doi [\mnras] {10.1093/mnras/stz1546},
  \href {https://ui.adsabs.harvard.edu/abs/2019MNRAS.487.3404B} {487, 3404}

\bibitem[\protect\citeauthoryear{{Baugh}}{{Baugh}}{2006}]{baugh2006}
{Baugh} C.~M.,  2006, \mn@doi [Reports on Progress in Physics]
  {10.1088/0034-4885/69/12/R02}, \href
  {https://ui.adsabs.harvard.edu/abs/2006RPPh...69.3101B} {69, 3101}

\bibitem[\protect\citeauthoryear{{Becerra}, {Greif}, {Springel}  \&
  {Hernquist}}{{Becerra} et~al.}{2015}]{becerra2015}
{Becerra} F.,  {Greif} T.~H.,  {Springel} V.,   {Hernquist} L.~E.,  2015,
  \mn@doi [\mnras] {10.1093/mnras/stu2284}, \href
  {https://ui.adsabs.harvard.edu/abs/2015MNRAS.446.2380B} {446, 2380}

\bibitem[\protect\citeauthoryear{{Becerra}, {Marinacci}, {Bromm}  \&
  {Hernquist}}{{Becerra} et~al.}{2018}]{becerra2018}
{Becerra} F.,  {Marinacci} F.,  {Bromm} V.,   {Hernquist} L.~E.,  2018, \mn@doi
  [\mnras] {10.1093/mnras/sty2210}, \href
  {https://ui.adsabs.harvard.edu/abs/2018MNRAS.480.5029B} {480, 5029}

\bibitem[\protect\citeauthoryear{{Begelman}, {Blandford}  \& {Rees}}{{Begelman}
  et~al.}{1980}]{begelman1980}
{Begelman} M.~C.,  {Blandford} R.~D.,   {Rees} M.~J.,  1980, \mn@doi [\nat]
  {10.1038/287307a0}, \href {http://adsabs.harvard.edu/abs/1980Natur.287..307B}
  {287, 307}

\bibitem[\protect\citeauthoryear{{Bellovary}, {Volonteri}, {Governato}, {Shen},
  {Quinn}  \& {Wadsley}}{{Bellovary} et~al.}{2011}]{bellovary2011}
{Bellovary} J.,  {Volonteri} M.,  {Governato} F.,  {Shen} S.,  {Quinn} T.,
  {Wadsley} J.,  2011, \mn@doi [\apj] {10.1088/0004-637X/742/1/13}, \href
  {https://ui.adsabs.harvard.edu/abs/2011ApJ...742...13B} {742, 13}

\bibitem[\protect\citeauthoryear{{Bhowmick}, {Blecha}, {Torrey}, {Kelley},
  {Vogelsberger}, {Nelson}, {Weinberger}  \& {Hernquist}}{{Bhowmick}
  et~al.}{2021}]{bhowmick2021}
{Bhowmick} A.~K.,  {Blecha} L.,  {Torrey} P.,  {Kelley} L.~Z.,  {Vogelsberger}
  M.,  {Nelson} D.,  {Weinberger} R.,   {Hernquist} L.,  2021, arXiv e-prints,
  \href {https://ui.adsabs.harvard.edu/abs/2021arXiv210706899B} {p.
  arXiv:2107.06899}

\bibitem[\protect\citeauthoryear{{Biava}, {Colpi}, {Capelo}, {Bonetti},
  {Volonteri}, {Tamfal}, {Mayer}  \& {Sesana}}{{Biava}
  et~al.}{2019}]{biava2019}
{Biava} N.,  {Colpi} M.,  {Capelo} P.~R.,  {Bonetti} M.,  {Volonteri} M.,
  {Tamfal} T.,  {Mayer} L.,   {Sesana} A.,  2019, \mn@doi [\mnras]
  {10.1093/mnras/stz1614}, \href
  {https://ui.adsabs.harvard.edu/abs/2019MNRAS.487.4985B} {487, 4985}

\bibitem[\protect\citeauthoryear{{Boco}, {Lapi}  \& {Danese}}{{Boco}
  et~al.}{2020}]{boco2020}
{Boco} L.,  {Lapi} A.,   {Danese} L.,  2020, \mn@doi [\apj]
  {10.3847/1538-4357/ab7446}, \href
  {https://ui.adsabs.harvard.edu/abs/2020ApJ...891...94B} {891, 94}

\bibitem[\protect\citeauthoryear{{Bondi}}{{Bondi}}{1952}]{bondi1952}
{Bondi} H.,  1952, \mn@doi [\mnras] {10.1093/mnras/112.2.195}, \href
  {https://ui.adsabs.harvard.edu/abs/1952MNRAS.112..195B} {112, 195}

\bibitem[\protect\citeauthoryear{{Bonoli}, {Mayer}  \& {Callegari}}{{Bonoli}
  et~al.}{2014}]{Bonoli2014}
{Bonoli} S.,  {Mayer} L.,   {Callegari} S.,  2014, \mn@doi [\mnras]
  {10.1093/mnras/stt1990}, \href
  {https://ui.adsabs.harvard.edu/abs/2014MNRAS.437.1576B} {437, 1576}

\bibitem[\protect\citeauthoryear{{Booth} \& {Schaye}}{{Booth} \&
  {Schaye}}{2009}]{booth2009}
{Booth} C.~M.,  {Schaye} J.,  2009, \mn@doi [\mnras]
  {10.1111/j.1365-2966.2009.15043.x}, \href
  {https://ui.adsabs.harvard.edu/abs/2009MNRAS.398...53B} {398, 53}

\bibitem[\protect\citeauthoryear{{Bortolas}, {Gualandris}, {Dotti}, {Spera}  \&
  {Mapelli}}{{Bortolas} et~al.}{2016}]{bortolas2016}
{Bortolas} E.,  {Gualandris} A.,  {Dotti} M.,  {Spera} M.,   {Mapelli} M.,
  2016, \mn@doi [\mnras] {10.1093/mnras/stw1372}, \href
  {https://ui.adsabs.harvard.edu/abs/2016MNRAS.461.1023B} {461, 1023}

\bibitem[\protect\citeauthoryear{{Bortolas}, {Mapelli}  \& {Spera}}{{Bortolas}
  et~al.}{2018}]{bortolas2018}
{Bortolas} E.,  {Mapelli} M.,   {Spera} M.,  2018, \mn@doi [\mnras]
  {10.1093/mnras/stx2795}, \href
  {https://ui.adsabs.harvard.edu/abs/2018MNRAS.474.1054B} {474, 1054}

\bibitem[\protect\citeauthoryear{{Bortolas}, {Capelo}, {Zana}, {Mayer},
  {Bonetti}, {Dotti}, {Davies}  \& {Madau}}{{Bortolas}
  et~al.}{2020}]{bortolas2020}
{Bortolas} E.,  {Capelo} P.~R.,  {Zana} T.,  {Mayer} L.,  {Bonetti} M.,
  {Dotti} M.,  {Davies} M.~B.,   {Madau} P.,  2020, arXiv e-prints, \href
  {https://ui.adsabs.harvard.edu/abs/2020arXiv200502409B} {p. arXiv:2005.02409}

\bibitem[\protect\citeauthoryear{{Boutsia}, {Grazian}, {Giallongo}, {Fiore}  \&
  {Civano}}{{Boutsia} et~al.}{2018}]{boutsia2018}
{Boutsia} K.,  {Grazian} A.,  {Giallongo} E.,  {Fiore} F.,   {Civano} F.,
  2018, \mn@doi [\apj] {10.3847/1538-4357/aae6c7}, \href
  {https://ui.adsabs.harvard.edu/abs/2018ApJ...869...20B} {869, 20}

\bibitem[\protect\citeauthoryear{{Bouwens} et~al.,}{{Bouwens}
  et~al.}{2012}]{bouwens2012}
{Bouwens} R.~J.,  et~al., 2012, \mn@doi [\apj] {10.1088/0004-637X/754/2/83},
  \href {https://ui.adsabs.harvard.edu/abs/2012ApJ...754...83B} {754, 83}

\bibitem[\protect\citeauthoryear{{Bouwens} et~al.,}{{Bouwens}
  et~al.}{2014}]{bouwens2014}
{Bouwens} R.~J.,  et~al., 2014, \mn@doi [\apj] {10.1088/0004-637X/795/2/126},
  \href {https://ui.adsabs.harvard.edu/abs/2014ApJ...795..126B} {795, 126}

\bibitem[\protect\citeauthoryear{{Bromm}}{{Bromm}}{2013}]{bromm2013}
{Bromm} V.,  2013, \mn@doi [Reports on Progress in Physics]
  {10.1088/0034-4885/76/11/112901}, \href
  {https://ui.adsabs.harvard.edu/abs/2013RPPh...76k2901B} {76, 112901}

\bibitem[\protect\citeauthoryear{Bromm \& Loeb}{Bromm \&
  Loeb}{2003}]{bromm2003}
Bromm V.,  Loeb A.,  2003, The Astrophysical Journal, 596, 34

\bibitem[\protect\citeauthoryear{{Callegari}, {Mayer}, {Kazantzidis}, {Colpi},
  {Governato}, {Quinn}  \& {Wadsley}}{{Callegari} et~al.}{2009}]{callegari2009}
{Callegari} S.,  {Mayer} L.,  {Kazantzidis} S.,  {Colpi} M.,  {Governato} F.,
  {Quinn} T.,   {Wadsley} J.,  2009, \mn@doi [\apjl]
  {10.1088/0004-637X/696/1/L89}, \href
  {http://adsabs.harvard.edu/abs/2009ApJ...696L..89C} {696, L89}

\bibitem[\protect\citeauthoryear{{Campanelli}, {Lousto}, {Zlochower}  \&
  {Merritt}}{{Campanelli} et~al.}{2007}]{campanelli2007}
{Campanelli} M.,  {Lousto} C.,  {Zlochower} Y.,   {Merritt} D.,  2007, \mn@doi
  [\apjl] {10.1086/516712}, \href
  {https://ui.adsabs.harvard.edu/abs/2007ApJ...659L...5C} {659, L5}

\bibitem[\protect\citeauthoryear{{Capelo}, {Volonteri}, {Dotti}, {Bellovary},
  {Mayer}  \& {Governato}}{{Capelo} et~al.}{2015}]{capelo2015}
{Capelo} P.~R.,  {Volonteri} M.,  {Dotti} M.,  {Bellovary} J.~M.,  {Mayer} L.,
   {Governato} F.,  2015, \mn@doi [\mnras] {10.1093/mnras/stu2500}, \href
  {https://ui.adsabs.harvard.edu/abs/2015MNRAS.447.2123C} {447, 2123}

\bibitem[\protect\citeauthoryear{{Chehade} et~al.,}{{Chehade}
  et~al.}{2018}]{chehade2018}
{Chehade} B.,  et~al., 2018, \mn@doi [\mnras] {10.1093/mnras/sty690}, \href
  {https://ui.adsabs.harvard.edu/abs/2018MNRAS.478.1649C} {478, 1649}

\bibitem[\protect\citeauthoryear{Chon \& Omukai}{Chon \&
  Omukai}{2020}]{chon2020}
Chon S.,  Omukai K.,  2020, Supermassive Star Formation via Super Competitive
  Accretion in Slightly Metal-enriched Clouds (\mn@eprint {arXiv} {2001.06491})

\bibitem[\protect\citeauthoryear{{Chon}, {Hirano}, {Hosokawa}  \&
  {Yoshida}}{{Chon} et~al.}{2016}]{chon2016}
{Chon} S.,  {Hirano} S.,  {Hosokawa} T.,   {Yoshida} N.,  2016, \mn@doi [\apj]
  {10.3847/0004-637X/832/2/134}, \href
  {https://ui.adsabs.harvard.edu/abs/2016ApJ...832..134C} {832, 134}

\bibitem[\protect\citeauthoryear{{Chon}, {Hosokawa}  \& {Omukai}}{{Chon}
  et~al.}{2021}]{chon2021}
{Chon} S.,  {Hosokawa} T.,   {Omukai} K.,  2021, \mn@doi [\mnras]
  {10.1093/mnras/stab061}, \href
  {https://ui.adsabs.harvard.edu/abs/2021MNRAS.502..700C} {502, 700}

\bibitem[\protect\citeauthoryear{{Cole}, {Lacey}, {Baugh}  \& {Frenk}}{{Cole}
  et~al.}{2000}]{cole2000}
{Cole} S.,  {Lacey} C.~G.,  {Baugh} C.~M.,   {Frenk} C.~S.,  2000, \mn@doi
  [\mnras] {10.1046/j.1365-8711.2000.03879.x}, \href
  {https://ui.adsabs.harvard.edu/abs/2000MNRAS.319..168C} {319, 168}

\bibitem[\protect\citeauthoryear{{Cole}, {Helly}, {Frenk}  \&
  {Parkinson}}{{Cole} et~al.}{2008}]{cole2008}
{Cole} S.,  {Helly} J.,  {Frenk} C.~S.,   {Parkinson} H.,  2008, \mn@doi
  [\mnras] {10.1111/j.1365-2966.2007.12516.x}, \href
  {https://ui.adsabs.harvard.edu/abs/2008MNRAS.383..546C} {383, 546}

\bibitem[\protect\citeauthoryear{{Colpi}}{{Colpi}}{2014}]{colpi2014}
{Colpi} M.,  2014, \mn@doi [\ssr] {10.1007/s11214-014-0067-1}, \href
  {http://adsabs.harvard.edu/abs/2014SSRv..183..189C} {183, 189}

\bibitem[\protect\citeauthoryear{{Dav{\'e}}, {Angl{\'e}s-Alc{\'a}zar},
  {Narayanan}, {Li}, {Rafieferantsoa}  \& {Appleby}}{{Dav{\'e}}
  et~al.}{2019}]{dave2019}
{Dav{\'e}} R.,  {Angl{\'e}s-Alc{\'a}zar} D.,  {Narayanan} D.,  {Li} Q.,
  {Rafieferantsoa} M.~H.,   {Appleby} S.,  2019, \mn@doi [\mnras]
  {10.1093/mnras/stz937}, \href
  {https://ui.adsabs.harvard.edu/abs/2019MNRAS.486.2827D} {486, 2827}

\bibitem[\protect\citeauthoryear{{Dayal}, {Ward}  \& {Cockell}}{{Dayal}
  et~al.}{2016}]{dayal2016}
{Dayal} P.,  {Ward} M.,   {Cockell} C.,  2016, arXiv e-prints, \href
  {https://ui.adsabs.harvard.edu/abs/2016arXiv160609224D} {p. arXiv:1606.09224}

\bibitem[\protect\citeauthoryear{{Dayal}, {Rossi}, {Shiralilou}, {Piana},
  {Choudhury}  \& {Volonteri}}{{Dayal} et~al.}{2019}]{dayal2019}
{Dayal} P.,  {Rossi} E.~M.,  {Shiralilou} B.,  {Piana} O.,  {Choudhury} T.~R.,
   {Volonteri} M.,  2019, \mn@doi [\mnras] {10.1093/mnras/stz897}, \href
  {https://ui.adsabs.harvard.edu/abs/2019MNRAS.486.2336D} {486, 2336}

\bibitem[\protect\citeauthoryear{{Dayal} et~al.,}{{Dayal}
  et~al.}{2020}]{dayal2020}
{Dayal} P.,  et~al., 2020, \mn@doi [\mnras] {10.1093/mnras/staa1138}, \href
  {https://ui.adsabs.harvard.edu/abs/2020MNRAS.495.3065D} {495, 3065}

\bibitem[\protect\citeauthoryear{{De Lucia}}{{De Lucia}}{2019}]{delucia2019}
{De Lucia} G.,  2019, \mn@doi [Galaxies] {10.3390/galaxies7020056}, \href
  {https://ui.adsabs.harvard.edu/abs/2019Galax...7...56D} {7, 56}

\bibitem[\protect\citeauthoryear{{De Rosa}, {Decarli}, {Walter}, {Fan},
  {Jiang}, {Kurk}, {Pasquali}  \& {Rix}}{{De Rosa} et~al.}{2011}]{derosa2011}
{De Rosa} G.,  {Decarli} R.,  {Walter} F.,  {Fan} X.,  {Jiang} L.,  {Kurk} J.,
  {Pasquali} A.,   {Rix} H.~W.,  2011, \mn@doi [\apj]
  {10.1088/0004-637X/739/2/56}, \href
  {https://ui.adsabs.harvard.edu/abs/2011ApJ...739...56D} {739, 56}

\bibitem[\protect\citeauthoryear{{DeGraf}, {Di Matteo}, {Khandai}, {Croft},
  {Lopez}  \& {Springel}}{{DeGraf} et~al.}{2012}]{degraf2012}
{DeGraf} C.,  {Di Matteo} T.,  {Khandai} N.,  {Croft} R.,  {Lopez} J.,
  {Springel} V.,  2012, \mn@doi [\mnras] {10.1111/j.1365-2966.2012.21294.x},
  \href {https://ui.adsabs.harvard.edu/abs/2012MNRAS.424.1892D} {424, 1892}

\bibitem[\protect\citeauthoryear{{DeGraf}, {Di Matteo}, {Treu}, {Feng}, {Woo}
  \& {Park}}{{DeGraf} et~al.}{2015}]{degraf2015}
{DeGraf} C.,  {Di Matteo} T.,  {Treu} T.,  {Feng} Y.,  {Woo} J.~H.,   {Park}
  D.,  2015, \mn@doi [\mnras] {10.1093/mnras/stv2002}, \href
  {https://ui.adsabs.harvard.edu/abs/2015MNRAS.454..913D} {454, 913}

\bibitem[\protect\citeauthoryear{Devecchi \& Volonteri}{Devecchi \&
  Volonteri}{2009}]{devecchi2009}
Devecchi B.,  Volonteri M.,  2009, The Astrophysical Journal, 694, 302

\bibitem[\protect\citeauthoryear{{Devecchi}, {Volonteri}, {Rossi}, {Colpi}  \&
  {Portegies Zwart}}{{Devecchi} et~al.}{2012}]{devecchi2012}
{Devecchi} B.,  {Volonteri} M.,  {Rossi} E.~M.,  {Colpi} M.,   {Portegies
  Zwart} S.,  2012, \mn@doi [\mnras] {10.1111/j.1365-2966.2012.20406.x}, \href
  {https://ui.adsabs.harvard.edu/abs/2012MNRAS.421.1465D} {421, 1465}

\bibitem[\protect\citeauthoryear{{Di Matteo}, {Springel}  \& {Hernquist}}{{Di
  Matteo} et~al.}{2005}]{dimatteo2005}
{Di Matteo} T.,  {Springel} V.,   {Hernquist} L.,  2005, \mn@doi [\nat]
  {10.1038/nature03335}, \href
  {https://ui.adsabs.harvard.edu/abs/2005Natur.433..604D} {433, 604}

\bibitem[\protect\citeauthoryear{{Di Matteo}, {Khandai}, {DeGraf}, {Feng},
  {Croft}, {Lopez}  \& {Springel}}{{Di Matteo} et~al.}{2012}]{dimatteo2012}
{Di Matteo} T.,  {Khandai} N.,  {DeGraf} C.,  {Feng} Y.,  {Croft} R.~A.~C.,
  {Lopez} J.,   {Springel} V.,  2012, \mn@doi [\apjl]
  {10.1088/2041-8205/745/2/L29}, \href
  {https://ui.adsabs.harvard.edu/abs/2012ApJ...745L..29D} {745, L29}

\bibitem[\protect\citeauthoryear{{Di Matteo}, {Croft}, {Feng}, {Waters}  \&
  {Wilkins}}{{Di Matteo} et~al.}{2017}]{dimatteo2017}
{Di Matteo} T.,  {Croft} R. A.~C.,  {Feng} Y.,  {Waters} D.,   {Wilkins} S.,
  2017, \mn@doi [\mnras] {10.1093/mnras/stx319}, \href
  {https://ui.adsabs.harvard.edu/abs/2017MNRAS.467.4243D} {467, 4243}

\bibitem[\protect\citeauthoryear{{Duncan} et~al.,}{{Duncan}
  et~al.}{2014}]{duncan2014}
{Duncan} K.,  et~al., 2014, \mn@doi [\mnras] {10.1093/mnras/stu1622}, \href
  {https://ui.adsabs.harvard.edu/abs/2014MNRAS.444.2960D} {444, 2960}

\bibitem[\protect\citeauthoryear{{Dunn}, {Holley-Bockelmann}  \&
  {Bellovary}}{{Dunn} et~al.}{2020}]{dunn2020}
{Dunn} G.,  {Holley-Bockelmann} K.,   {Bellovary} J.,  2020, \mn@doi [\apj]
  {10.3847/1538-4357/ab7cd2}, \href
  {https://ui.adsabs.harvard.edu/abs/2020ApJ...896...72D} {896, 72}

\bibitem[\protect\citeauthoryear{{Duras} et~al.,}{{Duras}
  et~al.}{2020}]{duras2020}
{Duras} F.,  et~al., 2020, \mn@doi [\aap] {10.1051/0004-6361/201936817}, \href
  {https://ui.adsabs.harvard.edu/abs/2020A&A...636A..73D} {636, A73}

\bibitem[\protect\citeauthoryear{{Eilers}, {Davies}  \& {Hennawi}}{{Eilers}
  et~al.}{2018}]{eilers2018}
{Eilers} A.-C.,  {Davies} F.~B.,   {Hennawi} J.~F.,  2018, \mn@doi [\apj]
  {10.3847/1538-4357/aad4fd}, \href
  {https://ui.adsabs.harvard.edu/abs/2018ApJ...864...53E} {864, 53}

\bibitem[\protect\citeauthoryear{{Eilers} et~al.,}{{Eilers}
  et~al.}{2020}]{eilers2020}
{Eilers} A.-C.,  et~al., 2020, \mn@doi [\apj] {10.3847/1538-4357/aba52e}, \href
  {https://ui.adsabs.harvard.edu/abs/2020ApJ...900...37E} {900, 37}

\bibitem[\protect\citeauthoryear{{Ellis} et~al.,}{{Ellis}
  et~al.}{2013}]{ellis2013}
{Ellis} R.~S.,  et~al., 2013, \mn@doi [\apjl] {10.1088/2041-8205/763/1/L7},
  \href {https://ui.adsabs.harvard.edu/abs/2013ApJ...763L...7E} {763, L7}

\bibitem[\protect\citeauthoryear{{Fan}, {Narayanan}, {Strauss}, {White},
  {Becker}, {Pentericci}  \& {Rix}}{{Fan} et~al.}{2002}]{fan2002}
{Fan} X.,  {Narayanan} V.~K.,  {Strauss} M.~A.,  {White} R.~L.,  {Becker}
  R.~H.,  {Pentericci} L.,   {Rix} H.-W.,  2002, \mn@doi [\aj]
  {10.1086/339030}, \href
  {https://ui.adsabs.harvard.edu/abs/2002AJ....123.1247F} {123, 1247}

\bibitem[\protect\citeauthoryear{{Fan} et~al.,}{{Fan} et~al.}{2003}]{fan2003}
{Fan} X.,  et~al., 2003, \mn@doi [\aj] {10.1086/368246}, \href
  {https://ui.adsabs.harvard.edu/abs/2003AJ....125.1649F} {125, 1649}

\bibitem[\protect\citeauthoryear{{Feng}, {Di-Matteo}, {Croft}, {Bird},
  {Battaglia}  \& {Wilkins}}{{Feng} et~al.}{2016}]{feng2016}
{Feng} Y.,  {Di-Matteo} T.,  {Croft} R.~A.,  {Bird} S.,  {Battaglia} N.,
  {Wilkins} S.,  2016, \mn@doi [\mnras] {10.1093/mnras/stv2484}, \href
  {https://ui.adsabs.harvard.edu/abs/2016MNRAS.455.2778F} {455, 2778}

\bibitem[\protect\citeauthoryear{{Ferrara}, {Salvadori}, {Yue}  \&
  {Schleicher}}{{Ferrara} et~al.}{2014}]{ferrara2014}
{Ferrara} A.,  {Salvadori} S.,  {Yue} B.,   {Schleicher} D.,  2014, \mn@doi
  [\mnras] {10.1093/mnras/stu1280}, \href
  {https://ui.adsabs.harvard.edu/abs/2014MNRAS.443.2410F} {443, 2410}

\bibitem[\protect\citeauthoryear{{Fiore} et~al.,}{{Fiore}
  et~al.}{2012}]{fiore2012}
{Fiore} F.,  et~al., 2012, \mn@doi [\aap] {10.1051/0004-6361/201117581}, \href
  {https://ui.adsabs.harvard.edu/abs/2012A&A...537A..16F} {537, A16}

\bibitem[\protect\citeauthoryear{{Fornasini}, {Civano}, {Fabbiano}, {Elvis},
  {Marchesi}, {Miyaji}  \& {Zezas}}{{Fornasini} et~al.}{2018}]{fornasini2018}
{Fornasini} F.~M.,  {Civano} F.,  {Fabbiano} G.,  {Elvis} M.,  {Marchesi} S.,
  {Miyaji} T.,   {Zezas} A.,  2018, \mn@doi [\apj] {10.3847/1538-4357/aada4e},
  \href {https://ui.adsabs.harvard.edu/abs/2018ApJ...865...43F} {865, 43}

\bibitem[\protect\citeauthoryear{{Giallongo} et~al.,}{{Giallongo}
  et~al.}{2019}]{giallongo2019}
{Giallongo} E.,  et~al., 2019, \mn@doi [\apj] {10.3847/1538-4357/ab39e1}, \href
  {https://ui.adsabs.harvard.edu/abs/2019ApJ...884...19G} {884, 19}

\bibitem[\protect\citeauthoryear{{Gonz{\'a}lez}, {Labb{\'e}}, {Bouwens},
  {Illingworth}, {Franx}  \& {Kriek}}{{Gonz{\'a}lez}
  et~al.}{2011}]{gonzalez2011}
{Gonz{\'a}lez} V.,  {Labb{\'e}} I.,  {Bouwens} R.~J.,  {Illingworth} G.,
  {Franx} M.,   {Kriek} M.,  2011, \mn@doi [\apjl]
  {10.1088/2041-8205/735/2/L34}, \href
  {https://ui.adsabs.harvard.edu/abs/2011ApJ...735L..34G} {735, L34}

\bibitem[\protect\citeauthoryear{{Grazian} et~al.,}{{Grazian}
  et~al.}{2015}]{grazian2015}
{Grazian} A.,  et~al., 2015, \mn@doi [\aap] {10.1051/0004-6361/201424750},
  \href {https://ui.adsabs.harvard.edu/abs/2015A&A...575A..96G} {575, A96}

\bibitem[\protect\citeauthoryear{{Graziani}, {Schneider}, {Ginolfi}, {Hunt},
  {Maio}, {Glatzle}  \& {Ciardi}}{{Graziani} et~al.}{2020}]{graziani2020}
{Graziani} L.,  {Schneider} R.,  {Ginolfi} M.,  {Hunt} L.~K.,  {Maio} U.,
  {Glatzle} M.,   {Ciardi} B.,  2020, \mn@doi [\mnras] {10.1093/mnras/staa796},
  \href {https://ui.adsabs.harvard.edu/abs/2020MNRAS.494.1071G} {494, 1071}

\bibitem[\protect\citeauthoryear{{Greene} et~al.,}{{Greene}
  et~al.}{2016}]{greene2016}
{Greene} J.~E.,  et~al., 2016, \mn@doi [\apjl] {10.3847/2041-8205/826/2/L32},
  \href {https://ui.adsabs.harvard.edu/abs/2016ApJ...826L..32G} {826, L32}

\bibitem[\protect\citeauthoryear{{Greif}, {Springel}, {White}, {Glover},
  {Clark}, {Smith}, {Klessen}  \& {Bromm}}{{Greif} et~al.}{2011}]{greif2011}
{Greif} T.~H.,  {Springel} V.,  {White} S. D.~M.,  {Glover} S. C.~O.,  {Clark}
  P.~C.,  {Smith} R.~J.,  {Klessen} R.~S.,   {Bromm} V.,  2011, \mn@doi [\apj]
  {10.1088/0004-637X/737/2/75}, \href
  {https://ui.adsabs.harvard.edu/abs/2011ApJ...737...75G} {737, 75}

\bibitem[\protect\citeauthoryear{{Griffin}, {Lacey}, {Gonzalez-Perez}, {Lagos},
  {Baugh}  \& {Fanidakis}}{{Griffin} et~al.}{2019}]{griffin2019}
{Griffin} A.~J.,  {Lacey} C.~G.,  {Gonzalez-Perez} V.,  {Lagos} C. d.~P.,
  {Baugh} C.~M.,   {Fanidakis} N.,  2019, \mn@doi [\mnras]
  {10.1093/mnras/stz1216}, \href
  {https://ui.adsabs.harvard.edu/abs/2019MNRAS.487..198G} {487, 198}

\bibitem[\protect\citeauthoryear{{Griffin}, {Lacey}, {Gonzalez-Perez}, {Lagos},
  {Baugh}  \& {Fanidakis}}{{Griffin} et~al.}{2020}]{griffin2020}
{Griffin} A.~J.,  {Lacey} C.~G.,  {Gonzalez-Perez} V.,  {Lagos} C. d.~P.,
  {Baugh} C.~M.,   {Fanidakis} N.,  2020, \mn@doi [\mnras]
  {10.1093/mnras/staa024}, \href
  {https://ui.adsabs.harvard.edu/abs/2020MNRAS.492.2535G} {492, 2535}

\bibitem[\protect\citeauthoryear{{Habouzit}, {Volonteri}  \&
  {Dubois}}{{Habouzit} et~al.}{2017}]{habouzit2017}
{Habouzit} M.,  {Volonteri} M.,   {Dubois} Y.,  2017, \mn@doi [\mnras]
  {10.1093/mnras/stx666}, \href
  {https://ui.adsabs.harvard.edu/abs/2017MNRAS.468.3935H} {468, 3935}

\bibitem[\protect\citeauthoryear{{Habouzit} et~al.,}{{Habouzit}
  et~al.}{2019}]{habouzit2019}
{Habouzit} M.,  et~al., 2019, \mn@doi [\mnras] {10.1093/mnras/stz102}, \href
  {https://ui.adsabs.harvard.edu/abs/2019MNRAS.484.4413H} {484, 4413}

\bibitem[\protect\citeauthoryear{{Habouzit} et~al.,}{{Habouzit}
  et~al.}{2020}]{habouzit2020}
{Habouzit} M.,  et~al., 2020, arXiv e-prints, \href
  {https://ui.adsabs.harvard.edu/abs/2020arXiv200610094H} {p. arXiv:2006.10094}

\bibitem[\protect\citeauthoryear{{Haemmerl{\'e}}, {Meynet}, {Mayer}, {Klessen},
  {Woods}  \& {Heger}}{{Haemmerl{\'e}} et~al.}{2019}]{haemmerle2019}
{Haemmerl{\'e}} L.,  {Meynet} G.,  {Mayer} L.,  {Klessen} R.~S.,  {Woods}
  T.~E.,   {Heger} A.,  2019, \mn@doi [\aap] {10.1051/0004-6361/201936716},
  \href {https://ui.adsabs.harvard.edu/abs/2019A&A...632L...2H} {632, L2}

\bibitem[\protect\citeauthoryear{{Haiman}, {Thoul}  \& {Loeb}}{{Haiman}
  et~al.}{1996}]{haiman1996}
{Haiman} Z.,  {Thoul} A.~A.,   {Loeb} A.,  1996, \mn@doi [\apj]
  {10.1086/177343}, \href
  {https://ui.adsabs.harvard.edu/abs/1996ApJ...464..523H} {464, 523}

\bibitem[\protect\citeauthoryear{{Heger} \& {Woosley}}{{Heger} \&
  {Woosley}}{2002}]{heger2002}
{Heger} A.,  {Woosley} S.~E.,  2002, \mn@doi [\apj] {10.1086/338487}, \href
  {https://ui.adsabs.harvard.edu/abs/2002ApJ...567..532H} {567, 532}

\bibitem[\protect\citeauthoryear{Hirano, Hosokawa, Yoshida, Umeda, Omukai,
  Chiaki  \& Yorke}{Hirano et~al.}{2014}]{hirano2014}
Hirano S.,  Hosokawa T.,  Yoshida N.,  Umeda H.,  Omukai K.,  Chiaki G.,
  Yorke H.~W.,  2014, The Astrophysical Journal, 781, 60

\bibitem[\protect\citeauthoryear{{Hirano}, {Zhu}, {Yoshida}, {Spergel}  \&
  {Yorke}}{{Hirano} et~al.}{2015}]{hirano2015}
{Hirano} S.,  {Zhu} N.,  {Yoshida} N.,  {Spergel} D.,   {Yorke} H.~W.,  2015,
  \mn@doi [\apj] {10.1088/0004-637X/814/1/18}, \href
  {https://ui.adsabs.harvard.edu/abs/2015ApJ...814...18H} {814, 18}

\bibitem[\protect\citeauthoryear{{Hirschmann}, {Dolag}, {Saro}, {Bachmann},
  {Borgani}  \& {Burkert}}{{Hirschmann} et~al.}{2014}]{hirschmann2014}
{Hirschmann} M.,  {Dolag} K.,  {Saro} A.,  {Bachmann} L.,  {Borgani} S.,
  {Burkert} A.,  2014, \mn@doi [\mnras] {10.1093/mnras/stu1023}, \href
  {https://ui.adsabs.harvard.edu/abs/2014MNRAS.442.2304H} {442, 2304}

\bibitem[\protect\citeauthoryear{{Hopkins}, {Richards}  \&
  {Hernquist}}{{Hopkins} et~al.}{2007}]{hopkins2007}
{Hopkins} P.~F.,  {Richards} G.~T.,   {Hernquist} L.,  2007, \mn@doi [\apj]
  {10.1086/509629}, \href
  {https://ui.adsabs.harvard.edu/abs/2007ApJ...654..731H} {654, 731}

\bibitem[\protect\citeauthoryear{{Hosokawa}, {Omukai}  \& {Yorke}}{{Hosokawa}
  et~al.}{2012}]{hosokawa2012}
{Hosokawa} T.,  {Omukai} K.,   {Yorke} H.~W.,  2012, \mn@doi [\apj]
  {10.1088/0004-637X/756/1/93}, \href
  {https://ui.adsabs.harvard.edu/abs/2012ApJ...756...93H} {756, 93}

\bibitem[\protect\citeauthoryear{{Hosokawa}, {Hirano}, {Kuiper}, {Yorke},
  {Omukai}  \& {Yoshida}}{{Hosokawa} et~al.}{2016}]{hosokawa2016}
{Hosokawa} T.,  {Hirano} S.,  {Kuiper} R.,  {Yorke} H.~W.,  {Omukai} K.,
  {Yoshida} N.,  2016, \mn@doi [\apj] {10.3847/0004-637X/824/2/119}, \href
  {https://ui.adsabs.harvard.edu/abs/2016ApJ...824..119H} {824, 119}

\bibitem[\protect\citeauthoryear{{Hoyle} \& {Lyttleton}}{{Hoyle} \&
  {Lyttleton}}{1941}]{hoyle1941}
{Hoyle} F.,  {Lyttleton} R.~A.,  1941, \mn@doi [\mnras]
  {10.1093/mnras/101.4.227}, \href
  {https://ui.adsabs.harvard.edu/abs/1941MNRAS.101..227H} {101, 227}

\bibitem[\protect\citeauthoryear{{Huang}, {Ni}, {Feng}  \& {Di Matteo}}{{Huang}
  et~al.}{2020}]{huang2020}
{Huang} K.-W.,  {Ni} Y.,  {Feng} Y.,   {Di Matteo} T.,  2020, \mn@doi [\mnras]
  {10.1093/mnras/staa1515}, \href
  {https://ui.adsabs.harvard.edu/abs/2020MNRAS.496....1H} {496, 1}

\bibitem[\protect\citeauthoryear{Inayoshi \& Tanaka}{Inayoshi \&
  Tanaka}{2015}]{inayoshi2015suppression}
Inayoshi K.,  Tanaka T.~L.,  2015, Monthly Notices of the Royal Astronomical
  Society, 450, 4350

\bibitem[\protect\citeauthoryear{{Inayoshi}, {Omukai}  \& {Tasker}}{{Inayoshi}
  et~al.}{2014}]{inayoshi2014}
{Inayoshi} K.,  {Omukai} K.,   {Tasker} E.,  2014, \mn@doi [\mnras]
  {10.1093/mnrasl/slu151}, \href
  {https://ui.adsabs.harvard.edu/abs/2014MNRAS.445L.109I} {445, L109}

\bibitem[\protect\citeauthoryear{{Inayoshi}, {Haiman}  \&
  {Ostriker}}{{Inayoshi} et~al.}{2016}]{inayoshi2016}
{Inayoshi} K.,  {Haiman} Z.,   {Ostriker} J.~P.,  2016, \mn@doi [\mnras]
  {10.1093/mnras/stw836}, \href
  {https://ui.adsabs.harvard.edu/abs/2016MNRAS.459.3738I} {459, 3738}

\bibitem[\protect\citeauthoryear{{Inayoshi}, {Visbal}  \& {Haiman}}{{Inayoshi}
  et~al.}{2019}]{Inayoshi2019}
{Inayoshi} K.,  {Visbal} E.,   {Haiman} Z.,  2019, arXiv e-prints, \href
  {https://ui.adsabs.harvard.edu/abs/2019arXiv191105791I} {p. arXiv:1911.05791}

\bibitem[\protect\citeauthoryear{{Inayoshi}, {Visbal}  \& {Haiman}}{{Inayoshi}
  et~al.}{2020}]{inayoshi2020}
{Inayoshi} K.,  {Visbal} E.,   {Haiman} Z.,  2020, \mn@doi [\araa]
  {10.1146/annurev-astro-120419-014455}, \href
  {https://ui.adsabs.harvard.edu/abs/2020ARA&A..58...27I} {58, 27}

\bibitem[\protect\citeauthoryear{{Johnson} \& {Haardt}}{{Johnson} \&
  {Haardt}}{2016}]{Johnson16}
{Johnson} J.~L.,  {Haardt} F.,  2016, \mn@doi [\pasa] {10.1017/pasa.2016.4},
  \href {http://adsabs.harvard.edu/abs/2016PASA...33....7J} {33, e007}

\bibitem[\protect\citeauthoryear{{Katz}, {Sijacki}  \& {Haehnelt}}{{Katz}
  et~al.}{2015}]{katz2015}
{Katz} H.,  {Sijacki} D.,   {Haehnelt} M.~G.,  2015, \mn@doi [\mnras]
  {10.1093/mnras/stv1048}, \href
  {https://ui.adsabs.harvard.edu/abs/2015MNRAS.451.2352K} {451, 2352}

\bibitem[\protect\citeauthoryear{{Kelly} \& {Merloni}}{{Kelly} \&
  {Merloni}}{2012}]{kelly2012}
{Kelly} B.~C.,  {Merloni} A.,  2012, \mn@doi [Advances in Astronomy]
  {10.1155/2012/970858}, \href
  {https://ui.adsabs.harvard.edu/abs/2012AdAst2012E...7K} {2012, 970858}

\bibitem[\protect\citeauthoryear{{Khandai}, {Di Matteo}, {Croft}, {Wilkins},
  {Feng}, {Tucker}, {DeGraf}  \& {Liu}}{{Khandai} et~al.}{2015}]{khandai2015}
{Khandai} N.,  {Di Matteo} T.,  {Croft} R.,  {Wilkins} S.,  {Feng} Y.,
  {Tucker} E.,  {DeGraf} C.,   {Liu} M.-S.,  2015, \mn@doi [\mnras]
  {10.1093/mnras/stv627}, \href
  {https://ui.adsabs.harvard.edu/abs/2015MNRAS.450.1349K} {450, 1349}

\bibitem[\protect\citeauthoryear{{Kistler}, {Y{\"u}ksel}, {Beacom}, {Hopkins}
  \& {Wyithe}}{{Kistler} et~al.}{2009}]{kistler2009}
{Kistler} M.~D.,  {Y{\"u}ksel} H.,  {Beacom} J.~F.,  {Hopkins} A.~M.,
  {Wyithe} J. S.~B.,  2009, \mn@doi [\apjl] {10.1088/0004-637X/705/2/L104},
  \href {https://ui.adsabs.harvard.edu/abs/2009ApJ...705L.104K} {705, L104}

\bibitem[\protect\citeauthoryear{{Labb{\'e}} et~al.,}{{Labb{\'e}}
  et~al.}{2013}]{labbe2013}
{Labb{\'e}} I.,  et~al., 2013, \mn@doi [\apjl] {10.1088/2041-8205/777/2/L19},
  \href {https://ui.adsabs.harvard.edu/abs/2013ApJ...777L..19L} {777, L19}

\bibitem[\protect\citeauthoryear{{Latif} \& {Ferrara}}{{Latif} \&
  {Ferrara}}{2016}]{latif2016a}
{Latif} M.~A.,  {Ferrara} A.,  2016, \mn@doi [\pasa] {10.1017/pasa.2016.41},
  \href {https://ui.adsabs.harvard.edu/abs/2016PASA...33...51L} {33, e051}

\bibitem[\protect\citeauthoryear{{Latif}, {Schleicher}, {Schmidt}  \&
  {Niemeyer}}{{Latif} et~al.}{2013}]{latif2013}
{Latif} M.~A.,  {Schleicher} D.~R.~G.,  {Schmidt} W.,   {Niemeyer} J.~C.,
  2013, \mn@doi [\mnras] {10.1093/mnras/stt1786}, \href
  {https://ui.adsabs.harvard.edu/abs/2013MNRAS.436.2989L} {436, 2989}

\bibitem[\protect\citeauthoryear{{Latif}, {Schleicher}  \& {Hartwig}}{{Latif}
  et~al.}{2016}]{latif2016b}
{Latif} M.~A.,  {Schleicher} D.~R.~G.,   {Hartwig} T.,  2016, \mn@doi [\mnras]
  {10.1093/mnras/stw297}, \href
  {https://ui.adsabs.harvard.edu/abs/2016MNRAS.458..233L} {458, 233}

\bibitem[\protect\citeauthoryear{{Lodato} \& {Natarajan}}{{Lodato} \&
  {Natarajan}}{2006}]{lodato2006}
{Lodato} G.,  {Natarajan} P.,  2006, \mn@doi [\mnras]
  {10.1111/j.1365-2966.2006.10801.x}, \href
  {https://ui.adsabs.harvard.edu/abs/2006MNRAS.371.1813L} {371, 1813}

\bibitem[\protect\citeauthoryear{{Lodato} \& {Natarajan}}{{Lodato} \&
  {Natarajan}}{2007}]{lodato2007}
{Lodato} G.,  {Natarajan} P.,  2007, \mn@doi [\mnras]
  {10.1111/j.1745-3933.2007.00304.x}, \href
  {https://ui.adsabs.harvard.edu/abs/2007MNRAS.377L..64L} {377, L64}

\bibitem[\protect\citeauthoryear{{Lupi}, {Colpi}, {Devecchi}, {Galanti}  \&
  {Volonteri}}{{Lupi} et~al.}{2014}]{lupi2014}
{Lupi} A.,  {Colpi} M.,  {Devecchi} B.,  {Galanti} G.,   {Volonteri} M.,  2014,
  \mn@doi [\mnras] {10.1093/mnras/stu1120}, \href
  {https://ui.adsabs.harvard.edu/abs/2014MNRAS.442.3616L} {442, 3616}

\bibitem[\protect\citeauthoryear{{Lupi}, {Volonteri}, {Decarli}, {Bovino},
  {Silk}  \& {Bergeron}}{{Lupi} et~al.}{2019}]{lupi2019}
{Lupi} A.,  {Volonteri} M.,  {Decarli} R.,  {Bovino} S.,  {Silk} J.,
  {Bergeron} J.,  2019, \mn@doi [\mnras] {10.1093/mnras/stz1959}, \href
  {https://ui.adsabs.harvard.edu/abs/2019MNRAS.488.4004L} {488, 4004}

\bibitem[\protect\citeauthoryear{{Lupi}, {Haiman}  \& {Volonteri}}{{Lupi}
  et~al.}{2021}]{lupi2021}
{Lupi} A.,  {Haiman} Z.,   {Volonteri} M.,  2021, \mn@doi [\mnras]
  {10.1093/mnras/stab692}, \href
  {https://ui.adsabs.harvard.edu/abs/2021MNRAS.503.5046L} {503, 5046}

\bibitem[\protect\citeauthoryear{{Madau} \& {Dickinson}}{{Madau} \&
  {Dickinson}}{2014}]{madau2014b}
{Madau} P.,  {Dickinson} M.,  2014, \mn@doi [\araa]
  {10.1146/annurev-astro-081811-125615}, \href
  {https://ui.adsabs.harvard.edu/abs/2014ARA&A..52..415M} {52, 415}

\bibitem[\protect\citeauthoryear{{Madau}, {Haardt}  \& {Dotti}}{{Madau}
  et~al.}{2014}]{madau2014a}
{Madau} P.,  {Haardt} F.,   {Dotti} M.,  2014, \mn@doi [\apjl]
  {10.1088/2041-8205/784/2/L38}, \href
  {https://ui.adsabs.harvard.edu/abs/2014ApJ...784L..38M} {784, L38}

\bibitem[\protect\citeauthoryear{{Mancini}, {Schneider}, {Graziani},
  {Valiante}, {Dayal}, {Maio}, {Ciardi}  \& {Hunt}}{{Mancini}
  et~al.}{2015}]{mancini2015}
{Mancini} M.,  {Schneider} R.,  {Graziani} L.,  {Valiante} R.,  {Dayal} P.,
  {Maio} U.,  {Ciardi} B.,   {Hunt} L.~K.,  2015, \mn@doi [\mnras]
  {10.1093/mnrasl/slv070}, \href
  {https://ui.adsabs.harvard.edu/abs/2015MNRAS.451L..70M} {451, L70}

\bibitem[\protect\citeauthoryear{{Mancini}, {Schneider}, {Graziani},
  {Valiante}, {Dayal}, {Maio}  \& {Ciardi}}{{Mancini}
  et~al.}{2016}]{mancini2016}
{Mancini} M.,  {Schneider} R.,  {Graziani} L.,  {Valiante} R.,  {Dayal} P.,
  {Maio} U.,   {Ciardi} B.,  2016, \mn@doi [\mnras] {10.1093/mnras/stw1783},
  \href {https://ui.adsabs.harvard.edu/abs/2016MNRAS.462.3130M} {462, 3130}

\bibitem[\protect\citeauthoryear{{Matsuoka} et~al.,}{{Matsuoka}
  et~al.}{2018}]{matsuoka2018}
{Matsuoka} Y.,  et~al., 2018, \mn@doi [\apj] {10.3847/1538-4357/aaee7a}, \href
  {https://ui.adsabs.harvard.edu/abs/2018ApJ...869..150M} {869, 150}

\bibitem[\protect\citeauthoryear{Matsuoka et~al.,}{Matsuoka
  et~al.}{2019}]{matsuoka2019}
Matsuoka Y.,  et~al., 2019, The Astrophysical Journal Letters, 872, L2

\bibitem[\protect\citeauthoryear{Mayer}{Mayer}{2018}]{mayer2018}
Mayer L.,  2018, arXiv preprint arXiv:1807.06243

\bibitem[\protect\citeauthoryear{{Mayer}}{{Mayer}}{2019}]{mayer2019b}
{Mayer} L.,  2019, {Super-Eddington accretion; flow regimes and conditions in
  high-z galaxies}.
pp 195--222, \mn@doi{10.1142/9789813227958_0011}

\bibitem[\protect\citeauthoryear{{Mayer} \& {Bonoli}}{{Mayer} \&
  {Bonoli}}{2019}]{mayer2019a}
{Mayer} L.,  {Bonoli} S.,  2019, \mn@doi [Reports on Progress in Physics]
  {10.1088/1361-6633/aad6a5}, \href
  {https://ui.adsabs.harvard.edu/abs/2019RPPh...82a6901M} {82, 016901}

\bibitem[\protect\citeauthoryear{{Mayer}, {Kazantzidis}, {Madau}, {Colpi},
  {Quinn}  \& {Wadsley}}{{Mayer} et~al.}{2007}]{mayer2007}
{Mayer} L.,  {Kazantzidis} S.,  {Madau} P.,  {Colpi} M.,  {Quinn} T.,
  {Wadsley} J.,  2007, \mn@doi [Science] {10.1126/science.1141858}, \href
  {https://ui.adsabs.harvard.edu/abs/2007Sci...316.1874M} {316, 1874}

\bibitem[\protect\citeauthoryear{{Mayer}, {Kazantzidis}, {Escala}  \&
  {Callegari}}{{Mayer} et~al.}{2010}]{mayer2010}
{Mayer} L.,  {Kazantzidis} S.,  {Escala} A.,   {Callegari} S.,  2010, \mn@doi
  [\nat] {10.1038/nature09294}, \href
  {https://ui.adsabs.harvard.edu/abs/2010Natur.466.1082M} {466, 1082}

\bibitem[\protect\citeauthoryear{{Mayer}, {Fiacconi}, {Bonoli}, {Quinn},
  {Ro{\v{s}}kar}, {Shen}  \& {Wadsley}}{{Mayer} et~al.}{2015}]{mayer2015}
{Mayer} L.,  {Fiacconi} D.,  {Bonoli} S.,  {Quinn} T.,  {Ro{\v{s}}kar} R.,
  {Shen} S.,   {Wadsley} J.,  2015, \mn@doi [\apj]
  {10.1088/0004-637X/810/1/51}, \href
  {https://ui.adsabs.harvard.edu/abs/2015ApJ...810...51M} {810, 51}

\bibitem[\protect\citeauthoryear{{Mazzucchelli} et~al.,}{{Mazzucchelli}
  et~al.}{2017}]{mazzucchelli2017}
{Mazzucchelli} C.,  et~al., 2017, \mn@doi [\apj] {10.3847/1538-4357/aa9185},
  \href {https://ui.adsabs.harvard.edu/abs/2017ApJ...849...91M} {849, 91}

\bibitem[\protect\citeauthoryear{{McAlpine}, {Bower}, {Harrison}, {Crain},
  {Schaller}, {Schaye}  \& {Theuns}}{{McAlpine} et~al.}{2017}]{mcalpine2017}
{McAlpine} S.,  {Bower} R.~G.,  {Harrison} C.~M.,  {Crain} R.~A.,  {Schaller}
  M.,  {Schaye} J.,   {Theuns} T.,  2017, \mn@doi [\mnras]
  {10.1093/mnras/stx658}, \href
  {https://ui.adsabs.harvard.edu/abs/2017MNRAS.468.3395M} {468, 3395}

\bibitem[\protect\citeauthoryear{{McAlpine}, {Bower}, {Rosario}, {Crain},
  {Schaye}  \& {Theuns}}{{McAlpine} et~al.}{2018}]{mcalpine2018}
{McAlpine} S.,  {Bower} R.~G.,  {Rosario} D.~J.,  {Crain} R.~A.,  {Schaye} J.,
   {Theuns} T.,  2018, \mn@doi [\mnras] {10.1093/mnras/sty2489}, \href
  {https://ui.adsabs.harvard.edu/abs/2018MNRAS.481.3118M} {481, 3118}

\bibitem[\protect\citeauthoryear{{McGreer}, {Fan}, {Jiang}  \& {Cai}}{{McGreer}
  et~al.}{2018}]{mcgreer2018}
{McGreer} I.~D.,  {Fan} X.,  {Jiang} L.,   {Cai} Z.,  2018, \mn@doi [\aj]
  {10.3847/1538-3881/aaaab4}, \href
  {https://ui.adsabs.harvard.edu/abs/2018AJ....155..131M} {155, 131}

\bibitem[\protect\citeauthoryear{{Merlin} et~al.,}{{Merlin}
  et~al.}{2019}]{merlin2019}
{Merlin} E.,  et~al., 2019, \mn@doi [\mnras] {10.1093/mnras/stz2615}, \href
  {https://ui.adsabs.harvard.edu/abs/2019MNRAS.490.3309M} {490, 3309}

\bibitem[\protect\citeauthoryear{{Merloni} \& {Heinz}}{{Merloni} \&
  {Heinz}}{2008}]{merloni2008}
{Merloni} A.,  {Heinz} S.,  2008, \mn@doi [\mnras]
  {10.1111/j.1365-2966.2008.13472.x}, \href
  {https://ui.adsabs.harvard.edu/abs/2008MNRAS.388.1011M} {388, 1011}

\bibitem[\protect\citeauthoryear{{Merloni} et~al.,}{{Merloni}
  et~al.}{2014}]{merloni2014}
{Merloni} A.,  et~al., 2014, \mn@doi [\mnras] {10.1093/mnras/stt2149}, \href
  {https://ui.adsabs.harvard.edu/abs/2014MNRAS.437.3550M} {437, 3550}

\bibitem[\protect\citeauthoryear{{Miyaji} et~al.,}{{Miyaji}
  et~al.}{2015}]{miyaji2015}
{Miyaji} T.,  et~al., 2015, \mn@doi [\apj] {10.1088/0004-637X/804/2/104}, \href
  {https://ui.adsabs.harvard.edu/abs/2015ApJ...804..104M} {804, 104}

\bibitem[\protect\citeauthoryear{{Mortlock} et~al.,}{{Mortlock}
  et~al.}{2011}]{mortlock2011}
{Mortlock} D.~J.,  et~al., 2011, \mn@doi [\nat] {10.1038/nature10159}, \href
  {https://ui.adsabs.harvard.edu/abs/2011Natur.474..616M} {474, 616}

\bibitem[\protect\citeauthoryear{{Natarajan}, {Pacucci}, {Ferrara}, {Agarwal},
  {Ricarte}, {Zackrisson}  \& {Cappelluti}}{{Natarajan}
  et~al.}{2017}]{natarajan2017}
{Natarajan} P.,  {Pacucci} F.,  {Ferrara} A.,  {Agarwal} B.,  {Ricarte} A.,
  {Zackrisson} E.,   {Cappelluti} N.,  2017, \mn@doi [\apj]
  {10.3847/1538-4357/aa6330}, \href
  {https://ui.adsabs.harvard.edu/abs/2017ApJ...838..117N} {838, 117}

\bibitem[\protect\citeauthoryear{{Niida} et~al.,}{{Niida}
  et~al.}{2020}]{niida2020}
{Niida} M.,  et~al., 2020, \mn@doi [\apj] {10.3847/1538-4357/abbe11}, \href
  {https://ui.adsabs.harvard.edu/abs/2020ApJ...904...89N} {904, 89}

\bibitem[\protect\citeauthoryear{{Oesch} et~al.,}{{Oesch}
  et~al.}{2014}]{oesch2014}
{Oesch} P.~A.,  et~al., 2014, \mn@doi [\apj] {10.1088/0004-637X/786/2/108},
  \href {https://ui.adsabs.harvard.edu/abs/2014ApJ...786..108O} {786, 108}

\bibitem[\protect\citeauthoryear{{Omukai}}{{Omukai}}{2001}]{omukai2001}
{Omukai} K.,  2001, \mn@doi [\apj] {10.1086/318296}, \href
  {https://ui.adsabs.harvard.edu/abs/2001ApJ...546..635O} {546, 635}

\bibitem[\protect\citeauthoryear{Omukai, Tsuribe, Schneider  \& Ferrara}{Omukai
  et~al.}{2005}]{omukai2005}
Omukai K.,  Tsuribe T.,  Schneider R.,   Ferrara A.,  2005, The Astrophysical
  Journal, 626, 627

\bibitem[\protect\citeauthoryear{Omukai, Schneider  \& Haiman}{Omukai
  et~al.}{2008}]{omukai2008}
Omukai K.,  Schneider R.,   Haiman Z.,  2008, The Astrophysical Journal, 686,
  801

\bibitem[\protect\citeauthoryear{{Onoue} et~al.,}{{Onoue}
  et~al.}{2019}]{onoue2019}
{Onoue} M.,  et~al., 2019, \mn@doi [\apj] {10.3847/1538-4357/ab29e9}, \href
  {https://ui.adsabs.harvard.edu/abs/2019ApJ...880...77O} {880, 77}

\bibitem[\protect\citeauthoryear{{Pacucci}, {Ferrara}, {Volonteri}  \&
  {Dubus}}{{Pacucci} et~al.}{2015}]{pacucci2015}
{Pacucci} F.,  {Ferrara} A.,  {Volonteri} M.,   {Dubus} G.,  2015, \mn@doi
  [\mnras] {10.1093/mnras/stv2196}, \href
  {https://ui.adsabs.harvard.edu/abs/2015MNRAS.454.3771P} {454, 3771}

\bibitem[\protect\citeauthoryear{{Pacucci}, {Ferrara}, {Grazian}, {Fiore},
  {Giallongo}  \& {Puccetti}}{{Pacucci} et~al.}{2016}]{pacucci2016}
{Pacucci} F.,  {Ferrara} A.,  {Grazian} A.,  {Fiore} F.,  {Giallongo} E.,
  {Puccetti} S.,  2016, \mn@doi [\mnras] {10.1093/mnras/stw725}, \href
  {https://ui.adsabs.harvard.edu/abs/2016MNRAS.459.1432P} {459, 1432}

\bibitem[\protect\citeauthoryear{Pacucci, Natarajan, Volonteri, Cappelluti  \&
  Urry}{Pacucci et~al.}{2017}]{pacucci2017}
Pacucci F.,  Natarajan P.,  Volonteri M.,  Cappelluti N.,   Urry C.~M.,  2017,
  arXiv preprint arXiv:1710.09375

\bibitem[\protect\citeauthoryear{{Parkinson}, {Cole}  \& {Helly}}{{Parkinson}
  et~al.}{2008}]{parkinson2008}
{Parkinson} H.,  {Cole} S.,   {Helly} J.,  2008, \mn@doi [\mnras]
  {10.1111/j.1365-2966.2007.12517.x}, \href
  {https://ui.adsabs.harvard.edu/abs/2008MNRAS.383..557P} {383, 557}

\bibitem[\protect\citeauthoryear{{Parsa}, {Dunlop}  \& {McLure}}{{Parsa}
  et~al.}{2018}]{parsa2018}
{Parsa} S.,  {Dunlop} J.~S.,   {McLure} R.~J.,  2018, \mn@doi [\mnras]
  {10.1093/mnras/stx2887}, \href
  {https://ui.adsabs.harvard.edu/abs/2018MNRAS.474.2904P} {474, 2904}

\bibitem[\protect\citeauthoryear{{Pezzulli}, {Valiante}  \&
  {Schneider}}{{Pezzulli} et~al.}{2016}]{pezzulli2016}
{Pezzulli} E.,  {Valiante} R.,   {Schneider} R.,  2016, \mn@doi [\mnras]
  {10.1093/mnras/stw505}, \href
  {https://ui.adsabs.harvard.edu/abs/2016MNRAS.458.3047P} {458, 3047}

\bibitem[\protect\citeauthoryear{{Pezzulli}, {Valiante}, {Orofino},
  {Schneider}, {Gallerani}  \& {Sbarrato}}{{Pezzulli}
  et~al.}{2017a}]{pezzulli2017a}
{Pezzulli} E.,  {Valiante} R.,  {Orofino} M.~C.,  {Schneider} R.,  {Gallerani}
  S.,   {Sbarrato} T.,  2017a, \mn@doi [\mnras] {10.1093/mnras/stw3243}, \href
  {http://adsabs.harvard.edu/abs/2017MNRAS.466.2131P} {466, 2131}

\bibitem[\protect\citeauthoryear{Pezzulli, Volonteri, Schneider  \&
  Valiante}{Pezzulli et~al.}{2017b}]{pezzulli2017b}
Pezzulli E.,  Volonteri M.,  Schneider R.,   Valiante R.,  2017b, Monthly
  Notices of the Royal Astronomical Society, 471, 589

\bibitem[\protect\citeauthoryear{{Pfister}, {Lupi}, {Capelo}, {Volonteri},
  {Bellovary}  \& {Dotti}}{{Pfister} et~al.}{2017}]{pfister2017}
{Pfister} H.,  {Lupi} A.,  {Capelo} P.~R.,  {Volonteri} M.,  {Bellovary} J.~M.,
    {Dotti} M.,  2017, \mn@doi [\mnras] {10.1093/mnras/stx1853}, \href
  {https://ui.adsabs.harvard.edu/abs/2017MNRAS.471.3646P} {471, 3646}

\bibitem[\protect\citeauthoryear{{Pfister}, {Volonteri}, {Dubois}, {Dotti}  \&
  {Colpi}}{{Pfister} et~al.}{2019}]{pfister2019}
{Pfister} H.,  {Volonteri} M.,  {Dubois} Y.,  {Dotti} M.,   {Colpi} M.,  2019,
  \mn@doi [\mnras] {10.1093/mnras/stz822}, \href
  {https://ui.adsabs.harvard.edu/abs/2019MNRAS.486..101P} {486, 101}

\bibitem[\protect\citeauthoryear{{Piana}, {Dayal}, {Volonteri}  \&
  {Choudhury}}{{Piana} et~al.}{2021}]{piana2021}
{Piana} O.,  {Dayal} P.,  {Volonteri} M.,   {Choudhury} T.~R.,  2021, \mn@doi
  [\mnras] {10.1093/mnras/staa3363}, \href
  {https://ui.adsabs.harvard.edu/abs/2021MNRAS.500.2146P} {500, 2146}

\bibitem[\protect\citeauthoryear{{Planck Collaboration} et~al.,}{{Planck
  Collaboration} et~al.}{2018}]{planck2018}
{Planck Collaboration} et~al., 2018, arXiv e-prints, \href
  {https://ui.adsabs.harvard.edu/abs/2018arXiv180706209P} {p. arXiv:1807.06209}

\bibitem[\protect\citeauthoryear{{Pons}, {McMahon}, {Simcoe}, {Banerji},
  {Hewett}  \& {Reed}}{{Pons} et~al.}{2019}]{pons2019}
{Pons} E.,  {McMahon} R.~G.,  {Simcoe} R.~A.,  {Banerji} M.,  {Hewett} P.~C.,
  {Reed} S.~L.,  2019, \mn@doi [\mnras] {10.1093/mnras/stz292}, \href
  {https://ui.adsabs.harvard.edu/abs/2019MNRAS.484.5142P} {484, 5142}

\bibitem[\protect\citeauthoryear{{Reed} et~al.,}{{Reed}
  et~al.}{2017}]{reed2017}
{Reed} S.~L.,  et~al., 2017, \mn@doi [\mnras] {10.1093/mnras/stx728}, \href
  {https://ui.adsabs.harvard.edu/abs/2017MNRAS.468.4702R} {468, 4702}

\bibitem[\protect\citeauthoryear{{Reed} et~al.,}{{Reed}
  et~al.}{2019}]{reed2019}
{Reed} S.~L.,  et~al., 2019, \mn@doi [\mnras] {10.1093/mnras/stz1341}, \href
  {https://ui.adsabs.harvard.edu/abs/2019MNRAS.487.1874R} {487, 1874}

\bibitem[\protect\citeauthoryear{{Regan} \& {Haehnelt}}{{Regan} \&
  {Haehnelt}}{2009}]{regan2009}
{Regan} J.~A.,  {Haehnelt} M.~G.,  2009, \mn@doi [\mnras]
  {10.1111/j.1365-2966.2009.14579.x}, \href
  {https://ui.adsabs.harvard.edu/abs/2009MNRAS.396..343R} {396, 343}

\bibitem[\protect\citeauthoryear{{Regan}, {Downes}, {Volonteri}, {Beckmann},
  {Lupi}, {Trebitsch}  \& {Dubois}}{{Regan} et~al.}{2019}]{regan2019}
{Regan} J.~A.,  {Downes} T.~P.,  {Volonteri} M.,  {Beckmann} R.,  {Lupi} A.,
  {Trebitsch} M.,   {Dubois} Y.,  2019, \mn@doi [\mnras]
  {10.1093/mnras/stz1045}, \href
  {https://ui.adsabs.harvard.edu/abs/2019MNRAS.486.3892R} {486, 3892}

\bibitem[\protect\citeauthoryear{{Regan}, {Haiman}, {Wise}, {O'Shea}  \&
  {Norman}}{{Regan} et~al.}{2020}]{regan2020}
{Regan} J.~A.,  {Haiman} Z.,  {Wise} J.~H.,  {O'Shea} B.~W.,   {Norman} M.~L.,
  2020, arXiv e-prints, \href
  {https://ui.adsabs.harvard.edu/abs/2020arXiv200614625R} {p. arXiv:2006.14625}

\bibitem[\protect\citeauthoryear{{Reines} \& {Volonteri}}{{Reines} \&
  {Volonteri}}{2015}]{reines2015}
{Reines} A.~E.,  {Volonteri} M.,  2015, \mn@doi [\apj]
  {10.1088/0004-637X/813/2/82}, \href
  {https://ui.adsabs.harvard.edu/abs/2015ApJ...813...82R} {813, 82}

\bibitem[\protect\citeauthoryear{{Reinoso}, {Schleicher}, {Fellhauer},
  {Klessen}  \& {Boekholt}}{{Reinoso} et~al.}{2018}]{reinoso2018}
{Reinoso} B.,  {Schleicher} D.~R.~G.,  {Fellhauer} M.,  {Klessen} R.~S.,
  {Boekholt} T.~C.~N.,  2018, \mn@doi [\aap] {10.1051/0004-6361/201732224},
  \href {https://ui.adsabs.harvard.edu/abs/2018A&A...614A..14R} {614, A14}

\bibitem[\protect\citeauthoryear{Reinoso, Schleicher, Fellhauer, Klessen,
  Boekholt, Vergara  \& Alister~Seguel}{Reinoso et~al.}{2019}]{reinoso2019}
Reinoso B.,  Schleicher D.,  Fellhauer M.,  Klessen R.,  Boekholt T.,  Vergara
  M.,   Alister~Seguel P.,  2019, Boletin de la Asociacion Argentina de
  Astronomia La Plata Argentina, 61, 154

\bibitem[\protect\citeauthoryear{{Ricarte} \& {Natarajan}}{{Ricarte} \&
  {Natarajan}}{2018a}]{ricarte2018a}
{Ricarte} A.,  {Natarajan} P.,  2018a, \mn@doi [\mnras]
  {10.1093/mnras/stx2851}, \href
  {https://ui.adsabs.harvard.edu/abs/2018MNRAS.474.1995R} {474, 1995}

\bibitem[\protect\citeauthoryear{{Ricarte} \& {Natarajan}}{{Ricarte} \&
  {Natarajan}}{2018b}]{ricarte2018b}
{Ricarte} A.,  {Natarajan} P.,  2018b, \mn@doi [\mnras]
  {10.1093/mnras/sty2448}, \href
  {https://ui.adsabs.harvard.edu/abs/2018MNRAS.481.3278R} {481, 3278}

\bibitem[\protect\citeauthoryear{Sakurai, Yoshida, Fujii  \& Hirano}{Sakurai
  et~al.}{2017}]{sakurai2017}
Sakurai Y.,  Yoshida N.,  Fujii M.~S.,   Hirano S.,  2017, Monthly Notices of
  the Royal Astronomical Society, 472, 1677

\bibitem[\protect\citeauthoryear{{Salvaterra}, {Haardt}, {Volonteri}  \&
  {Moretti}}{{Salvaterra} et~al.}{2012}]{salvaterra2012}
{Salvaterra} R.,  {Haardt} F.,  {Volonteri} M.,   {Moretti} A.,  2012, \mn@doi
  [\aap] {10.1051/0004-6361/201219965}, \href
  {https://ui.adsabs.harvard.edu/abs/2012A&A...545L...6S} {545, L6}

\bibitem[\protect\citeauthoryear{{Sassano}, {Schneider}, {Valiante},
  {Inayoshi}, {Chon}, {Omukai}, {Mayer}  \& {Capelo}}{{Sassano}
  et~al.}{2021}]{sassano2021}
{Sassano} F.,  {Schneider} R.,  {Valiante} R.,  {Inayoshi} K.,  {Chon} S.,
  {Omukai} K.,  {Mayer} L.,   {Capelo} P.~R.,  2021, arXiv e-prints, \href
  {https://ui.adsabs.harvard.edu/abs/2021arXiv210608330S} {p. arXiv:2106.08330}

\bibitem[\protect\citeauthoryear{{Schaye} et~al.,}{{Schaye}
  et~al.}{2015}]{schaye2015}
{Schaye} J.,  et~al., 2015, \mn@doi [\mnras] {10.1093/mnras/stu2058}, \href
  {https://ui.adsabs.harvard.edu/abs/2015MNRAS.446..521S} {446, 521}

\bibitem[\protect\citeauthoryear{{Schenker} et~al.,}{{Schenker}
  et~al.}{2013}]{schenker2013}
{Schenker} M.~A.,  et~al., 2013, \mn@doi [\apj] {10.1088/0004-637X/768/2/196},
  \href {https://ui.adsabs.harvard.edu/abs/2013ApJ...768..196S} {768, 196}

\bibitem[\protect\citeauthoryear{Schneider, Ferrara, Natarajan  \&
  Omukai}{Schneider et~al.}{2002}]{schneider2002}
Schneider R.,  Ferrara A.,  Natarajan P.,   Omukai K.,  2002, The Astrophysical
  Journal, 571, 30

\bibitem[\protect\citeauthoryear{Schneider, Omukai, Inoue  \&
  Ferrara}{Schneider et~al.}{2006}]{schneider2006}
Schneider R.,  Omukai K.,  Inoue A.~K.,   Ferrara A.,  2006, Monthly Notices of
  the Royal Astronomical Society, 369, 1437

\bibitem[\protect\citeauthoryear{{Schneider}, {Omukai}, {Bianchi}  \&
  {Valiante}}{{Schneider} et~al.}{2012}]{schneider2012b}
{Schneider} R.,  {Omukai} K.,  {Bianchi} S.,   {Valiante} R.,  2012, \mn@doi
  [\mnras] {10.1111/j.1365-2966.2011.19818.x}, \href
  {https://ui.adsabs.harvard.edu/abs/2012MNRAS.419.1566S} {419, 1566}

\bibitem[\protect\citeauthoryear{{Shakura} \& {Sunyaev}}{{Shakura} \&
  {Sunyaev}}{1973}]{shakura1973}
{Shakura} N.~I.,  {Sunyaev} R.~A.,  1973, \aap, \href
  {https://ui.adsabs.harvard.edu/abs/1973A&A....24..337S} {500, 33}

\bibitem[\protect\citeauthoryear{{Shankar}, {Weinberg}  \&
  {Miralda-Escud{\'e}}}{{Shankar} et~al.}{2009}]{shankar2009}
{Shankar} F.,  {Weinberg} D.~H.,   {Miralda-Escud{\'e}} J.,  2009, \mn@doi
  [\apj] {10.1088/0004-637X/690/1/20}, \href
  {https://ui.adsabs.harvard.edu/abs/2009ApJ...690...20S} {690, 20}

\bibitem[\protect\citeauthoryear{{Shankar}, {Crocce}, {Miralda-Escud{\'e}},
  {Fosalba}  \& {Weinberg}}{{Shankar} et~al.}{2010}]{shankar2010}
{Shankar} F.,  {Crocce} M.,  {Miralda-Escud{\'e}} J.,  {Fosalba} P.,
  {Weinberg} D.~H.,  2010, \mn@doi [\apj] {10.1088/0004-637X/718/1/231}, \href
  {https://ui.adsabs.harvard.edu/abs/2010ApJ...718..231S} {718, 231}

\bibitem[\protect\citeauthoryear{{Shankar} et~al.,}{{Shankar}
  et~al.}{2016}]{shankar2016}
{Shankar} F.,  et~al., 2016, \mn@doi [\mnras] {10.1093/mnras/stw678}, \href
  {https://ui.adsabs.harvard.edu/abs/2016MNRAS.460.3119S} {460, 3119}

\bibitem[\protect\citeauthoryear{{Shankar} et~al.,}{{Shankar}
  et~al.}{2020}]{shankar2020}
{Shankar} F.,  et~al., 2020, \mn@doi [Nature Astronomy]
  {10.1038/s41550-019-0949-y}, \href
  {https://ui.adsabs.harvard.edu/abs/2020NatAs...4..282S} {4, 282}

\bibitem[\protect\citeauthoryear{{Shao} et~al.,}{{Shao}
  et~al.}{2017}]{shao2017}
{Shao} Y.,  et~al., 2017, \mn@doi [\apj] {10.3847/1538-4357/aa826c}, \href
  {https://ui.adsabs.harvard.edu/abs/2017ApJ...845..138S} {845, 138}

\bibitem[\protect\citeauthoryear{{Shen} \& {Liu}}{{Shen} \&
  {Liu}}{2012}]{shen2012}
{Shen} Y.,  {Liu} X.,  2012, \mn@doi [\apj] {10.1088/0004-637X/753/2/125},
  \href {https://ui.adsabs.harvard.edu/abs/2012ApJ...753..125S} {753, 125}

\bibitem[\protect\citeauthoryear{{Shen} et~al.,}{{Shen}
  et~al.}{2011}]{shen2011}
{Shen} Y.,  et~al., 2011, \mn@doi [\apjs] {10.1088/0067-0049/194/2/45}, \href
  {https://ui.adsabs.harvard.edu/abs/2011ApJS..194...45S} {194, 45}

\bibitem[\protect\citeauthoryear{{Shen} et~al.,}{{Shen}
  et~al.}{2019}]{sheng2019}
{Shen} Y.,  et~al., 2019, \mn@doi [\apj] {10.3847/1538-4357/ab03d9}, \href
  {https://ui.adsabs.harvard.edu/abs/2019ApJ...873...35S} {873, 35}

\bibitem[\protect\citeauthoryear{{Shen}, {Hopkins}, {Faucher-Gigu{\`e}re},
  {Alexander}, {Richards}, {Ross}  \& {Hickox}}{{Shen} et~al.}{2020}]{shen2020}
{Shen} X.,  {Hopkins} P.~F.,  {Faucher-Gigu{\`e}re} C.-A.,  {Alexander} D.~M.,
  {Richards} G.~T.,  {Ross} N.~P.,   {Hickox} R.~C.,  2020, \mn@doi [\mnras]
  {10.1093/mnras/staa1381}, \href
  {https://ui.adsabs.harvard.edu/abs/2020MNRAS.495.3252S} {495, 3252}

\bibitem[\protect\citeauthoryear{{Sheth}, {Mo}  \& {Tormen}}{{Sheth}
  et~al.}{2001}]{sheth2001}
{Sheth} R.~K.,  {Mo} H.~J.,   {Tormen} G.,  2001, \mn@doi [\mnras]
  {10.1046/j.1365-8711.2001.04006.x}, \href
  {https://ui.adsabs.harvard.edu/abs/2001MNRAS.323....1S} {323, 1}

\bibitem[\protect\citeauthoryear{{Sijacki}, {Springel}, {Di Matteo}  \&
  {Hernquist}}{{Sijacki} et~al.}{2007}]{sijacki2007}
{Sijacki} D.,  {Springel} V.,  {Di Matteo} T.,   {Hernquist} L.,  2007, \mn@doi
  [\mnras] {10.1111/j.1365-2966.2007.12153.x}, \href
  {https://ui.adsabs.harvard.edu/abs/2007MNRAS.380..877S} {380, 877}

\bibitem[\protect\citeauthoryear{{Sijacki}, {Vogelsberger}, {Genel},
  {Springel}, {Torrey}, {Snyder}, {Nelson}  \& {Hernquist}}{{Sijacki}
  et~al.}{2015}]{sijacki2015}
{Sijacki} D.,  {Vogelsberger} M.,  {Genel} S.,  {Springel} V.,  {Torrey} P.,
  {Snyder} G.~F.,  {Nelson} D.,   {Hernquist} L.,  2015, \mn@doi [\mnras]
  {10.1093/mnras/stv1340}, \href
  {https://ui.adsabs.harvard.edu/abs/2015MNRAS.452..575S} {452, 575}

\bibitem[\protect\citeauthoryear{{S{\k{a}}dowski}}{{S{\k{a}}dowski}}{2009}]{sadowski2009}
{S{\k{a}}dowski} A.,  2009, \mn@doi [\apjs] {10.1088/0067-0049/183/2/171},
  \href {https://ui.adsabs.harvard.edu/abs/2009ApJS..183..171S} {183, 171}

\bibitem[\protect\citeauthoryear{{Somerville}, {Hopkins}, {Cox}, {Robertson}
  \& {Hernquist}}{{Somerville} et~al.}{2008}]{somerville2008}
{Somerville} R.~S.,  {Hopkins} P.~F.,  {Cox} T.~J.,  {Robertson} B.~E.,
  {Hernquist} L.,  2008, \mn@doi [\mnras] {10.1111/j.1365-2966.2008.13805.x},
  \href {https://ui.adsabs.harvard.edu/abs/2008MNRAS.391..481S} {391, 481}

\bibitem[\protect\citeauthoryear{{Song} et~al.,}{{Song}
  et~al.}{2016}]{song2016}
{Song} M.,  et~al., 2016, \mn@doi [\apj] {10.3847/0004-637X/825/1/5}, \href
  {https://ui.adsabs.harvard.edu/abs/2016ApJ...825....5S} {825, 5}

\bibitem[\protect\citeauthoryear{{Souza Lima}, {Mayer}, {Capelo}, {Bortolas}
  \& {Quinn}}{{Souza Lima} et~al.}{2020}]{lima2020}
{Souza Lima} R.,  {Mayer} L.,  {Capelo} P.~R.,  {Bortolas} E.,   {Quinn} T.~R.,
   2020, arXiv e-prints, \href
  {https://ui.adsabs.harvard.edu/abs/2020arXiv200313789S} {p. arXiv:2003.13789}

\bibitem[\protect\citeauthoryear{{Springel} et~al.,}{{Springel}
  et~al.}{2005}]{springel2005}
{Springel} V.,  et~al., 2005, \mn@doi [\nat] {10.1038/nature03597}, \href
  {https://ui.adsabs.harvard.edu/abs/2005Natur.435..629S} {435, 629}

\bibitem[\protect\citeauthoryear{{Stacy}, {Bromm}  \& {Lee}}{{Stacy}
  et~al.}{2016}]{stacy2016}
{Stacy} A.,  {Bromm} V.,   {Lee} A.~T.,  2016, \mn@doi [\mnras]
  {10.1093/mnras/stw1728}, \href
  {https://ui.adsabs.harvard.edu/abs/2016MNRAS.462.1307S} {462, 1307}

\bibitem[\protect\citeauthoryear{{Stark}, {Schenker}, {Ellis}, {Robertson},
  {McLure}  \& {Dunlop}}{{Stark} et~al.}{2013}]{stark2013}
{Stark} D.~P.,  {Schenker} M.~A.,  {Ellis} R.,  {Robertson} B.,  {McLure} R.,
  {Dunlop} J.,  2013, \mn@doi [\apj] {10.1088/0004-637X/763/2/129}, \href
  {https://ui.adsabs.harvard.edu/abs/2013ApJ...763..129S} {763, 129}

\bibitem[\protect\citeauthoryear{{Sugimura}, {Matsumoto}, {Hosokawa}, {Hirano}
  \& {Omukai}}{{Sugimura} et~al.}{2020}]{sugimura2020}
{Sugimura} K.,  {Matsumoto} T.,  {Hosokawa} T.,  {Hirano} S.,   {Omukai} K.,
  2020, arXiv e-prints, \href
  {https://ui.adsabs.harvard.edu/abs/2020arXiv200200012S} {p. arXiv:2002.00012}

\bibitem[\protect\citeauthoryear{{Suh}, {Civano}, {Trakhtenbrot}, {Shankar},
  {Hasinger}, {Sanders}  \& {Allevato}}{{Suh} et~al.}{2020}]{suh2020}
{Suh} H.,  {Civano} F.,  {Trakhtenbrot} B.,  {Shankar} F.,  {Hasinger} G.,
  {Sanders} D.~B.,   {Allevato} V.,  2020, \mn@doi [\apj]
  {10.3847/1538-4357/ab5f5f}, \href
  {https://ui.adsabs.harvard.edu/abs/2020ApJ...889...32S} {889, 32}

\bibitem[\protect\citeauthoryear{{Takeo}, {Inayoshi}, {Ohsuga}, {Takahashi}  \&
  {Mineshige}}{{Takeo} et~al.}{2018}]{takeo2018}
{Takeo} E.,  {Inayoshi} K.,  {Ohsuga} K.,  {Takahashi} H.~R.,   {Mineshige} S.,
   2018, \mn@doi [\mnras] {10.1093/mnras/sty264}, \href
  {https://ui.adsabs.harvard.edu/abs/2018MNRAS.476..673T} {476, 673}

\bibitem[\protect\citeauthoryear{{Tamburello}, {Capelo}, {Mayer}, {Bellovary}
  \& {Wadsley}}{{Tamburello} et~al.}{2017}]{tamburello2017}
{Tamburello} V.,  {Capelo} P.~R.,  {Mayer} L.,  {Bellovary} J.~M.,   {Wadsley}
  J.~W.,  2017, \mn@doi [\mnras] {10.1093/mnras/stw2561}, \href
  {https://ui.adsabs.harvard.edu/abs/2017MNRAS.464.2952T} {464, 2952}

\bibitem[\protect\citeauthoryear{{Tamfal}, {Capelo}, {Kazantzidis}, {Mayer},
  {Potter}, {Stadel}  \& {Widrow}}{{Tamfal} et~al.}{2018}]{tamfal2018}
{Tamfal} T.,  {Capelo} P.~R.,  {Kazantzidis} S.,  {Mayer} L.,  {Potter} D.,
  {Stadel} J.,   {Widrow} L.~M.,  2018, \mn@doi [\apjl]
  {10.3847/2041-8213/aada4b}, \href
  {https://ui.adsabs.harvard.edu/abs/2018ApJ...864L..19T} {864, L19}

\bibitem[\protect\citeauthoryear{{Tanaka} \& {Haiman}}{{Tanaka} \&
  {Haiman}}{2009}]{tanaka2009}
{Tanaka} T.,  {Haiman} Z.,  2009, \mn@doi [\apj]
  {10.1088/0004-637X/696/2/1798}, \href
  {https://ui.adsabs.harvard.edu/abs/2009ApJ...696.1798T} {696, 1798}

\bibitem[\protect\citeauthoryear{{Tenneti}, {Di Matteo}, {Croft}, {Garcia}  \&
  {Feng}}{{Tenneti} et~al.}{2018}]{tenneti2018}
{Tenneti} A.,  {Di Matteo} T.,  {Croft} R.,  {Garcia} T.,   {Feng} Y.,  2018,
  \mn@doi [\mnras] {10.1093/mnras/stx2788}, \href
  {https://ui.adsabs.harvard.edu/abs/2018MNRAS.474..597T} {474, 597}

\bibitem[\protect\citeauthoryear{{Tenneti}, {Wilkins}, {Di Matteo}, {Croft}  \&
  {Feng}}{{Tenneti} et~al.}{2019}]{tenneti2019}
{Tenneti} A.,  {Wilkins} S.~M.,  {Di Matteo} T.,  {Croft} R. A.~C.,   {Feng}
  Y.,  2019, \mn@doi [\mnras] {10.1093/mnras/sty3161}, \href
  {https://ui.adsabs.harvard.edu/abs/2019MNRAS.483.1388T} {483, 1388}

\bibitem[\protect\citeauthoryear{{Thomas}, {Dav{\'e}}, {Angl{\'e}s-Alc{\'a}zar}
   \& {Jarvis}}{{Thomas} et~al.}{2019}]{Thomas2019}
{Thomas} N.,  {Dav{\'e}} R.,  {Angl{\'e}s-Alc{\'a}zar} D.,   {Jarvis} M.,
  2019, \mn@doi [\mnras] {10.1093/mnras/stz1703}, \href
  {https://ui.adsabs.harvard.edu/abs/2019MNRAS.487.5764T} {487, 5764}

\bibitem[\protect\citeauthoryear{{Tremmel}, {Karcher}, {Governato},
  {Volonteri}, {Quinn}, {Pontzen}, {Anderson}  \& {Bellovary}}{{Tremmel}
  et~al.}{2017}]{tremmel2017}
{Tremmel} M.,  {Karcher} M.,  {Governato} F.,  {Volonteri} M.,  {Quinn} T.~R.,
  {Pontzen} A.,  {Anderson} L.,   {Bellovary} J.,  2017, \mn@doi [\mnras]
  {10.1093/mnras/stx1160}, \href
  {https://ui.adsabs.harvard.edu/abs/2017MNRAS.470.1121T} {470, 1121}

\bibitem[\protect\citeauthoryear{{Ueda}, {Akiyama}, {Hasinger}, {Miyaji}  \&
  {Watson}}{{Ueda} et~al.}{2014}]{ueda2014}
{Ueda} Y.,  {Akiyama} M.,  {Hasinger} G.,  {Miyaji} T.,   {Watson} M.~G.,
  2014, \mn@doi [\apj] {10.1088/0004-637X/786/2/104}, \href
  {https://ui.adsabs.harvard.edu/abs/2014ApJ...786..104U} {786, 104}

\bibitem[\protect\citeauthoryear{Valiante, Schneider, Salvadori  \&
  Bianchi}{Valiante et~al.}{2011}]{valiante2011}
Valiante R.,  Schneider R.,  Salvadori S.,   Bianchi S.,  2011, Monthly Notices
  of the Royal Astronomical Society, 416, 1916

\bibitem[\protect\citeauthoryear{{Valiante}, {Schneider}, {Maiolino},
  {Salvadori}  \& {Bianchi}}{{Valiante} et~al.}{2012}]{valiante2012}
{Valiante} R.,  {Schneider} R.,  {Maiolino} R.,  {Salvadori} S.,   {Bianchi}
  S.,  2012, \mn@doi [\mnras] {10.1111/j.1745-3933.2012.01345.x}, \href
  {https://ui.adsabs.harvard.edu/abs/2012MNRAS.427L..60V} {427, L60}

\bibitem[\protect\citeauthoryear{Valiante, Schneider, Salvadori  \&
  Gallerani}{Valiante et~al.}{2014}]{valiante2014}
Valiante R.,  Schneider R.,  Salvadori S.,   Gallerani S.,  2014, Monthly
  Notices of the Royal Astronomical Society, 444, 2442

\bibitem[\protect\citeauthoryear{Valiante, Schneider, Volonteri  \&
  Omukai}{Valiante et~al.}{2016}]{valiante2016}
Valiante R.,  Schneider R.,  Volonteri M.,   Omukai K.,  2016, Monthly Notices
  of the Royal Astronomical Society, 457, 3356

\bibitem[\protect\citeauthoryear{{Valiante}, {Agarwal}, {Habouzit}  \&
  {Pezzulli}}{{Valiante} et~al.}{2017}]{valiante2017}
{Valiante} R.,  {Agarwal} B.,  {Habouzit} M.,   {Pezzulli} E.,  2017, \mn@doi
  [\pasa] {10.1017/pasa.2017.25}, \href
  {https://ui.adsabs.harvard.edu/abs/2017PASA...34...31V} {34, e031}

\bibitem[\protect\citeauthoryear{{Valiante}, {Schneider}, {Graziani}  \&
  {Zappacosta}}{{Valiante} et~al.}{2018a}]{valiante2018statistics}
{Valiante} R.,  {Schneider} R.,  {Graziani} L.,   {Zappacosta} L.,  2018a,
  \mn@doi [\mnras] {10.1093/mnras/stx3028}, \href
  {https://ui.adsabs.harvard.edu/abs/2018MNRAS.474.3825V} {474, 3825}

\bibitem[\protect\citeauthoryear{{Valiante}, {Schneider}, {Zappacosta},
  {Graziani}, {Pezzulli}  \& {Volonteri}}{{Valiante}
  et~al.}{2018b}]{valiante2018observability}
{Valiante} R.,  {Schneider} R.,  {Zappacosta} L.,  {Graziani} L.,  {Pezzulli}
  E.,   {Volonteri} M.,  2018b, \mn@doi [\mnras] {10.1093/mnras/sty213}, \href
  {https://ui.adsabs.harvard.edu/abs/2018MNRAS.476..407V} {476, 407}

\bibitem[\protect\citeauthoryear{{Valiante} et~al.,}{{Valiante}
  et~al.}{2020}]{valiante2020}
{Valiante} R.,  et~al., 2020, \mn@doi [\mnras] {10.1093/mnras/staa3395}, \href
  {https://ui.adsabs.harvard.edu/abs/2020MNRAS.500.4095V} {500, 4095}

\bibitem[\protect\citeauthoryear{{Vito} et~al.,}{{Vito}
  et~al.}{2018}]{vito2018}
{Vito} F.,  et~al., 2018, \mn@doi [\mnras] {10.1093/mnras/stx2486}, \href
  {https://ui.adsabs.harvard.edu/abs/2018MNRAS.473.2378V} {473, 2378}

\bibitem[\protect\citeauthoryear{{Vogelsberger} et~al.,}{{Vogelsberger}
  et~al.}{2014}]{vogelsberger2014}
{Vogelsberger} M.,  et~al., 2014, \mn@doi [\mnras] {10.1093/mnras/stu1536},
  \href {https://ui.adsabs.harvard.edu/abs/2014MNRAS.444.1518V} {444, 1518}

\bibitem[\protect\citeauthoryear{Volonteri}{Volonteri}{2010}]{volonteri2010}
Volonteri M.,  2010, The Astronomy and Astrophysics Review, 18, 279

\bibitem[\protect\citeauthoryear{{Volonteri} \& {Natarajan}}{{Volonteri} \&
  {Natarajan}}{2009}]{volonteri2009}
{Volonteri} M.,  {Natarajan} P.,  2009, \mn@doi [\mnras]
  {10.1111/j.1365-2966.2009.15577.x}, \href
  {https://ui.adsabs.harvard.edu/abs/2009MNRAS.400.1911V} {400, 1911}

\bibitem[\protect\citeauthoryear{{Volonteri}, {Lodato}  \&
  {Natarajan}}{{Volonteri} et~al.}{2008}]{volonteri2008}
{Volonteri} M.,  {Lodato} G.,   {Natarajan} P.,  2008, \mn@doi [\mnras]
  {10.1111/j.1365-2966.2007.12589.x}, \href
  {https://ui.adsabs.harvard.edu/abs/2008MNRAS.383.1079V} {383, 1079}

\bibitem[\protect\citeauthoryear{{Volonteri}, {Dubois}, {Pichon}  \&
  {Devriendt}}{{Volonteri} et~al.}{2016}]{volonteri2016}
{Volonteri} M.,  {Dubois} Y.,  {Pichon} C.,   {Devriendt} J.,  2016, \mn@doi
  [\mnras] {10.1093/mnras/stw1123}, \href
  {https://ui.adsabs.harvard.edu/abs/2016MNRAS.460.2979V} {460, 2979}

\bibitem[\protect\citeauthoryear{{Wang} et~al.,}{{Wang}
  et~al.}{2018}]{wang2018}
{Wang} F.,  et~al., 2018, \mn@doi [\apjl] {10.3847/2041-8213/aaf1d2}, \href
  {https://ui.adsabs.harvard.edu/abs/2018ApJ...869L...9W} {869, L9}

\bibitem[\protect\citeauthoryear{{Wang} et~al.,}{{Wang}
  et~al.}{2020}]{wang2020}
{Wang} F.,  et~al., 2020, \mn@doi [\apj] {10.3847/1538-4357/ab8c45}, \href
  {https://ui.adsabs.harvard.edu/abs/2020ApJ...896...23W} {896, 23}

\bibitem[\protect\citeauthoryear{{Wang} et~al.,}{{Wang}
  et~al.}{2021}]{wang2021}
{Wang} F.,  et~al., 2021, \mn@doi [\apjl] {10.3847/2041-8213/abd8c6}, \href
  {https://ui.adsabs.harvard.edu/abs/2021ApJ...907L...1W} {907, L1}

\bibitem[\protect\citeauthoryear{{Weinberger} et~al.,}{{Weinberger}
  et~al.}{2017}]{weinberger2017}
{Weinberger} R.,  et~al., 2017, \mn@doi [\mnras] {10.1093/mnras/stw2944}, \href
  {https://ui.adsabs.harvard.edu/abs/2017MNRAS.465.3291W} {465, 3291}

\bibitem[\protect\citeauthoryear{{Weingartner} \& {Draine}}{{Weingartner} \&
  {Draine}}{2001}]{weingartner2001}
{Weingartner} J.~C.,  {Draine} B.~T.,  2001, \mn@doi [\apj] {10.1086/318651},
  \href {https://ui.adsabs.harvard.edu/abs/2001ApJ...548..296W} {548, 296}

\bibitem[\protect\citeauthoryear{{Willott} et~al.,}{{Willott}
  et~al.}{2010a}]{willott2010a}
{Willott} C.~J.,  et~al., 2010a, \mn@doi [\aj] {10.1088/0004-6256/139/3/906},
  \href {https://ui.adsabs.harvard.edu/abs/2010AJ....139..906W} {139, 906}

\bibitem[\protect\citeauthoryear{{Willott} et~al.,}{{Willott}
  et~al.}{2010b}]{willott2010b}
{Willott} C.~J.,  et~al., 2010b, \mn@doi [\aj] {10.1088/0004-6256/140/2/546},
  \href {https://ui.adsabs.harvard.edu/abs/2010AJ....140..546W} {140, 546}

\bibitem[\protect\citeauthoryear{{Wise}, {Turk}  \& {Abel}}{{Wise}
  et~al.}{2008}]{wise2008}
{Wise} J.~H.,  {Turk} M.~J.,   {Abel} T.,  2008, \mn@doi [\apj]
  {10.1086/588209}, \href
  {https://ui.adsabs.harvard.edu/abs/2008ApJ...682..745W} {682, 745}

\bibitem[\protect\citeauthoryear{{Wise}, {Regan}, {O'Shea}, {Norman}, {Downes}
  \& {Xu}}{{Wise} et~al.}{2019}]{wise2019}
{Wise} J.~H.,  {Regan} J.~A.,  {O'Shea} B.~W.,  {Norman} M.~L.,  {Downes}
  T.~P.,   {Xu} H.,  2019, \mn@doi [\nat] {10.1038/s41586-019-0873-4}, \href
  {https://ui.adsabs.harvard.edu/abs/2019Natur.566...85W} {566, 85}

\bibitem[\protect\citeauthoryear{{Woods} et~al.,}{{Woods}
  et~al.}{2019}]{woods2019}
{Woods} T.~E.,  et~al., 2019, \mn@doi [\pasa] {10.1017/pasa.2019.14}, \href
  {https://ui.adsabs.harvard.edu/abs/2019PASA...36...27W} {36, e027}

\bibitem[\protect\citeauthoryear{{Wu} et~al.,}{{Wu} et~al.}{2015}]{wu2015}
{Wu} X.-B.,  et~al., 2015, \mn@doi [\nat] {10.1038/nature14241}, \href
  {https://ui.adsabs.harvard.edu/abs/2015Natur.518..512W} {518, 512}

\bibitem[\protect\citeauthoryear{{Yang} et~al.,}{{Yang}
  et~al.}{2020}]{yang2020}
{Yang} J.,  et~al., 2020, \mn@doi [\apjl] {10.3847/2041-8213/ab9c26}, \href
  {https://ui.adsabs.harvard.edu/abs/2020ApJ...897L..14Y} {897, L14}

\bibitem[\protect\citeauthoryear{{Yue}, {Ferrara}, {Salvaterra}, {Xu}  \&
  {Chen}}{{Yue} et~al.}{2013}]{bin2013}
{Yue} B.,  {Ferrara} A.,  {Salvaterra} R.,  {Xu} Y.,   {Chen} X.,  2013,
  \mn@doi [\mnras] {10.1093/mnras/stt826}, \href
  {https://ui.adsabs.harvard.edu/abs/2013MNRAS.433.1556Y} {433, 1556}

\bibitem[\protect\citeauthoryear{{Zhang}, {Behroozi}, {Volonteri}, {Silk},
  {Fan}, {Hopkins}, {Yang}  \& {Aird}}{{Zhang} et~al.}{2021}]{zhang2021}
{Zhang} H.,  {Behroozi} P.,  {Volonteri} M.,  {Silk} J.,  {Fan} X.,  {Hopkins}
  P.~F.,  {Yang} J.,   {Aird} J.,  2021, arXiv e-prints, \href
  {https://ui.adsabs.harvard.edu/abs/2021arXiv210510474Z} {p. arXiv:2105.10474}

\bibitem[\protect\citeauthoryear{{Zhu}, {Li}, {Li}, {Maji}, {Yajima},
  {Schneider}  \& {Hernquist}}{{Zhu} et~al.}{2020}]{zhu2020}
{Zhu} Q.,  {Li} Y.,  {Li} Y.,  {Maji} M.,  {Yajima} H.,  {Schneider} R.,
  {Hernquist} L.,  2020, arXiv e-prints, \href
  {https://ui.adsabs.harvard.edu/abs/2020arXiv201201458Z} {p. arXiv:2012.01458}

\bibitem[\protect\citeauthoryear{{de Bennassuti}, {Schneider}, {Valiante}  \&
  {Salvadori}}{{de Bennassuti} et~al.}{2014}]{debennassuti2014}
{de Bennassuti} M.,  {Schneider} R.,  {Valiante} R.,   {Salvadori} S.,  2014,
  \mn@doi [\mnras] {10.1093/mnras/stu1962}, \href
  {https://ui.adsabs.harvard.edu/abs/2014MNRAS.445.3039D} {445, 3039}

\bibitem[\protect\citeauthoryear{{de Bennassuti}, {Salvadori}, {Schneider},
  {Valiante}  \& {Omukai}}{{de Bennassuti} et~al.}{2017}]{debennassuti2017}
{de Bennassuti} M.,  {Salvadori} S.,  {Schneider} R.,  {Valiante} R.,
  {Omukai} K.,  2017, \mn@doi [\mnras] {10.1093/mnras/stw2687}, \href
  {https://ui.adsabs.harvard.edu/abs/2017MNRAS.465..926D} {465, 926}

\makeatother
\end{thebibliography}


\bsp	
\label{lastpage}
\end{document}